\documentclass[12pt]{article}
\pdfoutput=1

\usepackage[dvipsnames]{xcolor} \usepackage[a4paper]{geometry}
\usepackage[english]{babel}
\usepackage{cite,enumerate,enumitem,booktabs,float,graphicx}

\usepackage[T1]{fontenc}
\usepackage[utf8]{inputenc}
\usepackage{lmodern}

\usepackage{url}

\makeatletter
\g@addto@macro\bfseries{\boldmath}
\makeatother

\usepackage{bm,amsmath,amssymb,tensor,mathtools,braket}
\numberwithin{equation}{section}

\usepackage{tikz}
\usetikzlibrary{cd}

\usepackage{caption}
\usepackage{subcaption}

\usepackage[pdftex]{hyperref}
\definecolor{dark-blue}{rgb}{0.15,0.15,0.4}
\hypersetup{
pdftitle   = {The Geometry of Gravitational Radiation},
  pdfkeywords = {Carrollian geometry, celestial holography, Carrollian holography, holographic reconstruction, asymptotically flat spacetimes, holographic renormalisation},
  pdfauthor  = {Jelle Hartong, Emil Have, Vijay Nenmeli, Gerben Oling},
  colorlinks,
  linkcolor={dark-blue},
  citecolor={blue},
  linktoc=page
}

\newcommand{\os}[2]{\overset{(#1)}{#2}{}}

\title{\textbf{
    The Geometry of Gravitational Radiation
  }
}
\date{}
\author{Jelle Hartong}

\usepackage{authblk}

\author{Jelle Hartong}
\affil{%
  School of Mathematics and Maxwell Institute for Mathematical Sciences,
  \protect\\
  University of Edinburgh,
  Peter Guthrie Tait Road,
  Edinburgh EH9 3FD, UK
}
\date{}
\begin{document}

\maketitle
\thispagestyle{empty}

\begin{abstract}
    We consider 4-dimensional asymptotically flat vacuum spacetimes near future null infinity endowed with the most general allowable Carroll geometry. We show that the near-boundary radial expansion (to a certain order) can be organised in terms of connections that can be obtained by gauging the conformal Carroll algebra. The only non-vanishing curvatures in this gauging procedure are those that are associated with the special conformal generators and we will refer to these as the $K$-curvatures. The vanishing of these $K$-curvatures defines an asymptotic vacuum spacetime and we use this to construct a boundary (i.e. Carroll) covariant expression for the vacuum (soft) shear in terms of two boundary Carroll scalar fields. The $K$-curvatures transform in a hierarchical fashion into one another under Carroll boosts. This leads to a classification of 4 types of spacetimes: vacuum, strongly and weakly non-radiative, and radiative spacetimes. It is shown that the $K$-curvatures correspond to 5 of the 10 Weyl tensor components at leading order in their $1/r$ expansion. We furthermore observe that one of the $K$-curvatures is equal to the recently found Carroll boost anomaly. Finally, we show that the Bondi loss equations for the boundary energy-momentum-news complex can be cast into a form involving another energy-momentum tensor with vanishing energy flux that is traceless and whose non-conservation is entirely captured by the $K$-curvatures. The BMS currents can be obtained by contracting this latter energy-momentum tensor with a Carroll conformal Killing vector.
\end{abstract}

    \newpage
\tableofcontents

\section{Introduction}%
\label{sec:intro}

An important aspect of gravitational wave physics is described by the Bondi loss equations. These equations govern the loss of mass and angular momentum due to the emission of gravitational waves. The Bondi loss equations are part of the Bondi--Sachs formalism~\cite{Bondi:1962px,Sachs:1962wk,Sachs:1962zza,Newman:1961qr}, and feature when we study the metric and the Einstein equations for 4-dimensional asymptotically flat spacetimes near future null infinity $\mathcal{I}^+$.

Famously, Bondi, van der Burg, Metzner and Sachs showed that one can construct an infinite number of charges that form the BMS algebra whose time dependence is governed by the Bondi loss equations \cite{Bondi:1962px,Sachs:1962wk,Sachs:1962zza}. In the original BMS work the algebra consisted of supertranslations and the Lorentz algebra $\mathfrak{so}(1,3)$.
This was later enlarged to include superrotations to what is called the extended BMS algebra by Barnich and Troessaert \cite{Barnich:2009se,Barnich:2010eb,Barnich:2011mi,Barnich:2010ojg}.

It is known that future null infinity is a (conformal) Carroll manifold \cite{Duval:2014lpa,Duval:2014uva}. This observation was used to define Carroll-covariant descriptions of 3-dimensional asymptotically flat 
spacetimes and subsequently to use this to define a boundary energy-momentum tensor in \cite{Hartong:2015usd}. Going to 4 dimensions it is then natural to suggest that we can define a boundary energy-momentum tensor by using the near-boundary expansion of the metric and by adding an appropriate set of boundary terms to the Einstein--Hilbert action, and that this boundary energy-momentum tensor contains the Bondi mass and angular momentum aspects. 
The works \cite{Hartong:2025jpp,Hartong:2026rbr} make this statement precise. In section \ref{subsec:reviewEMTnews} we review this construction. Apart from a boundary energy-momentum tensor there is also news which is the response to varying the shear. Together we refer to this as the energy-momentum-news complex. The Bondi loss equations follow from the diffeomorphism Ward identity that is obeyed by this energy-momentum-news complex. It was also shown in~\cite{Fiorucci:2025twa} that the Bondi loss equations, in standard Bondi--Sachs coordinates, can be written in the form of a diffeomorphism Ward identity of some Carrollian field theory.
    However, this work does not contain a bulk derivation of a boundary energy-momentum tensor.

The essential ingredients of this construction involve Carroll geometric data whose responses define the boundary energy-momentum tensor. However the presence of gravitational radiation means that the boundary description also contains the shear tensor. It is well-known that the shear contains both vacuum (soft) and radiative (hard) parts. It was however not known how to make this split in a Carroll covariant manner. In the present paper we solve this problem.

The gauging of the Carroll algebra can be used to define Carroll geometry \cite{Hartong:2015xda}. Future null infinity is a conformal Carroll manifold and so it is natural to gauge the conformal Carroll algebra (which is isomorphic to the Poincar\'e algebra). We will do this in section \ref{sec:gauging} and show that this can be used to account for the boundary geometry and the shear, as well as certain subleading orders in the near-boundary expansion of the metric (see section \ref{sec:confcarbulk}). Certain important properties of the shear, from the gauging perspective, are captured by curvatures associated with the special conformal generators and we will call these $K$-curvatures. These $K$-curvatures vanish if and only if the shear is equal to the vacuum shear. We use this combined with our covariant boundary description to construct an explicit and manifest Carroll-covariant expression for the vacuum (soft) shear.

Covariant descriptions of future null infinity and the physics of radiation goes back to the classical works of Geroch and Ashtekar \cite{Geroch:1977big,Ashtekar:1978zz,Ashtekar:2023wfn}. Also the Newman--Penrose (NP) formalism \cite{Newman:1961qr} can be used to describe any choice of boundary geometry. The Carroll approach in a way combines these two ways of thinking\footnote{It uses nullbeine in the bulk just like in the NP formalism but it does not use other (spatial) vielbeine. It also employs Weyl covariant derivatives which are similar to certain derivative operators used in the NP formalism. It treats the boundary covariantly like in the works of Geroch and Ashtekar but it does so in the language of non-Lorentzian geometry.} and it adds to this a more central role for the Carroll boosts enabling us to use the full framework of non-Lorentzian geometries. It is this that allowed us to find explicit covariant expressions of the vacuum shear whose associated news tensor is vacuum news which in turn is related to the (trace-free part of the) Geroch tensor\footnote{The shear transforms with an inhomogeneous term. This term is the transformation of the Carroll-covariant shear tensor under the Carroll boosts. The vacuum shear transforms with the same shift term so that $C_{\mu\nu}-C^{\text{vac}}_{\mu\nu}$ is Carroll boost invariant. When gauge fixing the boundary geometry to the standard Bondi coordinates the Carroll boosts are related to transformations of retarded time.}.

Equipped with this covariant notion of vacuum shear we can construct $C_{\mu\nu}-C^{\text{vac}}_{\mu\nu}$ which describes non-vacuum shear. The above mentioned $K$-curvatures only depend on $C_{\mu\nu}-C^{\text{vac}}_{\mu\nu}$. In section \ref{sec:KcurvBondi} we show that the Bondi loss equations can be rewritten in terms of another boundary energy-momentum tensor such that its non-conservation is described by these $K$-curvatures. It is with respect to this other boundary energy-momentum tensor that the BMS currents are most easily defined. 
This new boundary energy-momentum tensor\footnote{The current level of understanding of this alternative energy-momentum tensor is via a rewriting of the Bondi loss equations. Understanding it as a boundary energy-momentum tensor that results from responses to certain variations of boundary sources is something that is currently a work in progress \cite{Hartong:2026WIP}.} for the case of standard Bondi--Sachs coordinates is expected to be related to what in the literature is sometimes referred to as the `good prescription' for the BMS charges \cite{Donnay:2023mrd}.

\subsubsection*{Note added}

This work was first presented at the Galileo Galilei Institute in June 2025 (as part of the programme: From Asymptotic Symmetries to Flat Holography: Theoretical Aspects and Observable Consequences) \cite{HartongGGI2025} as well as at the Corfu Summer Institute's Workshop on Quantum Gravity and Strings in September 2025. During the final stages of writing this manuscript, the preprint \cite{Fiorucci:2026nyg} appeared which has some overlap with this preprint.

\section{Review of Carroll-covariant Bondi--Sachs gauge}\label{sec:reviewCarcovBSgauge}

In this section we review the expansion of the metric near future null infinity. We will allow for a general Carroll geometry at $\mathcal{I}^+$ and present the asymptotic solution in Carroll-covariant Bondi--Sachs gauge. We will also discuss the role of bulk diffeomorphisms and their action on the boundary.

\subsection{Metric near $\mathcal{I}^+$ and gauge choice}\label{subsec:Carcovgauge}

In this paper we only consider 4-dimensional vacuum spacetimes. 
We know from \cite{Hartong:2025jpp,Hartong:2026rbr} that the metric near future null infinity in Carroll-covariant Bondi--Sachs (BS) gauge, which in 4 dimensions reads
\begin{equation}\label{eq:CarcovBSgauge}
    g_{rr}=0\,,\qquad \Gamma^\rho_{rr}=0\,,\qquad \Gamma^\rho_{\rho r}=2r^{-1}\,,
\end{equation}
can be written as
\begin{equation}
    ds^2=-2e^\beta\tau_\mu dx^\mu dr+g_{\mu\nu}dx^\mu dx^\nu\,,
\end{equation}
where 
\begin{equation}\label{eq:gtoPi}
    g_{\mu\nu}=-e^{2\beta}S\tau_\mu\tau_\nu+\Pi_{\mu\nu}\,.
\end{equation}
A bulk point has coordinates $(r,x^\mu)$ where $x^\mu$ corresponds to a point on the conformal boundary $\mathcal{I}^+$. The fields $\beta, S, \Pi_{\mu\nu}$ are bulk fields and so depend on the radial coordinate $r$. The 1-form $\tau_\mu(x)$ is a boundary object that should be viewed as fixed and arbitrary. The tensor $\Pi_{\mu\nu}$ has signature $(0,1,1)$. For the inverse metric we have
\begin{equation}
    g^{rr}=S\,,\qquad g^{r\mu}=U^\mu\,,\qquad g^{\mu\nu}=\Pi^{\mu\nu}\,,
\end{equation}
where $\Pi^{\mu\nu}$ has signature $(0,1,1)$. We have the relations
\begin{align}
    &\Pi^{\mu\rho}\Pi_{\rho\nu}-U^\mu V_\nu=\delta^\mu_\nu\,,\\
    & U^\mu V_\mu=-1\,,\qquad U^\mu\Pi_{\mu\nu}=0\,,\qquad V_\mu\Pi^{\mu\nu}=0\,,\label{eq:propsUVPiinvPi}
\end{align}
where $V_\mu=e^\beta\tau_\mu$.

The bulk fields are assumed to expand such that\footnote{Using \eqref{eq:gtoPi} this leads to an expansion for $\Pi_{\mu\nu}$ which similar to $g_{\mu\nu}$ takes the form 
\begin{equation}
    \Pi_{\mu\nu} = r^2 h_{\mu\nu}+r\os{-1}{\Pi}_{\mu\nu}+\os{0}{\Pi}_{\mu\nu}+r^{-1}\log r \os{1,1}{\Pi}_{\mu\nu}+r^{-1}\os{1}{\Pi}_{\mu\nu}+\cdots\,.
\end{equation}} 
\begin{eqnarray}
    \beta & = & r^{-2}\os{2}{\beta}+\cdots\,,\label{eq:betaexp}\\
    S & = & rK+\os{0}{S}+r^{-1}\os{1}{S}+\cdots\,,\label{eq:Sexp}\\
    g_{\mu\nu} & = & r^2 h_{\mu\nu}+r\os{-1}{g}_{\mu\nu}+\os{0}{g}_{\mu\nu}+r^{-1}\log r \os{1,1}{g}_{\mu\nu}+r^{-1}\os{1}{g}_{\mu\nu}+\cdots\,,\label{eq:gexp}
\end{eqnarray}
where the dots denote terms that are subleading and where $K$ is defined further below. They are assumed to be of the form $r^{-n}\log^m r$ for suitable $n,m$. For the inverse objects $U^\mu$ and $\Pi^{\mu\nu}$ we have at leading order 
\begin{eqnarray}
    U^\mu & = & v^\mu+\mathcal{O}(r^{-1})\,,\\
    \Pi^{\mu\nu} & = & r^{-2}h^{\mu\nu}+\mathcal{O}(r^{-3})\,,
\end{eqnarray}
where $v^\mu$ and $h^{\mu\nu}$ can be constructed from $\tau_\mu$ and $h_{\mu\nu}$ by using 
\begin{align}
    &h^{\mu\rho}h_{\rho\nu}-v^\mu \tau_\nu=\delta^\mu_\nu\,,\label{eq:completeness}\\
    & v^\mu \tau_\mu=-1\,,\qquad v^\mu h_{\mu\nu}=0\,,\qquad \tau_\mu h^{\mu\nu}=0\,.\label{eq:vinvh}
\end{align}

We will often use the notation
\begin{equation}
    h^\mu_\nu=h^{\mu\rho}h_{\nu\rho}\,.
\end{equation}
A tensor is called spatial if all of its contractions with $v^\mu$ and $\tau_\mu$ are zero. For such tensors we raise and lower indices with $h^{\mu\nu}$ and $h_{\mu\nu}$.

The vacuum Einstein equations at leading order in the radial expansion do not allow us to have an arbitrary Carroll geometry on $\mathcal{I}^+$ because they enforce the following constraint
\begin{equation}\label{eq:Kconstraint}
    K_{\mu\nu}=\frac{1}{2}Kh_{\mu\nu}\,,
\end{equation}
where we defined 
\begin{equation}
    K_{\mu\nu}=-\frac{1}{2}\mathcal{L}_v h_{\mu\nu}\,,\qquad\text{and $K=h^{\mu\nu}K_{\mu\nu}$}\,.
\end{equation}
In here $\mathcal{L}_v$ denotes the Lie derivative along $v^\mu$. In \cite{Hartong:2026rbr} it is shown that this constraint is equivalent to writing 
\begin{equation}\label{eq:metrich}
    h_{\mu\nu}dx^\mu dx^\nu=e^{2\varphi}dX^a dX^a\,,
\end{equation}
where $\varphi$ and $X^a$ with $a=1,2$ are arbitrary scalar fields. It follows that
\begin{equation}\label{eq:Kinvarphi}
    K=-2\mathcal{L}_v\varphi\,.
\end{equation}

The leading order vacuum Einstein equations furthermore tell us that $\beta$ has to be at least order $r^{-2}$ which is why we do not write an order $r^{-1}$ term in \eqref{eq:betaexp} and that $S=rK+\mathcal{O}(1)$. Finally, these equations also tell us that 
\begin{equation}\label{eq:g-1}
    \os{-1}{g}_{\mu\nu} = -K\tau_\mu\tau_\nu-\tau_\mu a_\nu-\tau_\nu a_\mu+C_{\mu\nu}\,,
\end{equation}
where we defined
\begin{equation}\label{eq:defa}
    a_\mu=\mathcal{L}_v\tau_\mu\,,
\end{equation}
and where $C_{\mu\nu}$ is spatial and symmetric trace-free (STF), but otherwise arbitrary. This is the shear tensor.

\subsection{Gauge transformations}

The bulk diffeomorphisms that preserve the Carroll-covariant BS gauge are generated by vector fields  $\xi^M=(\xi^r,\xi^\mu)$ of the form
\begin{eqnarray}
    \xi^r & = & r\Lambda_D-\frac{1}{2}e^{-1}\partial_\mu\left(e \lambda^\mu\right)+\mathcal{O}(r^{-1})\,,\\
    \xi^\mu & = & \chi^\mu+r^{-1}\lambda^\mu+\mathcal{O}(r^{-2})\,,
\end{eqnarray}
where $\lambda^\mu=h^{\mu\nu}\lambda_\nu$ and $v^\mu\lambda_\mu=0$. In here we defined $e=\text{det}\,(\tau_\mu\,,e^a_\mu)$.
These transformations act on the boundary fields $\tau_\mu$, $h_{\mu\nu}$ and $C_{\mu\nu}$ as 
\begin{eqnarray}
    \delta\tau_\mu & = & \mathcal{L}_\chi\tau_\mu+\Lambda_D\tau_\mu+\lambda_\mu\,,\label{eq:tottrafotau}\\
    \delta h_{\mu\nu} & = & \mathcal{L}_\chi h_{\mu\nu}+2\Lambda_D h_{\mu\nu}\,,\label{eq:tottrafoh}\\
    \delta C_{\mu\nu} & = & \mathcal{L}_\chi C_{\mu\nu}+\Lambda_D C_{\mu\nu}+\Delta_{\mu\nu}\,,\label{eq:trafoshear}
\end{eqnarray}
where $\lambda_\mu$ is any spatial 1-form, and where $\Delta_{\mu\nu}$ is spatial, STF and given by
\begin{equation}\label{eq:Delta}
    \Delta_{\mu\nu}=2h^\rho_{\langle\mu}h^\sigma_{\nu\rangle}\left(\mathcal{D}_\rho +a_\rho\right)\lambda_{\sigma}\,.
\end{equation}
Angle brackets are used to denote the STF part.
For the inverse objects $v^\mu$ and $h^{\mu\nu}$ we have the transformations
\begin{eqnarray}
    \delta v^\mu & = & \mathcal{L}_\chi v^\mu-\Lambda_D v^\mu\,,\label{eq:trafov}\\
    \delta h^{\mu\nu} & = & \mathcal{L}_\chi h^{\mu\nu}-2\Lambda_D h^{\mu\nu}+v^\mu\lambda^\nu+v^\nu\lambda^\mu\,.\label{eq:trafoinvh}
\end{eqnarray}
These transformations correspond to boundary diffeomorphisms generated by $\chi^\mu$, boundary Weyl transformation ($\Lambda_D$) and Carroll boosts ($\lambda_\mu$). A BMS$_4$ transformation is one for which
\begin{equation}
    \delta\tau_\mu=0\,,\qquad\delta h_{\mu\nu}=0\,.
\end{equation}

The boundary Carroll metric $h_{\mu\nu}$ must be of the form \eqref{eq:metrich}.
The fields $\varphi$ and $X^a$ suffer from an ambiguity which is a transformation that leaves $h_{\mu\nu}$ invariant. Infinitesimally this transformation is given by
\begin{eqnarray}
    \delta_{\text{am}} X^a & = & -H^a(X)\,,\label{eq:confisom1}\\
    \delta_{\text{am}}\varphi & = & \frac{1}{2}\frac{\partial H^a}{\partial X^a}\,,\label{eq:confisom2}
\end{eqnarray}
where $H^a(X)$ obeys the Cauchy--Riemann differential equations which can be written as
\begin{equation}
    \delta_{ac}\frac{\partial H^c}{\partial X^b}+\delta_{bc}\frac{\partial H^c}{\partial X^a}-\delta_{ab}\frac{\partial H^c}{\partial X^c}=0\,.
\end{equation}
We then obtain that $\delta_{\text{am}} h_{\mu\nu}=0$. 
If we define $Z=\frac{1}{\sqrt{2}}\left(X^1+iX^2\right)$ then the ambiguity can be written (as a finite transformation) as
\begin{equation}
    h_{\mu\nu}dx^\mu dx^\nu=2e^{2\varphi(x)}dZd\bar Z=2e^{2\phi(x)}dWd\bar W\,,
\end{equation}
where 
\begin{eqnarray}
    W & = & f(Z)\,,\\
    e^{2\phi(x)}f'(Z)\bar f'(\bar Z) & = & e^{2\varphi(x)}\,.\label{eq:varphiprime}
\end{eqnarray}
In other words the ambiguity is a conformal isometry that acts on the $X^a$ field space. 

The total gauge transformation of $h_{\mu\nu}$ and the above ambiguity means that the total gauge transformations of $\varphi$ and $X^a$ are 
\begin{eqnarray}
    \delta X^a & = & \mathcal{L}_\chi X^a-H^a(X)\,,\\
    \delta\varphi & = & \mathcal{L}_\chi\varphi+\Lambda_D+\frac{1}{2}\partial_a H^a(X)\,.\label{eq:totgaugetrafovarphi}
\end{eqnarray}

\subsection{Boundary covariant derivatives}\label{subsec:bdrygeom}

In order to solve the Einstein equations it is very convenient to use  an affine connection on the boundary that allows us to define covariant derivatives. By affine connection we mean a connection that transforms appropriately under diffeomorphisms. We will allow it transform non-trivially with respect to Carroll boosts. 

We will denote our choice of boundary connection by $\mathcal{C}^\rho_{\mu\nu}$ and take it to be
\begin{equation}\label{eq:conn}
    \mathcal{C}^\rho_{\mu\nu}=-\frac{1}{2}v^\rho\left(\partial_\mu\tau_\nu+\partial_\nu\tau_\mu+a_\mu\tau_\nu+a_\nu\tau_\mu\right)+\frac{1}{2}h^{\rho\sigma}\left(\partial_\mu h_{\nu\sigma}+\partial_\nu h_{\mu\sigma}-\partial_\sigma h_{\mu\nu}\right)\,.
\end{equation}
The associated covariant derivative will be denoted by $\mathcal{D}_\mu$ whose curvature tensor we denote by $\mathcal{R}_{\mu\nu\rho}{}^\sigma$.
We refer to appendix \ref{app:bdryRiem} for properties of the boundary curvature tensor. This connection has the following properties:
\begin{equation}\label{eq:propconn}
    \begin{array}{rl}
       \text{no torsion:}  & \mathcal{C}^\rho_{[\mu\nu]}=0\,, \\
       \text{volume form compatible:}  & \mathcal{C}^\rho_{\rho\nu}=e^{-1}\partial_\nu e\,,
    \end{array}
\end{equation}
where we remind the reader that $e=\text{det}\,(\tau_\mu\,, e^a_\mu)$ with $e^a_\mu$ for $a=1,2$ spatial vielbeine in terms of which we can write the metric $h_{\mu\nu}=\delta_{ab}e^a_\mu e^b_\nu$. Due to the signature of $h_{\mu\nu}$ this is always possible. For example using \eqref{eq:metrich} we can take $e^a_\mu dx^\mu=e^\varphi dX^a$.

The Carroll metric data is $h_{\mu\nu}$ and $v^\mu$ and the connection is not Carroll metric compatible. Instead we have 
\begin{eqnarray}
    \mathcal{D}_\rho h_{\mu\nu} & = & -\frac{1}{2}K\left(\tau_\mu h_{\nu\rho}+\tau_\nu h_{\mu\rho}\right)\,,\label{eq:covDh}\\
    \mathcal{D}_\mu v^\nu & = & -\frac{1}{2}K h_\mu^\nu\,,\label{eq:covDv}
\end{eqnarray}
so that the non-metric compatibility is captured by $K$. The conditions we imposed thus far fix the connection to be \eqref{eq:conn} up to a term of the form $v^\rho e^a_\mu e^b_\nu X_{ab}$ where $X_{ab}$ is symmetric. We will take this term to be zero or what is the same we will demand that 
\begin{equation}
    h^\rho_{(\mu} h^\sigma_{\nu)} \mathcal{D}_\rho\tau_\sigma=0\,.
\end{equation}
The connection \eqref{eq:conn} is such that we have
\begin{eqnarray}
    \mathcal{D}_\mu\tau_\nu & = & \frac{1}{2}F_{\mu\nu}-\tau_\mu a_\nu\,,\label{eq:covDtau}\\
    \mathcal{D}_\rho h^{\mu\nu} & = & -\tau_\rho\left(v^\mu a^\nu+v^\nu a^\mu\right)-\frac{1}{2}v^\mu F^\nu{}_\rho-\frac{1}{2}v^\nu F^\mu{}_\rho\,,\label{eq:covDinvh}
\end{eqnarray}
where we defined 
\begin{equation}\label{eq:defF}
    F_{\mu\nu}=h^\rho_\mu h^\sigma_\nu\left(\partial_\rho\tau_\sigma-\partial_\sigma\tau_\rho\right)\,.
\end{equation}
A useful property to note is
\begin{equation}
    \mathcal{D}_\rho h^{\rho\nu}=a^\nu\,.
\end{equation}

The antisymmetric tensor $F_{\mu\nu}$ measures the extent to which $\tau_\mu$ is not hypersurface orthogonal as the vanishing of $F_{\mu\nu}$ is equivalent to the Frobenius condition $\tau\wedge d\tau=0$. It is useful to note that in form notation
\begin{equation}\label{eq:dtau}
    d\tau=a\wedge \tau+F\,.
\end{equation}

The choice to have zero torsion means that we cannot have Carroll metric compatibility as well. There is a trade off here. We chose to work with a connection that has no torsion as this makes the Bianchi identities for the Riemann curvature tensor easier and it makes covariantisation (in the sense of replacing ordinary derivatives with covariant derivatives) easier. Using completeness \eqref{eq:completeness} we can always decompose any tensor into purely spatial tensors and lower rank spatial tensors/scalars multiplying $\tau_\mu$ or $v^\mu$. Hence, we only need to define covariant derivatives that act on spatial tensors. All other derivatives can be written in terms of $\mathcal{L}_v$ acting on spatial tensors/scalars and as spatial tensors/scalars multiplying boundary objects such as $K$, $a_\mu$ and $F_{\mu\nu}$. This is because $\mathcal{D}_\mu v^\nu$ and $\mathcal{D}_\mu\tau_\nu$ can be written in terms of exterior derivatives and Lie derivatives along $v^\mu$ and because $v^\rho\mathcal{D}_\rho$ acting on a spatial tensor can always be written as $\mathcal{L}_v$ acting on that same spatial tensor plus terms involving $K$ times the spatial tensor in question.

\subsection{Near-boundary expansion}\label{subsec:nearbdryexp}

The near-boundary expansion is taken to be of the form \eqref{eq:betaexp}, \eqref{eq:Sexp} and \eqref{eq:gexp} where $h_{\mu\nu}$ obeys the constraint \eqref{eq:Kconstraint}. Solving the Einstein equations beyond leading order in the radial expansion we find \cite{Hartong:2025jpp,Hartong:2026rbr},
\begin{equation}
    \os{2}{\beta} = \frac{1}{16}\left(F^2-C^2\right)\,,
\end{equation}
and
\begin{eqnarray}
    \os{0}{g}_{\mu\nu} & = & -\left(\os{0}{S}-a^2\right)\tau_\mu\tau_\nu-\tau_\mu \os{0}{P}_\nu-\tau_\nu \os{0}{P}_\mu+h^\rho_\mu h^\sigma_\nu\os{0}{\Pi}_{\rho\sigma}\,,\label{eq:g0}\\
    \os{1,1}{g}_{\mu\nu} & = &  \frac{2}{3}\tau_\mu h^{\rho\sigma}\mathcal{D}{}_\rho D_{\sigma\nu}+\frac{2}{3}\tau_\nu h^{\rho\sigma}\mathcal{D}{}_\rho D_{\sigma\mu}\,,\label{eq:g1,1}
\end{eqnarray}
where $a^2=h^{\mu\nu}a_\mu a_\nu$ and where 
\begin{eqnarray}
    \os{0}{S} & = &  \frac{1}{2}\mathcal{R}+\frac{3}{2}\mathcal{D}_\mu a^\mu\,,\label{eq:S0}\\
    \os{0}{P}_\mu & = &  -\frac{1}{2}\left(\mathcal{D}_\rho-2a_\rho\right)C^\rho{}_\mu+\frac{1}{2}\mathcal{D}_\rho F^\rho{}_\mu\,,\label{eq:P0}\\
    h^\rho_\mu h^\sigma_\nu\os{0}{\Pi}_{\rho\sigma} & = & D_{\mu\nu}+\frac{1}{2}F_{\mu\rho} C^\rho{}_\nu+\frac{1}{4}h_{\mu\nu}C^2\,,
\end{eqnarray}where $\mathcal{R}$ is the boundary Ricci scalar defined in 
\eqref{eq:StoRicscalar} and using \eqref{eq:metrich} it can be shown to be given by \eqref{eq:Ricscalar}. Furthermore, $D_{\mu\nu}$ is any spatial and STF tensor obeying
\begin{equation}
    \mathcal{L}_v D_{\mu\nu}=0\,.
\end{equation}

At order $r^{-1}$ in the expansion of $g_{\mu\nu}$ we find certain components of the boundary energy-momentum tensor (EMT). We use the decomposition
\begin{equation}
    \os{1}{g}_{\mu\nu}=-\left(\os{1}{S}+2K\,\os{2}{\beta}-v^\rho v^\sigma\os{1}{\Pi}_{\rho\sigma}\right)\tau_\mu\tau_\nu-2\tau_{(\mu}h^\rho_{\nu)} v^\sigma \os{1}{\Pi}_{\rho\sigma}+h^\rho_\mu h^\sigma_\nu \os{1}{\Pi}_{\rho\sigma}\,,
\end{equation}
where 
\begin{align}
\begin{split}
    v^\rho v^\sigma \os{1}{\Pi}_{\rho\sigma} &= \, 2v^\rho a^\sigma \os{0}{\Pi}_{\rho\sigma}-a^\rho a^\sigma C_{\rho\sigma}\,,\\
    h^{\rho\sigma} \os{1}{\Pi}_{\rho\sigma}&=  h^{\mu\rho}h^{\nu\sigma}C_{\mu\nu}D_{\rho\sigma}\,.
\end{split}
\end{align}
In the expression for $\os{1}{g}_{\mu\nu}$ the objects $\os{1}{S}$ and $P^\rho_{\mu} v^\sigma \os{1}{\Pi}_{\rho\sigma}$ are free (except that they have to obey a PDE which is essentially the Bondi loss equation) and form part of the boundary EMT. Finally, the object
$h^\rho_{\langle \mu} h^\sigma_{\nu\rangle} \os{1}{\Pi}_{\rho\sigma}$ will be of no concern to us as it does not affect the boundary EMT. It is a solution to an evolution equation of the form 
\begin{equation}
    \mathcal{L}_v\left(h^\kappa_{\langle \rho} h^\lambda_{\sigma\rangle} \os{1}{\Pi}_{\kappa\lambda}\right)=\cdots\,,
\end{equation}
where the dots are known terms that contain lower order coefficients in the expansion of $g_{\mu\nu}$.

\section{Gauging the conformal Carroll algebra}\label{sec:gauging}

Section 2 provided a geometric perspective on the boundary and the near boundary expansion in terms of (conformal) Carroll geometry. In this section we will show that there is an algebraic formulation of the near boundary expansion (at least for the first few orders in the radial expansion). This algebraic perspective has to do with the gauging procedure for certain spacetime algebras. It will turn out that for our purposes the conformal Carroll algebra provides the relevant algebraic structure. 

This algebra has been gauged in \cite{Bergshoeff:2024ilz} in the context of Carroll gravity. Here we will have a different application in mind as we will gauge the conformal Carroll algebra with the objective to reconstruct (part of) the asymptotic solution space discussed in the previous section. This way of thinking was also adopted in \cite{Herfray:2020rvq,Herfray:2021xyp,Herfray:2021qmp}. The main difference from that body of work that is relevant for this paper is that we will not fix boundary coordinates. 

\subsection{The conformal Carroll algebra}

The conformal Carroll algebra has generators $H, P_a, B_a, J_{ab}, D, K, K_a$ where $a=1,\cdots,d$. On a $(d+1)$-dimensional flat Carroll manifold these have the following interpretation: time translations ($H$), space translations ($P_a$), Carroll boosts ($B_a$), rotations ($J_{ab}=-J_{ba}$), scale transformations ($D$), temporal special conformal transformations ($K$) and spatial special conformal transformations ($K_a$). The conformal Carroll algebra is 
\begin{equation}
    \begin{array}{rclrclrcl}
     \left[P_a\,,B_b\right] & = & \delta_{ab}H\,,&\quad\left[D\,, H\right] & = & -H\,,&\quad\left[D\,, P_a\right] & = & -P_a\,,\\
     \left[D\,, K\right] & = & K\,,&\quad\left[D\,, K_a\right] & = & K_a\,,&\quad \left[K\,, P_a\right] & = & -B_a\,,\\
      \left[K_a\,, H\right] & = & -B_a\,,&\quad\left[K_a\,, P_b\right] & = & -\delta_{ab}D+J_{ab}\,,&\quad  \left[K_a\,, B_b\right] & = & -\delta_{ab}K\,,
\end{array}
\end{equation}
to which we add the usual brackets with the rotations
\begin{eqnarray}
    \left[J_{ab}\,,J_{cd}\right] & = & \delta_{ac}J_{bd}-\delta_{bc}J_{ad}-\delta_{ad}J_{bc}+\delta_{bd}J_{ac}\,,\\
    \left[X_a\,,J_{bc}\right] & = & -\delta_{ab}X_c+\delta_{ac}X_b\,,
\end{eqnarray}
for $X_a=P_a, B_a, K_a$. This agrees with equation (2.13) of \cite{Afshar:2024llh}. We will start our discussion for general dimension $d$ but from section \ref{subsec:curvcon1} onward we will choose $d=2$.

\subsection{Connections, curvatures, and gauge transformations}

We define the Lie algebra-valued connection 
\begin{equation}\label{eq:Liealgebraconnection}
    A_\mu = H\tau_\mu+P_a e^a_\mu+B_a \omega_\mu{}^a+\frac{1}{2}J_{ab}\omega_\mu{}^{ab}+D b_\mu+Kf_\mu+K_a f_\mu{}^a\,,
\end{equation}
where the rotation connection $\omega_\mu{}^{ab}$ is antisymmetric in $a$ and $b$.
The objects of interest in the gauging process are the curvature
\begin{equation}
    F_{\mu\nu}=\partial_\mu A_\nu-\partial_\nu A_\mu+\left[A_\mu\,,A_\nu\right]\,,
\end{equation}
and the gauge transformations\footnote{We have split the conformal Carroll algebra into the subspaces spanned by $\mathfrak{m}=\text{span}\{H,P_a\}$ and by $\mathfrak{h}=\text{span}\{B_a, J_{ab}, D, K, K_a\}$. This is like a coset split of the conformal Carroll algebra with respect to the subalgebra spanned by $\mathfrak{h}$. The gauging procedure is related to the Cartan geometry modelled on the homogeneous space $G/H$ where the Lie algebra of $G$ is the conformal Carroll algebra and the Lie algebra of $H$ is $\mathfrak{h}$. We refer to \cite{Figueroa-OFarrill:2022mcy} for more details on the relation between the gauging procedure and Cartan geometry. Note that we are not working with the adjoint gauge transformations with respect to the whole conformal Carroll algebra, only with respect to $\mathfrak{h}$ and that we assume that $A_\mu$ transforms as a tensor under general coordinate transformations.}
\begin{eqnarray}
    \delta A_\mu & = & \mathcal{L}_\chi A_\mu+\partial_\mu\Sigma+\left[A_\mu\,,\Sigma\right]\,,\label{eq:deltabar}\\
    \delta F_{\mu\nu} & = & \mathcal{L}_\chi F_{\mu\nu}+\left[F_{\mu\nu}\,,\Sigma\right]\,,
\end{eqnarray}
where the Lie derivatives are along the vector $\chi^\mu$ and where 
\begin{equation}\label{eq:Sigma}
    \Sigma=B_a \lambda^a+\frac{1}{2}J_{ab}\lambda^{ab}+D \Lambda_D+K\sigma+K_a \sigma^a\,,
\end{equation}
with $\lambda^{ab}=-\lambda^{ba}$.
We will also need the Bianchi identities
\begin{equation}
    \partial_{[\mu}F_{\nu\rho]}+\left[A_{[\mu}\,,F_{\nu\rho]}
    \right]=0\,.
\end{equation}
We expand $F_{\mu\nu}$ in the Lie algebra basis as
\begin{eqnarray}
    F_{\mu\nu}& =& HR(H)_{\mu\nu}+P_a R(P)_{\mu\nu}{}^a+B_a R(B)_{\mu\nu}{}^a+\frac{1}{2}J_{ab} R(J)_{\mu\nu}{}^{ab}\nonumber\\
    &&+D R(D)_{\mu\nu}+K R(K)_{\mu\nu}+K_a R(K)_{\mu\nu}{}^a\,.\label{eq:LieexpF}
\end{eqnarray}

In components the gauge transformations of the gauge fields are
\begin{eqnarray}
    \delta \tau_\mu & = & \mathcal{L}_\chi\tau_\mu+\Lambda_D\tau_\mu+\lambda^a e^a_\mu\,,\label{eq:trafotau}\\
    \delta e^a_\mu & = & \mathcal{L}_\chi e^a_\mu+\Lambda_D e^a_\mu+\lambda^{ab}e^b_\mu\,,\label{eq:trafoe}\\
    \delta\omega_\mu{}^a & = & \mathcal{L}_\chi\omega_\mu{}^{a}+\partial_\mu\lambda^{a}+\lambda^{ab}\omega_\mu{}^b-\lambda^b\omega_{\mu}{}^{ab}+\sigma e^a_\mu+\sigma^a\tau_\mu\,,\label{eq:gaugetrafoboostcon}\\
    \delta\omega_\mu{}^{ab} & = & \mathcal{L}_\chi\omega_\mu{}^{ab}+\partial_\mu\lambda^{ab}+\lambda^{ac}\omega_\mu{}^{cb}-\lambda^{bc}\omega_\mu{}^{ca}-\sigma^a e^b_\mu+\sigma^b e^a_\mu\,,\label{eq:traforotcon}\\
    \delta b_\mu & = & \mathcal{L}_\chi b_\mu+\partial_\mu\Lambda_D+\sigma^a e^a_\mu\,,\label{eq:trafoWeylcon}\\
    \delta f_\mu & = & \mathcal{L}_\chi f_\mu+\partial_\mu\sigma+\sigma b_\mu-\Lambda_D f_\mu-\lambda^a f_\mu{}^a+\sigma^a\omega_\mu{}^a\,,\label{eq:deltaf}\\
    \delta f_\mu{}^a & = & \mathcal{L}_\chi f_\mu{}^a+\partial_\mu\sigma^a+\lambda^{ab}f_\mu{}^b-\sigma^b\omega_\mu{}^{ab}+\sigma^a b_\mu-\Lambda_D f_\mu{}^a\,.
\end{eqnarray}
Furthermore, the components of the field strength are
\begin{eqnarray}
    R(H)_{\mu\nu} & = & \partial_\mu\tau_\nu-\partial_\nu\tau_\mu-\omega_\mu{}^a e^a_\nu+\omega_\nu{}^a e^a_\mu- b_\mu\tau_\nu+ b_\nu\tau_\mu\,,\\
    R(P)_{\mu\nu}{}^a & = & \partial_\mu e^a_\nu-\partial_\nu e^a_\mu-\omega_\mu{}^{ab}e^b_\nu+\omega_\nu{}^{ab}e^b_\mu-b_\mu e^a_\nu+b_\nu e^a_\mu\,,\\
    R(B)_{\mu\nu}{}^a & = & \partial_\mu\omega_\nu{}^a-\partial_\nu\omega_\mu{}^a-\omega_\mu{}^c\omega_{\nu}{}^{ca}+\omega_\nu{}^c\omega_{\mu}{}^{ca}\nonumber\\
    &&+e^a_\mu f_\nu-e^a_\nu f_\mu-f_\mu{}^a\tau_\nu+f_\nu{}^a\tau_\mu\,,\label{eq:RB}\\
    R(J)_{\mu\nu}{}^{ab} & = & \partial_\mu\omega_\nu{}^{ab}-\partial_\nu\omega_\mu{}^{ab}+\omega_\mu{}^{ca}\omega_\nu{}^{cb}-\omega_\nu{}^{ca}\omega_\mu{}^{cb}\nonumber\\
    &&-e_\mu^b f_\nu{}^a+e_\mu^a f_\nu{}^b+e_\nu^b f_\mu{}^a-e_\nu^a f_\mu{}^b\,,\label{eq:RJ}\\
    R(D)_{\mu\nu} & = & \partial_\mu b_\nu-\partial_\nu b_\mu+e^a_\mu f_\nu{}^a-e^a_\nu f_\mu{}^a\,,\label{eq:RD}\\
    R(K)_{\mu\nu} & = & \partial_\mu f_\nu-\partial_\nu f_\mu-f_\mu{}^a\omega_\nu{}^a+f_\nu{}^a\omega_\mu{}^a+b_\mu f_\nu-b_\nu f_\mu\,,\\
    R(K)_{\mu\nu}{}^a & = & \partial_\mu f_\nu{}^a-\partial_\nu f_\mu{}^a-\omega_\mu{}^{ac}f_\nu{}^c+\omega_\nu{}^{ac}f_\mu{}^c+b_\mu f_\nu{}^a-b_\nu f_\mu{}^a\,,
\end{eqnarray}
and they transform as 
\begin{eqnarray}
    \delta R(H)_{\mu\nu} & = & \mathcal{L}_\chi R(H)_{\mu\nu}+\Lambda_D R(H)_{\mu\nu}+\lambda^a R(P)_{\mu\nu}{}^a\,,\label{eq:deltaRH}\\
    \delta R(P)_{\mu\nu}{}^a & = & \mathcal{L}_\chi R(P)_{\mu\nu}{}^a+\Lambda_D R(P)_{\mu\nu}{}^a+\lambda^{ab}R(P)_{\mu\nu}{}^b\,,\label{eq:deltaRP}\\
    \delta R(B)_{\mu\nu}{}^a & = & \mathcal{L}_\chi R(B)_{\mu\nu}{}^a+\lambda^{ab}R(B)_{\mu\nu}{}^b-\lambda^b R(J)_{\mu\nu}{}^{ab}\nonumber\\
    &&+\sigma R(P)_{\mu\nu}{}^a+\sigma^a R(H)_{\mu\nu}\,,\\
    \delta R(J)_{\mu\nu}{}^{ab} & = & \mathcal{L}_\chi R(J)_{\mu\nu}{}^{ab}+\lambda^{ac}R(J)_{\mu\nu}{}^{cb}-\lambda^{bc}R(J)_{\mu\nu}{}^{ca}\nonumber\\
    &&-\sigma^a R(P)_{\mu\nu}{}^{b}+\sigma^b R(P)_{\mu\nu}{}^{a}\,,\\
    \delta R(D)_{\mu\nu} & = & \mathcal{L}_\chi R(D)_{\mu\nu}+\sigma^a R(P)_{\mu\nu}{}^a\,,\\
    \delta R(K)_{\mu\nu} & = & \mathcal{L}_\chi R(K)_{\mu\nu}+\sigma R(D)_{\mu\nu}-\Lambda_D R(K)_{\mu\nu}\nonumber\\
    &&-\lambda^a R(K)_{\mu\nu}{}^a+\sigma^a R(B)_{\mu\nu}{}^a\,,\label{eq:deltaRKtemp}\\
    \delta R(K)_{\mu\nu}{}^a & = & \mathcal{L}_\chi R(K)_{\mu\nu}{}^a+\lambda^{ab}R(K)_{\mu\nu}{}^b-\sigma^b R(J)_{\mu\nu}{}^{ab}\nonumber\\
    &&+\sigma^a R(D)_{\mu\nu}-\Lambda_D R(K)_{\mu\nu}{}^a\,,\label{eq:deltaRK}
\end{eqnarray}
and satisfy the Bianchi identities
\begin{eqnarray}
0 & = & \partial_{[\mu}R(H)_{\nu\rho]}+e^a_{[\mu}R(B){}_{\nu\rho]}{}^a-\omega_{[\mu}{}^a R(P){}_{\nu\rho]}{}^a+\tau_{[\mu}R(D)_{\nu\rho]}-b_{[\mu}R(H)_{\nu\rho]}\,,\label{eq:DRH}\\ 
0 & = & \partial_{[\mu}R(P)_{\nu\rho]}{}^a-e^c_{[\mu}R(J){}_{\nu\rho]}{}^{ca}+\omega_{[\mu}{}^{ca}R(P){}_{\nu\rho]}{}^{c}+e^a_{[\mu}R(D)_{\nu\rho]}-b_{[\mu}R(P)_{\nu\rho]}{}^a \,,\nonumber\\
&&\label{eq:DRP}\\
0 & = & \partial_{[\mu}R(B)_{\nu\rho]}{}^a-\omega_{[\mu}{}^cR(J){}_{\nu\rho]}{}^{ca}+\omega_{[\mu}{}^{ca}R(B){}_{\nu\rho]}{}^{c}-f_{[\mu}{}^aR(H)_{\nu\rho]}+\tau_{[\mu}R(K)_{\nu\rho]}{}^a\nonumber\\
&&-f_{[\mu}R(P)_{\nu\rho]}{}^a +e^a_{[\mu}R(K){}_{\nu\rho]}\,,\label{eq:DRB}\\
0 & = & \partial_{[\mu}R(J){}_{\nu\rho]}{}^{ab}+e^a_{[\mu}R(K)_{\nu\rho]}{}^b-e^b_{[\mu}R(K){}_{\nu\rho]}{}^a+f_{[\mu}{}^a R(P){}_{\nu\rho]}{}^b-f_{[\mu}{}^b R(P){}_{\nu\rho]}{}^a\nonumber\\
&&-\omega_{[\mu}{}^{cb}R(J){}_{\nu\rho}{}^{ca}+\omega_{[\mu}{}^{ca}R(J){}_{\nu\rho]}{}^{cb}\,,\label{eq:DRJ}\\
0 & = & \partial_{[\mu}R(D)_{\nu\rho]}+e^a_{[\mu}R(K)_{\nu\rho]}{}^a-f_{[\mu}{}^a R(P){}_{\nu\rho]}{}^a\,,\label{eq:DRD}\\
0 & = & \partial_{[\mu}R(K){}_{\nu\rho]}+\omega_{[\mu}{}^a R(K){}_{\nu\rho]}{}^a-f_{[\mu}{}^a R(B){}_{\nu\rho]}{}^a+b_{[\mu}R(K){}_{\nu\rho]}-f_{[\mu}R(D){}_{\nu\rho]}\,,\nonumber\\
&&\label{eq:DRK}\\
0 & = & \partial_{[\mu}R(K){}_{\nu\rho]}{}^a+\omega_{[\mu}{}^{ca}R(K){}_{\nu\rho]}{}^c-f_{[\mu}{}^c R(J){}_{\nu\rho]}{}^{ca}+b_{[\mu}R(K)_{\nu\rho]}{}^a-f_{[\mu}{}^a R(D){}_{\nu\rho]}\,.\nonumber\\
&&\label{eq:DRL}
\end{eqnarray}

We will assume that $\tau_\mu$ and $e^a_\mu$ are nowhere vanishing and that $(\tau_\mu\,,e^a_\mu)$ is invertible with inverse $(-v^\mu\,,e^\mu_a)$ given by
\begin{equation}
    v^\mu\tau_\mu=-1\,,\qquad v^\mu e_\mu^a=0\,,\qquad e^\mu_a\tau_\mu=0\,,\qquad e^\mu_a e_\mu^b=\delta^b_a\,.
\end{equation}
Using \eqref{eq:trafotau} and \eqref{eq:trafoe} we deduce that $v^\mu$ and $e^\mu_a$ transform as
\begin{eqnarray}
    \delta v^\mu & = & \mathcal{L}_\chi v^\mu-\Lambda_D v^\mu\,,\label{eq:gaugetrafov}\\
    \delta e^\mu_a & = & \mathcal{L}_\chi e^\mu_a-\Lambda_D e^\mu_a+\lambda^a v^\mu+\lambda^{ab}e^\mu_b\,.\label{eq:gaugetrafoinve}
\end{eqnarray}
We will often use the objects $h_{\mu\nu}=\delta_{ab}e^a_\mu e^b_\nu$ and $h^{\mu\nu}=\delta^{ab}e^\mu_a e^\nu_b$.

We introduce a symmetric affine connection $\Gamma^\rho_{\mu\nu}$, that (for now) we will  assume to be invariant under all the local transformations in $\Sigma$ (see equation \eqref{eq:Sigma}), via a covariant derivative $D_\mu$ that acts on $\tau_\mu$ and $e^a_\mu$ as 
\begin{eqnarray}
    D_\mu e^a_\nu & = & \partial_\mu e^a_\nu-\Gamma^\rho_{\mu\nu}e_\rho^a-\omega_\mu{}^{ab}e^b_\nu-b_\mu e^a_\nu\,,\label{eq:covcalDe}\\
    D_\mu \tau_\nu & = & \partial_\mu \tau_\nu-\Gamma^\rho_{\mu\nu}\tau_\rho-\omega_\mu{}^{a}e^a_\nu-b_\mu \tau_\nu\,.\label{eq:covcalDtau}
\end{eqnarray}
For the inverse vielbeine we have
\begin{eqnarray}
    D_\mu e^{\nu a} & = & \partial_\mu e^{\nu a}+\Gamma^\nu_{\mu\rho}e^{\rho a}-\omega_\mu{}^{ab}e^{\nu b}-\omega_\mu{}^a v^\nu+b_\mu e^{\nu a}\,,\label{eq:covcalDinve}\\
    D_\mu v^\nu & = & \partial_\mu v^\nu+\Gamma^\nu_{\mu\rho}v^\rho+b_\mu v^\nu\,.\label{eq:covcalDv}
\end{eqnarray}
These covariant derivatives transform covariantly with respect to general coordinate transformations and with respect to the gauge transformations in $\Sigma$. Later we will work with other affine connections that are not invariant under all gauge transformations in $\Sigma$. 
The covariant derivatives transform as
\begin{eqnarray}
    \delta D_\mu e^a_\nu & = & \mathcal{L}_\chi D_\mu e^a_\nu+\Lambda_D D_\mu e^a_\nu+\lambda^{ab}D_\mu e^b_\nu+\sigma^a h_{\mu\nu}-\sigma^b\left(e^a_\mu e^b_\nu+e^b_\mu e^a_\nu\right)\,,\label{eq:trafocalDe}\\
    \delta D_\mu \tau_\nu & = & \mathcal{L}_\chi D_\mu \tau_\nu+\Lambda_D D_\mu \tau_\nu+\lambda^a D_\mu e^a_\nu -\sigma h_{\mu\nu}-\sigma^a\left(\tau_\mu e_\nu^a+\tau_\nu e_\mu^a\right)\,.\label{eq:trafocalDtau}
\end{eqnarray}
In particular they are designed such that the gauge parameters $\Lambda_D$, $\lambda^a$, $\sigma$ and $\sigma^a$ appear without derivatives.

The assumption that $\Gamma^\rho_{\mu\nu}$ is invariant under all the transformations in $\Sigma$, as well as the above transformations leads us to define the second covariant derivatives as
\begin{eqnarray}
    D_\rho D_\mu e^a_\nu & = & \partial_\rho\left(D_\mu e^a_\nu\right)-\Gamma^\sigma_{\rho\mu}D_\sigma e^a_\nu-\Gamma^\sigma_{\rho\nu}D_\mu e^a_\sigma-\omega_\rho{}^{ab}D_\mu e^b_\nu-b_\rho D_\mu e^a_\nu\nonumber\\
    &&-f_\rho{}^a h_{\mu\nu}+f_\rho{}^b\left(e^a_\mu e^b_\nu+e^a_\nu e^b_\mu\right)\,,\\
    D_\rho D_\mu \tau_\nu & = & \partial_\rho D_\mu \tau_\nu-\Gamma_{\rho\mu}^\sigma D_\sigma \tau_\nu-\Gamma_{\rho\nu}^\sigma D_\mu \tau_\sigma-b_\rho D_\mu \tau_\nu-\omega_\rho{}^a D_\mu e^a_\nu\nonumber\\
    &&+f_\rho h_{\mu\nu}+f_\rho{}^a\left(\tau_\mu e_\nu^a+\tau_\nu e_\mu^a\right)\,.
\end{eqnarray}
It then follows that
\begin{eqnarray}
    \left[D_\rho\,,D_\mu\right]e_\nu^a & = & R_{\rho\mu\nu}{}^\sigma e_\sigma^a-R(J)_{\rho\mu}{}^{ab}e^b_\nu-R(D)_{\rho\mu}e^a_\nu\,,\label{eq:calDcomme}\\
    \left[D_\rho\,,D_\mu\right]\tau_\nu & = & R_{\rho\mu\nu}{}^\sigma \tau_\sigma-R(B)_{\rho\mu}{}^a e^a_\nu-R(D)_{\rho\mu}\tau_\nu\,,\label{eq:calDcommtau}
\end{eqnarray}
where $R_{\rho\mu\nu}{}^\sigma$ is the curvature tensor associated with $\Gamma^\rho_{\mu\nu}$, i.e.
\begin{equation}\label{eq:curvten}
    R_{\rho\mu\nu}{}^\sigma=-\partial_\rho\Gamma_{\mu\nu}^\sigma+\partial_\mu\Gamma_{\rho\nu}^\sigma-\Gamma_{\rho\lambda}^\sigma\Gamma^\lambda_{\mu\nu}+\Gamma_{\mu\lambda}^\sigma\Gamma^\lambda_{\rho\nu}\,.
\end{equation}

The assumption that $\Gamma^\rho_{\mu\nu}$ is invariant under the gauge transformations in $\Sigma$ was only used to motivate the definitions of $D_\mu e^a_\nu$ and $D_\mu \tau_\nu$ as well as their double covariant derivatives. It is in no way a necessary assumption and we will not in fact work with such affine connections further below. So to be explicit when $\Gamma^\rho_{\mu\nu}$ is not invariant under some of the gauge transformations in $\Sigma$ we will still use the same expressions for the covariant derivatives $D_\mu e^a_\nu$ and $D_\mu \tau_\nu$ as well as their double covariant derivatives. Equations such as \eqref{eq:calDcomme} and \eqref{eq:calDcommtau} are identities and so they are true for any choice of $\Gamma^\rho_{\mu\nu}$.

\subsection{Curvature constraints, part 1}\label{subsec:curvcon1}

For the remainder of this section we will work with $d=2$. 

So far in our gauging procedure the connections in \eqref{eq:Liealgebraconnection} are all independent. The goal is to make contact with the fields in the asymptotic expansion of the metric near $\mathcal{I}^+$ which in Carroll-covariant BS gauge consist of $\tau_\mu$, $h_{\mu\nu}=\delta_{ab}e^a_\mu e_\nu^b$ and $C_{\mu\nu}$. We will make contact with these fields by first imposing appropriate curvature constraints. These will involve setting some of the curvatures in \eqref{eq:LieexpF} equal to zero. The purpose of doing so is to use these curvature constraints to algebraically solve for some of the gauge fields (such as $\omega_\mu{}^{ab}$) in terms of fields we wish to keep free (such as $\tau_\mu$ and $e^a_\mu$). The full set of curvature constraints should be gauge invariant so that when using \eqref{eq:deltaRH}--\eqref{eq:deltaRK} the variation of the constraints vanishes as well. Once this has been achieved we end up with a new set of independent fields. We will see that  some of these fields can be gauge fixed. This will ultimately lead to the desired field content whose properties match with the boundary fields $\tau_\mu, h_{\mu\nu}, C_{\mu\nu}$ we discussed in the previous section.

We will start by imposing the curvature constraints
\begin{equation}\label{eq:curvconstraint1and2}
    R(H)_{\mu\nu}=0\,,\qquad R(P)_{\mu\nu}{}^a=0\,.
\end{equation}
These constraints are preserved by the gauge transformations \eqref{eq:deltaRH} and \eqref{eq:deltaRP}, i.e. 
\begin{equation}\label{eq:varconst=0}
    \delta R(H)_{\mu\nu}=0\,,\qquad \delta R(P)_{\mu\nu}{}^a=0\,.
\end{equation}
We will use the curvature constraints \eqref{eq:curvconstraint1and2} to solve for $\omega_\mu{}^a$ and $\omega_\mu{}^{ab}$. Let us first consider $R(H)_{\mu\nu}=0$. By using completeness of the vielbeine we can write
\begin{equation}\label{eq:decompboostcon}
    \omega_\mu{}^a=S^{ab}e^b_\mu+S^a\tau_\mu\,,
\end{equation}
where we can decompose
\begin{equation}
S^{ab}=S^{\langle ab\rangle}+\frac{1}{2}S\delta^{ab}+S^{[ab]}\,.
\end{equation}
Angle brackets denote the symmetric trace-free (STF) part and square brackets the antisymmetric part. Solving $R(H)_{\mu\nu}=0$ for $\omega_\mu{}^a$ leads to 
\begin{equation}\label{eq:Cboostcon}
    \omega_\mu{}^a=S^{\langle ab\rangle}e^b_\mu+\frac{1}{2}Se^a_\mu-\frac{1}{2}F^{ab}e^b_\mu+(b_\rho-a_\rho) e^\rho_a\tau_\mu\,,
\end{equation}
where $F_{ab}=e^\mu_a e^\nu_b F_{\mu\nu}$. We remind the reader that $a_\mu$ and $F_{\mu\nu}$ are defined in \eqref{eq:defa} and \eqref{eq:defF}.
Comparing with \eqref{eq:decompboostcon} we read off that 
\begin{eqnarray}
    S^a & = & (b_\rho-a_\rho)e^\rho_a\,,\label{eq:Sa}\\
    S^{[ab]} & = & -\frac{1}{2}F^{ab}\,.
\end{eqnarray}

From the gauge transformation \eqref{eq:gaugetrafoboostcon} we can see that $S$ and $e^\rho_a b_\rho$ are pure gauge as we can use the gauge parameters $\sigma$ and $\sigma^a$ to fix them to any value. 
The STF part of $S^{ab}$ will not be fixed by any of the curvature constraints and we will see that this corresponds to the shear. This is in agreement with various observations that have been made in the literature \cite{Ashtekar:1981hw,Korovin:2017xqu,Herfray:2021qmp,Nguyen:2022zgs,Baulieu:2025itt,Fiorucci:2025twa,Hartong:2025jpp}.

Next, we consider $R(P)_{\mu\nu}{}^a=0$ for $d=2$. By contracting it with $v^\mu e^\nu_b$ and decomposing the $a,b$ indices into an antisymmetric, trace and STF part we learn that 
\begin{eqnarray}
    v^\mu\omega_\mu{}^{ab} & = & -\frac{1}{2}\left(e^\mu_a\mathcal{L}_v e^b_\mu-e^\mu_b\mathcal{L}_v e^a_\mu\right)\,,\\
    v^\mu b_\mu & = & -\frac{1}{2}K\,,\label{eq:vb}\\
    K_{\mu\nu} & = & \frac{1}{2}K h_{\mu\nu}\,,
\end{eqnarray}
where the latter equation follows from the STF part after contracting it with $e^a_\mu e^b_\nu$. We thus see that we can reproduce the constraint \eqref{eq:Kconstraint} from the gauging perspective.

Next we consider $e^\mu_c e^\nu_d R(P){}_{\mu\nu}{}^a=0$. By cyclically permuting the indices we find
\begin{equation}
    e^\mu_c\omega_\mu{}^{ab}=-\frac{1}{2}\left(-E_{cba}+E_{bac}-E_{acb}\right)+e^\mu_b b_\mu\delta_{ac}-e^\mu_a b_\mu\delta_{cb}\,,
\end{equation}
where we defined
\begin{equation}\label{eq:Eabc}
    E_{cda}=e^\mu_c e^\nu_d\left(\partial_\mu e_\nu^a-\partial_\nu e_\mu^a\right)\,.
\end{equation}
By contracting $e^\mu_c\omega_\mu{}^{ab}$ with $e^c_\rho$ we obtain
\begin{eqnarray}\label{eq:rotationcon}
    \omega_\mu{}^{ab} & = & -\frac{1}{2} e_\mu^c\left(-E_{cba}+E_{bac}-E_{acb}\right)+\frac{1}{2}\tau_\mu \left(e^\rho_a\mathcal{L}_v e^b_\rho-e^\rho_b\mathcal{L}_v e^a_\rho\right)\nonumber\\
    &&-e^\rho_a b_\rho e^b_\mu+e^\rho_b b_\rho e^a_\mu\,.
\end{eqnarray}
We have thus been able to solve for all components of $\omega_\mu{}^{ab}$.

Before we imposed the curvature constraints the gauge transformations of $\omega_\mu{}^a$ and $\omega_\mu{}^{ab}$ were given in \eqref{eq:gaugetrafoboostcon} and \eqref{eq:traforotcon}. We have checked that after the curvature constraints have been solved for, the gauge transformations of $\omega_\mu{}^{ab}$ is still given by  \eqref{eq:traforotcon}. We will next show that we can furthermore assign a gauge transformation to $S^{(ab)}$ (round brackets denote the symmetric part), the undetermined part of the Carroll boost connection, such that the gauge transformation of $\omega_\mu{}^a$ is likewise unaffected\footnote{This is because the curvature constraints \eqref{eq:curvconstraint1and2} are fixed under the gauge transformation as shown in \eqref{eq:varconst=0}. We point this out because in the supergravity literature one often starts the gauging procedure with the adjoint algebra of gauge transformations (as opposed to \eqref{eq:deltabar}) and in that case a curvature constraint such as the one imposed in \eqref{eq:curvconstraint1and2} is not fixed by the adjoint transformations which leads to a new algebra of gauge transformations that acts on the reduced set of fields obtained by solving the curvature constraints. That new algebra is precisely the one of \eqref{eq:deltabar}.} and thus still given by \eqref{eq:gaugetrafoboostcon}. We will see that $2S_{\langle\mu\nu\rangle}:=2S^{\langle ab\rangle}e^a_\mu e^b_\nu$ transforms exactly like the shear tensor.

If we take the solution for $\omega_\mu{}^a$ and compute its gauge transformation we obtain
\begin{eqnarray}
    \delta \omega_{\mu}{}^a & = & \mathcal{L}_\chi\omega_{\mu}{}^a+\partial_\mu\lambda^a -\lambda^b\omega_\mu{}^{ab}+\lambda^{ab}\omega_\mu{}^b+\sigma^a\tau_\mu+\sigma e^a_\mu\nonumber\\
    &&+e^b_\mu\left[\delta S^{(ab)}-\lambda^{ac}S^{(cb)}-\lambda^{bc}S^{(ac)}+\Lambda_D S^{(ab)}-\sigma\delta^{ab}\right.\nonumber\\
    &&\left.-\frac{1}{2}\lambda_a e^\mu_b a_\mu-\frac{1}{2}\lambda_b e^\mu_a a_\mu+\lambda^\rho b_\rho\delta^{ab}-\frac{1}{2}e^\rho_a e^\sigma_b \mathcal{L}_\lambda h_{\rho\sigma}\right]\,.
\end{eqnarray}
The first line reproduces the transformation \eqref{eq:gaugetrafoboostcon} from before imposing the curvature constraint. If we demand that the gauge transformation of the dependent gauge field is the same as before we imposed the curvature constraint then we find the following transformation for $S_{(\mu\nu)}:=S^{(ab)}e^a_\mu e^b_\nu$,
\begin{eqnarray}
    \delta S_{(\mu\nu)} & = & \mathcal{L}_\chi S_{(\mu\nu)}+\Lambda_D S_{(\mu\nu)}+\sigma h_{\mu\nu}-\lambda^\rho b_\rho h_{\mu\nu}\nonumber\\
    &&+\frac{1}{2}\lambda_{\mu} a_\nu+\frac{1}{2}\lambda_{\nu} a_\mu+\frac{1}{2}h^\rho_\mu h^\sigma_\nu\mathcal{L}_\lambda h_{\rho\sigma}\,.
\end{eqnarray}
The trace and STF parts of $S_{\mu\nu}$ then transform as 
\begin{eqnarray}
    \delta\left(h^{\mu\nu}S_{\mu\nu}\right) & = & \mathcal{L}_\chi\left(h^{\mu\nu}S_{\mu\nu}\right)-\Lambda_D h^{\mu\nu}S_{\mu\nu}+2\left(\sigma-\lambda^\rho b_\rho\right)+\lambda^\rho a_\rho+\frac{1}{2}h^{\rho\sigma}\mathcal{L}_\lambda h_{\rho\sigma}\,,\nonumber\\
    &&\\
    \delta S_{\langle\mu\nu\rangle} & = & \mathcal{L}_\chi S_{\langle\mu\nu\rangle}+\Lambda_D S_{\langle\mu\nu\rangle}+\frac{1}{2}h^\rho_{\langle\mu}h^\sigma_{\nu\rangle}\left(\mathcal{L}_\lambda h_{\rho\sigma}+2\lambda_\rho a_\sigma\right)\,.
\end{eqnarray}
As advertised the latter transformation takes the same form as the transformation of the shear \eqref{eq:trafoshear}. We thus identify the shear with twice the STF part of $S_{\mu\nu}$, i.e.
\begin{equation}\label{eq:shearalgebra}
    C_{\mu\nu}=2S^{\langle ab\rangle}e^a_\mu e^b_\nu\,.
\end{equation}
We also see that the trace part of $S_{\mu\nu}$ is pure gauge as it undergoes a shift transformation containing the gauge parameter $\sigma$.

When imposing curvature constraints some of the Bianchi identities for the curvatures become algebraic which tells us that some other curvatures are now algebraically related to each other. Consider equations
\eqref{eq:DRH} and \eqref{eq:DRP}. If we use the curvature constraints $R(H)_{\mu\nu}=0$ and $R(P)_{\mu\nu}{}^a=0$ then \eqref{eq:DRH} and \eqref{eq:DRP} simplify to
\begin{eqnarray}
0 & = & e^a_{[\mu}R(B){}_{\nu\rho]}{}^a+\tau_{[\mu}R(D)_{\nu\rho]}\,,\\ 
0 & = & -e^c_{[\mu}R(J){}_{\nu\rho]}{}^{ca}+e^a_{[\mu}R(D)_{\nu\rho]}\,.
\end{eqnarray}
If we contract this with $v^\mu e^\nu_b e^\rho_d$ we find
\begin{eqnarray}
    0 & = & v^\mu R(B){}_{\mu bd}-v^\mu R(B){}_{\mu db}-R(D)_{bd}\,,\label{eq:curvconstraint3}\\
    0 & = & -v^\mu R(J){}_{\mu bda}+v^\mu R(J){}_{\mu dba}+\delta^a_d v^\mu R(D)_{\mu b}-\delta^a_b v^\mu R(D)_{\mu d}\,,\label{eq:curvconstraint4}
\end{eqnarray}
where we raise and lower $a,b,c,\ldots$ indices with a Kronecker delta\footnote{Because of this we are sometimes a bit cavalier with the index placement of the $a,b,c,\ldots$ indices.} and where e.g. we write $R(B){}_{\mu bd}=R(B){}_{\mu\nu d} e^\nu_b$. If we trace the latter equation over $a$ and $d$ we obtain
\begin{equation}\label{eq:curvconstraint5}
    v^\mu R(D){}_{\mu b}+v^\mu R(J){}_{\mu cbc}=0\,.
\end{equation}
Under a gauge transformation \eqref{eq:curvconstraint3} transforms into \eqref{eq:curvconstraint5} whereas 
\eqref{eq:curvconstraint4} is inert. Using \eqref{eq:curvconstraint5} we can rewrite \eqref{eq:curvconstraint4} as
\begin{equation}
    v^\mu R(J){}_{\mu bda}-v^\mu R(J){}_{\mu dba}+\delta^a_d v^\mu R(J){}_{\mu cbc}-\delta^a_b v^\mu R(J){}_{\mu cdc}=0\,.
\end{equation}
For $d=2$ this equation is identically satisfied, so the relevant equations are \eqref{eq:curvconstraint3} and \eqref{eq:curvconstraint5}.

\subsection{Choice of affine connection}

To make progress we next want to compute the part of the $J$ and $B$ curvatures, i.e. $R(J)_{\mu\nu}{}^{ab}$ and $R(B)_{\mu\nu}{}^a$, that depend on the $\omega_\mu{}^a$ and $\omega_\mu{}^{ab}$ connections and compare this with expressions in the asymptotic expansion of the metric. This would then inform us what curvature constraints we need to impose on the $R(J)_{\mu\nu}{}^{ab}$ and $R(B)_{\mu\nu}{}^a$ curvatures that allow us to solve for $f_\mu$ and $f_\mu{}^a$ which are so far unfixed. To this end we will write the curvatures in \eqref{eq:RB},  \eqref{eq:RJ} and \eqref{eq:RD} as follows 
\begin{eqnarray}
R(B)_{\mu\nu}{}^a & = & {\bar R}(B)_{\mu\nu}{}^a+e^a_\mu f_\nu-e^a_\nu f_\mu-f_\mu{}^a\tau_\nu+f_\nu{}^a\tau_\mu\,,\\
    R(J)_{\mu\nu}{}^{ab} & = & {\bar R}(J)_{\mu\nu}{}^{ab}-e_\mu^b f_\nu{}^a+e_\mu^a f_\nu{}^b+e_\nu^b f_\mu{}^a-e_\nu^a f_\mu{}^b\,,\\
    R(D)_{\mu\nu} & = & \bar R(D)_{\mu\nu}+e^a_\mu f_\nu{}^a-e^a_\nu f_\mu{}^a\,,
\end{eqnarray}
where we defined 
\begin{eqnarray}
    \bar R(B)_{\mu\nu}{}^a & = & \partial_\mu\omega_\nu{}^a-\partial_\nu\omega_\mu{}^a-\omega_\mu{}^c\omega_{\nu}{}^{ca}+\omega_\nu{}^c\omega_{\mu}{}^{ca}\,,\label{eq:barRB}\\
    \bar R(J)_{\mu\nu}{}^{ab} & = & \partial_\mu\omega_\nu{}^{ab}-\partial_\nu\omega_\mu{}^{ab}+\omega_\mu{}^{ca}\omega_\nu{}^{cb}-\omega_\nu{}^{ca}\omega_\mu{}^{cb}\,,\label{eq:barRJ}\\
    \bar R(D)_{\mu\nu} & = & \partial_\mu b_\nu-\partial_\nu b_\mu\,.\label{eq:barRD}
\end{eqnarray}
For $d=2$ the last two terms in the barred $J$ curvature cancel due to the Abelian nature of the rotations.
In order to compute $\bar R(B)_{\mu\nu}{}^a$ and 
$\bar R(J)_{\mu\nu}{}^{ab}$, which are second order in derivatives of the Carroll metric data, it will be helpful to have an explicit affine connection $\Gamma^\rho_{\mu\nu}$.

We can find an explicit expression for the affine connection by imposing a vielbein postulate (VP). These are expressions which set $D_\mu e^a_\nu$ and $D_\mu \tau_\nu$, defined in \eqref{eq:covcalDe} and \eqref{eq:covcalDtau}, equal to something known.
A VP is thus an equation of the form 
\begin{eqnarray}
    D_\mu e_\nu^a & = & -X^\rho{}_{\mu\nu}e_\rho^a\,,\label{eq:De=X}\\
    D_\mu \tau_\nu & = & -X^\rho{}_{\mu\nu}\tau_\rho\,,
\end{eqnarray}
where $X^\rho{}_{\mu\nu}$ is a given tensor. If we solve these equations using \eqref{eq:Cboostcon} and \eqref{eq:rotationcon} we obtain
\begin{eqnarray}
\Gamma^\rho_{\mu\nu} & = & \mathcal{C}{}^\rho_{\mu\nu}-\delta^\rho_\mu b_\nu-\delta^\rho_\nu b_\mu+h^{\rho\sigma}b_\sigma h_{\mu\nu}-\frac{1}{2}K v^\rho\tau_\mu\tau_\nu\nonumber\\
&&+\frac{1}{2}v^\rho C_{\mu\nu}-\frac{1}{2}S v^\rho h_{\mu\nu}+X^\rho{}_{\mu\nu}\,.\label{eq:GammaandX}
\end{eqnarray}
where $\mathcal{C}^\rho_{\mu\nu}$ is given in \eqref{eq:conn}. The covariant derivative associated with $\Gamma^\rho_{\mu\nu}$ will be denoted by $\nabla_\mu$.

For different calculations different choices of $X^\rho{}_{\mu\nu}$ can be convenient. At the end of the day the goal is to find expressions for the barred curvatures given in equations \eqref{eq:barRB}, \eqref{eq:barRJ} and \eqref{eq:barRD} in terms of objects that are covariant with respect to the connection $\mathcal{C}^\rho_{\mu\nu}$. This is because in the work \cite{Hartong:2025jpp,Hartong:2026rbr} this connection was used and so in order to compare with the near-boundary expansion of section \ref{subsec:nearbdryexp} this is what we need to achieve.

\subsection{Computing the $B$ curvature}

We start by computing $\bar R(B)_{\mu\nu}{}^a$ which we remind the reader is defined in \eqref{eq:barRB}. In order to do so we will use the decomposition of the Carroll boost connection given in \eqref{eq:decompboostcon}. First of all we derive
\begin{eqnarray}
    \bar R(B)_{\mu\nu}{}^a & = & \partial_\mu\omega_\nu{}^a-\partial_\nu\omega_\mu{}^a-\omega_\mu{}^c\omega_{\nu}{}^{ca}+\omega_\nu{}^c\omega_{\mu}{}^{ca}\nonumber\\
    & = & \left(\partial_\mu S^{ab}+b_\mu S^{ab}-\omega_\mu{}^{ac}S^{cb}-\omega_\mu{}^{bc}S^{ac}+\omega_\mu{}^b S^a\right)e_\nu^b\nonumber\\
    &&+\left(\partial_\mu S^a+b_\mu S^a-\omega_\mu{}^{ac}S^c\right)\tau_\nu-\left(\mu\leftrightarrow\nu\right)\,,\label{eq:barRBintermed}
\end{eqnarray}
where we used \eqref{eq:curvconstraint1and2} and where we remind the reader that $S^a$ is defined in \eqref{eq:Sa} and that 
\begin{equation}\label{eq:Sab}
    S^{ab}=\frac{1}{2}C^{ab}+\frac{1}{2}S\delta^{ab}-\frac{1}{2}F^{ab}\,,
\end{equation}
where we used \eqref{eq:shearalgebra}.

Let us define
\begin{equation}
    Z_\mu{}^a=\partial_\mu S^a+b_\mu S^a-\omega_\mu{}^{ac}S^c\,.
\end{equation}
A useful result is 
\begin{eqnarray}\label{eq:idomegatonabla}
    Z_\mu{}^a & = & e^{\rho a}\left(\nabla_\mu Y_\rho-Y_\sigma e^\sigma_b D_\mu e^b_\rho\right)\nonumber\\
    & = & e^{\rho a}\left(\mathcal{D}_\mu Y_\rho+b_\mu Y_\rho+b_\rho Y_\mu-h_{\mu\rho}Y^\sigma b_\sigma\right)\,,
\end{eqnarray}
where, inspired by \eqref{eq:Sa}, we defined
\begin{eqnarray}
    Y_\mu & = & b_\mu-\tilde b_\mu=-v^\rho\omega_\rho{}^a e^a_\mu\,,\label{eq:defYmu}
\end{eqnarray}
in which we defined
\begin{equation}\label{eq:tildeb}
    \tilde b_\mu=a_\mu+\frac{1}{2}K\tau_\mu\,,
\end{equation}
so that we can write $S^a=Y_\rho e^{\rho a}$. We note that $Y_\mu$ is spatial since we know that $v^\mu b_\mu=-\frac{1}{2}K$. The rewriting of \eqref{eq:idomegatonabla} initially uses \eqref{eq:covcalDinve} which is then converted to $D_\mu e^b_\rho$ using completeness of the vielbeine. 
Using \eqref{eq:De=X} and \eqref{eq:GammaandX} we see that \eqref{eq:idomegatonabla} is independent of $X^\rho{}_{\mu\nu}$. The second equality follows from \eqref{eq:GammaandX}.

Let us define
\begin{equation}
    Z_\mu{}^{ab}=\partial_\mu S^{ab}+b_\mu S^{ab}-\omega_\mu{}^{ac}S^{cb}-\omega_\mu{}^{bc}S^{ac}\,,
\end{equation}
as well as 
\begin{equation}
    S^{ab}=e^a_\rho e^b_\sigma S^{\rho\sigma}\,,
\end{equation}
where we take $S^{\rho\sigma}$ to be spatial and $S^{ab}$ is given in \eqref{eq:Sab}.
We can then
similarly show that
\begin{eqnarray}
    Z_\mu{}^{ab} & = & e^a_\rho e^b_\sigma\left(\nabla_\mu S^{\rho\sigma}-X_{\mu\lambda}^\sigma S^{\rho\lambda}-X_{\mu\lambda}^\rho S^{\lambda\sigma}+3b_\mu S^{\rho\sigma}\right)\\
    & = & e^a_\rho e^b_\sigma\left(\mathcal{D}_\mu S^{\rho\sigma}+b_\mu S^{\rho\sigma}\right)-e^b_\mu S^{ac}b_c+b^b S^{ac}e^c_\mu-e^a_\mu S^{cb}b_c+b^a S^{cb}e^c_\mu\,,\nonumber
\end{eqnarray}
where we used spatial vielbein components such as $b_a=e^\rho_a b_\rho$.
Again we see that the result is independent of $X^\rho{}_{\mu\nu}$.

If we contract $\bar R(B)_{\mu\nu}{}^a$ with $v^\mu e^\nu_b$ we find
\begin{eqnarray}
    v^\mu e^\nu_b \bar R(B)_{\mu\nu}{}^a& = & e^\mu_a e^\nu_b\left[-\frac{1}{2}N_{\mu\nu}+\frac{1}{2}h_{\mu\nu}\left(\mathcal{L}_v-\frac{1}{2}K\right)S-\frac{1}{2}\left(\mathcal{L}_v+\frac{1}{2}K\right)F_{\mu\nu}\right.\nonumber\\
    &&\left.+\mathcal{D}_\nu Y_\mu+2b_\sigma h^\sigma_{\langle\nu}Y_{\mu\rangle}-Y_\mu Y_\nu\right]\,,\label{eq:vbarRBab}
\end{eqnarray}
where $\mathcal{D}_\mu$ is the covariant derivative associated with $\mathcal{C}^\rho_{\mu\nu}$ and where we defined the news tensor
\begin{equation}\label{eq:newsdef}
    N_{\mu\nu}=-\left(\mathcal{L}_v+\frac{1}{2}K\right)C_{\mu\nu}\,.
\end{equation}
In obtaining \eqref{eq:vbarRBab} from \eqref{eq:barRBintermed} we used \eqref{eq:rotationcon}.

The antisymmetric part of \eqref{eq:vbarRBab} obeys
\begin{equation}
    v^\mu\bar R(B)_{\mu ba}-v^\mu\bar R(B)_{\mu ab}+e^\mu_a e^\nu_b\bar R(D)_{\mu\nu}=0\,.
\end{equation}
In fact this equation is still true if we remove the bar on the curvatures as follows from 
\eqref{eq:RB} and \eqref{eq:RD}. 
This can also be understood by referring to the Bianchi identity \eqref{eq:curvconstraint3}.

Similarly we can show that
\begin{eqnarray}
    h^\rho_\mu e^\sigma_a\bar R(B)_{\rho\sigma}{}^a & = & -\frac{1}{2}\left(\mathcal{D}_\rho-2a_\rho\right)C^\rho{}_\mu+\frac{1}{2}\mathcal{D}_\rho F^\rho{}_\mu+\frac{1}{2}h^\rho_\mu\left(\partial_\rho+b_\rho\right)S\nonumber\\
    &&+\frac{1}{2}Y_\nu\left(C^\nu{}_\mu-F^\nu{}_\mu\right)\,.\label{eq:spatialtracebarRB}
\end{eqnarray}
For $d=2$ we can use the identity
\begin{equation}\label{eq:d=2Id}
    \bar R(B)_{ab}{}^c=e^\mu_a e^\nu_b\bar R(B)_{\mu\nu}{}^c=-\delta^c_a\bar R(B)_{bd}{}^d+\delta^c_b\bar R(B)_{ad}{}^d\,,
\end{equation}
so that $h^\rho_\mu e^\sigma_a\bar R(B)_{\rho\sigma}{}^a$ is enough to know $\bar R(B)_{ab}{}^c$ or similarly $h^\rho_\mu h^\sigma_\nu \bar R(B)_{\rho\sigma}{}^a$. 

We have thus managed to express the barred $B$ curvature in terms of Carroll geometric quantities (including the shear) and the connection $\mathcal{C}^\rho_{\mu\nu}$. These expressions also contain the independent objects $b_\mu$ and $S=S^{aa}$, but these will be gauge fixed further below.

\subsection{Computing the $J$ curvature}

We next turn to the barred $J$ curvature. For $d=2$ we can use the identity
\begin{equation}
    E_{cba}+E_{acb}=E_{abc}\,,
\end{equation}
where $E_{abc}$ is defined in \eqref{eq:Eabc}, so that the rotation connection \eqref{eq:rotationcon} can be written as
\begin{equation}
    \omega_\mu{}^{ab}=e_\mu^c E_{abc}+\tau_\mu\Omega_{ab}-b_a e^b_\mu+b_b e^a_\mu\,,
\end{equation}
where 
\begin{equation}
    \Omega_{ab}=\frac{1}{2}\left(e^\rho_a\mathcal{L}_v e^b_\rho-e^\rho_b\mathcal{L}_v e^a_\rho\right)\,.
\end{equation}
Furthermore, for $d=2$ the barred $J$ curvature \eqref{eq:barRJ} simplifies to 
\begin{equation}
    \bar R(J)_{\mu\nu}{}^{ab}=\partial_\mu\omega_\nu{}^{ab}-\partial_\nu\omega_\mu{}^{ab}\,.
\end{equation}

In this subsection will use a VP \eqref{eq:De=X} with 
\begin{equation}\label{eq:singleVP}
    D_\mu e^a_\nu=\partial_\mu e_\nu^a-\Gamma_{\mu\nu}^\rho e^a_\rho-\omega_\mu{}^{ab}e^b_\nu-b_\mu e^a_\nu=0\,,
\end{equation}
but with $D_\mu\tau_\nu$ arbitrary. However this is fine since $D_\mu\tau_\nu$ drops out of the calculations. For such an affine connection, using \eqref{eq:calDcomme} with $f_\mu$ and $f_\mu{}^a$ set to zero, we have
\begin{equation}\label{eq:RiemanntobarRJ}
    R_{\rho\mu\nu}{}^\sigma e_\sigma^a=\bar R(J)_{\rho\mu}{}^{ab}e^b_\nu+\bar R(D)_{\rho\mu}e^a_\nu\,.
\end{equation}
This equation is an identity that can also be independently verified by using \eqref{eq:curvten}, the Ricci identity and \eqref{eq:singleVP}.
The affine connection $\Gamma^\rho_{\mu\nu}$ is torsion-free and so the curvature tensor obeys
\begin{equation}
    R_{[\rho\mu\nu]}{}^\sigma=0\,.
\end{equation}
This implies
\begin{equation}\label{eq:Idone}
    v^\rho e^\nu_c R_{\rho\sigma\nu}{}^\sigma=e^\nu_c R_{\nu\sigma\rho}{}^\sigma v^\rho-e^\nu_c v^\rho R_{\nu\rho\sigma}{}^\sigma\,,
\end{equation}
where
\begin{equation}
    R_{\rho\nu\sigma}{}^\sigma=-\partial_\rho\Gamma^\sigma_{\nu\sigma}+\partial_\nu\Gamma^\sigma_{\rho\sigma}\,.
\end{equation}
We next compute the three terms in \eqref{eq:Idone} one by one. Starting with the left-hand side we can write
\begin{equation}\label{eq:LHS}
    v^\rho e^\nu_c R_{\rho\sigma\nu}{}^\sigma=v^\rho e^\nu_c e^\mu_a e^a_\sigma R_{\rho\mu\nu}{}^\sigma=v^\rho\bar R(J)_{\rho a}{}^{ac}+v^\rho\bar R(D)_{\rho c}\,.
\end{equation}
The first and the second term on the right-hand side are
\begin{eqnarray}
    e^\nu_c R_{\nu\sigma\rho}{}^\sigma v^\rho & = & -e^\nu_c\left[\nabla_\nu\,,\nabla_\sigma\right]v^\sigma\,,\\
    -e^\nu_c v^\rho R_{\nu\rho\sigma}{}^\sigma & = & -e^\nu_c v^\rho\left(\partial_\rho\Gamma^\sigma_{\nu\sigma}-\partial_\nu\Gamma^\sigma_{\rho\sigma}\right)\,,
\end{eqnarray}
where 
\begin{eqnarray}
    \nabla_\mu v^\nu & = & -b_\mu v^\nu+v^\nu v^\rho D_\mu\tau_\rho\,,\\
    \Gamma^\rho_{\rho\nu} & = & e^{-1}\partial_\nu e-3b_\nu+v^\rho D_\rho \tau_\nu\,.
\end{eqnarray}
A straightforward calculation shows that the right-hand side of \eqref{eq:Idone} is equal to 
\begin{equation}
     e^\nu_c R_{\nu\sigma\rho}{}^\sigma v^\rho-e^\nu_c v^\rho R_{\nu\rho\sigma}{}^\sigma=2v^\rho e^\sigma_c \bar R(D)_{\rho\sigma}\,.
\end{equation}
Combining this with \eqref{eq:LHS} we obtain
\begin{equation}\label{eq:vbarRJ}
    v^\rho\bar R(J)_{\rho aab}=v^\rho\bar R(D)_{\rho b}\,.
\end{equation}
Using a $d=2$ identity similar to \eqref{eq:d=2Id} this gives an equation for $v^\rho\bar R(J)_{\rho abc}$ which can be shown to be the same as what follows from the Bianchi identity \eqref{eq:curvconstraint5} because the $f_\mu{}^a$ connection drops out of \eqref{eq:curvconstraint5}.

What is left is to compute $h^\rho_\mu h^\sigma_\nu\bar R(J)_{\rho\sigma}{}^{ab}$. In $d=2$ it is enough to compute $\bar R(J)_{ab}{}^{ab}$. Using \eqref{eq:RiemanntobarRJ} we can write 
\begin{equation}
    \bar R(J)_{ab}{}^{ab}=h^{\mu\nu}h^\rho_\sigma R_{\rho\mu\nu}{}^\sigma\,.
\end{equation}
This means that we need to use \eqref{eq:singleVP} to compute the right-hand side. Equation \eqref{eq:singleVP} implies that
\begin{equation}
    \Gamma^\rho_{\mu\nu}=\mathcal{C}^\rho_{\mu\nu}-b_\mu h_\nu^\rho-b_\nu h_\mu^\rho+h^{\rho\sigma}b_\sigma h_{\mu\nu}+v^\rho X_{\mu\nu}\,,
\end{equation}
where $X_{\mu\nu}$ is any symmetric tensor. It can be shown that 
\begin{equation}\label{eq:doubletracebarRJ}
    \bar R(J)_{ab}{}^{ab}=h^{\mu\nu}h^\rho_\sigma R_{\rho\mu\nu}{}^\sigma=-2\left(\overset{(0)}{S}-a^2\right)-2h^{\mu\nu}\mathcal{D}_\mu\left(b_\nu-\tilde b_\nu\right)\,,
\end{equation}
where we used \eqref{eq:S0}. This calculation is essentially a question of finding the relation between two curvature tensors for two connections that differ by a tensor.

\subsection{Curvature constraints, part 2: $K$-curvatures}

We will next impose curvature constraints that will fix $f_\mu$ and $f_\mu{}^a$ and nothing more than that. The curvature constraints that do exactly that are
\begin{equation}\label{eq:curvRBJ=0}
    R(B)_{\mu\nu}{}^a=0\,,\qquad R(J)_{\mu\nu}{}^{ab}=0\,.
\end{equation}
For $d=2$ these are 9 components and we can use them to solve for $f_\mu$ and $f_\mu{}^a$ which are also 9 components. The Bianchi identities imply that with these curvature constraints we have 
\begin{equation}\label{eq:curvRD=0}
R(D)_{\mu\nu}=0\,. 
\end{equation}
Solving the curvature constraints \eqref{eq:curvRBJ=0} leads to
\begin{eqnarray}
    v^\mu f_\mu{}^a & = & -v^\mu\bar R(J)_{\mu b}{}^{ab}\,,\label{eq:vfa}\\
    e^\mu_a f_\mu{}^a & = & -\frac{1}{2}\bar R(J)_{ab}{}^{ab}\,,\label{eq:efa}\\
    \delta^a_b v^\mu f_\mu+e^\mu_b f_\mu{}^a & = & v^\mu\bar R(B)_{\mu b}{}^a\,,\label{eq:vf}\\
    e^\mu_b f_\mu & = & \bar R(B)_{ba}{}^a\,,\label{eq:ef}
\end{eqnarray}
where the terms on the right-hand side have been computed above. The gauge transformations of the dependent connections $f_\mu$ and $f_\mu{}^a$ is the same as the gauge transformation we had before imposing these extra curvature constraints.

Together with \eqref{eq:curvconstraint1and2}
 we are thus setting all the curvatures to zero except those associated with the special conformal generators $K$ and $K_a$. Using the gauge transformations of the curvatures in \eqref{eq:deltaRH}--\eqref{eq:deltaRK} we can readily check that this set of curvature constraints is closed under the gauge transformations. 
The complete set of curvature constraints is thus to set all curvatures to zero except those associated with the special conformal generators $K$ and $K_a$, i.e. $R(K)_{\mu\nu}$ and $R(K)_{\mu\nu}{}^a$. We will collectively refer to these as the $K$-curvatures. 

These $K$-curvatures have to obey the Bianchi identities \eqref{eq:DRB}--\eqref{eq:DRL}. The first three of these are algebraic Bianchi identities because they read
\begin{eqnarray}
0 & = & \tau_{[\mu}R(K)_{\nu\rho]}{}^a +e^a_{[\mu}R(K){}_{\nu\rho]}\,,\\
0 & = & e^a_{[\mu}R(K)_{\nu\rho]}{}^b-e^b_{[\mu}R(K){}_{\nu\rho]}{}^a\,,\\
0 & = & e^a_{[\mu}R(K)_{\nu\rho]}{}^a\,.
\end{eqnarray}
These tell us that $v^\mu R(K)_{\mu ab}$ (where we lowered the third index with a Kronecker delta) is STF in $a$ and $b$ and that $R(K)_{ab}{}^c$ is fixed in terms of $v^\mu R(K)_{\mu c}$, specifically
\begin{equation}
R(K)_{ab}{}^c=-\delta^c_a v^\mu R(K)_{\mu b}+\delta^c_b v^\mu R(K)_{\mu a}\,.
\end{equation}
These algebraic Bianchi identities do not fix $R(K)_{ab}$. For $d=2$ we thus have 5 independent $K$-curvatures components: $v^\mu R(K)_{\mu a}$, $R(K)_{ab}$ and $v^\mu R(K)_{\mu a}{}^b$. There are also differential Bianchi identities relating the $K$-curvatures but we will not need those.

Using equations \eqref{eq:deltaRKtemp} and \eqref{eq:deltaRK} as well as the transformations of the inverse vielbeine given in \eqref{eq:gaugetrafov} and \eqref{eq:gaugetrafoinve} we can show that
\begin{eqnarray}
    \delta R(K)_{ab} & = & \mathcal{L}_\chi R(K)_{ab}-3\Lambda_D R(K)_{ab}\nonumber\\
    &&+2\lambda_a v^\mu R(K)_{\mu b}-2\lambda_b v^\mu R(K)_{\mu a}\,,\\
    \delta\left(v^\mu R(K)_{\mu a}\right) & = & \mathcal{L}_\chi\left(v^\mu R(K)_{\mu a}\right)-3\Lambda_D v^\mu R(K)_{\mu a}\nonumber\\
    &&-\lambda^b v^\mu R(K)_{\mu a b}\,,\\ 
    \delta\left(v^\mu R(K)_{\mu ab}\right) & = & \mathcal{L}_\chi\left(v^\mu R(K)_{\mu ab}\right)-3\Lambda_D v^\mu R(K)_{\mu ab}\nonumber\\
    &&+\lambda_{ac}v^\mu R(K)_{\mu cb}+\lambda_{bc}v^\mu R(K)_{\mu ac}\,.
\end{eqnarray}
It is interesting to focus on the transformations of the $K$-curvatures under Carroll boosts with parameter $\lambda^a$. We see that under Carroll boosts we have the sequence
\begin{equation}\label{eq:CarrollboostKcurv}
    R(K)_{ab}\rightarrow v^\mu R(K)_{\mu a}\rightarrow v^\mu R(K)_{\mu ab}\rightarrow 0\,.
\end{equation}
We will refer to this as the $K$-curvature hierarchy. It means that we can consider 4 types of gauge invariant $K$-curvature constraints, namely
\begin{equation}\label{eq:hierarchyofrad}
\begin{array}{cccccccccc}
\hspace{-.5cm} i). & h^\mu_\rho h^\nu_\sigma R(K)_{\mu\nu} & = & 0\,, &  v^\mu R(K)_{\mu\nu}  & = & 0\,,&  v^\mu  e^a_\sigma R(K)_{\mu\nu}{}^a  & = & 0\\
\hspace{-.5cm}ii). & h^\mu_\rho h^\nu_\sigma R(K)_{\mu\nu} & \neq  & 0\,, &  v^\mu  R(K)_{\mu\nu}  & = & 0\,,&  v^\mu  e^a_\sigma R(K)_{\mu\nu}{}^a  & = & 0\\
\hspace{-.5cm}iii). & h^\mu_\rho h^\nu_\sigma R(K)_{\mu\nu} & \neq & 0\,, &  v^\mu  R(K)_{\mu\nu}  & \neq & 0\,,&  v^\mu  e^a_\sigma R(K)_{\mu\nu}{}^a  & = & 0\\
\hspace{-.5cm}iv). & h^\mu_\rho h^\nu_\sigma R(K)_{\mu\nu} & \neq  & 0\,, &  v^\mu R(K)_{\mu\nu}  & \neq & 0\,,&  v^\mu  e^a_\sigma R(K)_{\mu\nu}{}^a  & \neq & 0
\end{array}
\end{equation}
Here instead of working with spatial vielbein projections we work with spatial $h^\mu_\nu$ projections.

\subsection{Gauge fixing and $K$-curvature expressions}

The totality of all the curvature constraints have allowed us to solve for $f_\mu$, $f_\mu{}^a$, $v^\mu b_\mu$, $\omega_\mu{}^{ab}$, $\omega_\mu{}^a$ in terms of $\tau_\mu$, $e_\mu^a$, $C_{\mu\nu}$, $S$, $h^\rho_\mu b_\rho$ and derivatives thereof. We will next show that we can gauge fix $S$ and $h^\rho_\mu b_\rho$. One way to do this is to observe that
\begin{eqnarray}
    -v^\rho\omega_\rho{}^a e^a_\mu & = & b_\mu-\tilde b_\mu\,,\\
    e^\mu_a\omega_\mu{}^a & = & S\,,
\end{eqnarray}
where $\tilde b_\mu$ is given in \eqref{eq:tildeb} and furthermore that 
\begin{eqnarray}
 \hspace{-1cm}   \delta\left(v^\rho\omega_\rho{}^a\right) & = & \mathcal{L}_\chi\left(v^\rho\omega_\rho{}^a\right)-\Lambda_Dv^\rho\omega_\rho{}^a+\lambda^{ab}v^\rho\omega_\rho{}^b\nonumber\\
 &&+v^\mu \left(\partial_\mu\lambda^a-\omega_\mu{}^{ab}\lambda^b\right)-\sigma^a\,,\\
 \hspace{-1cm}   \delta\left(e^\mu_a\omega_\mu{}^a\right) & = & \mathcal{L}_\chi\left(e^\mu_a\omega_\mu{}^a\right)-\Lambda_D e^\mu_a\omega_\mu{}^a+\lambda_a v^\mu\omega_\mu{}^a\nonumber\\
 &&+e^\mu_a \left(\partial_\mu\lambda^a-\omega_\mu{}^{ab}\lambda^b\right)+2\sigma\,,
\end{eqnarray}
where we used \eqref{eq:gaugetrafoboostcon}. We can use the $\sigma$ and $\sigma^a$ transformations to choose the gauge
\begin{equation}\label{eq:gaugechoice}
    S=0\,,\qquad b_\mu=\tilde b_\mu\,,
\end{equation}
where we remind the reader that before we fixed the gauge we already knew that $v^\mu b_\mu=v^\mu\tilde b_\mu$.
An alternative choice is to make the gauge choice\footnote{This is the choice made in section 3.2 of \cite{Hartong:2026rbr}. The choice \eqref{eq:gaugechoice} is alluded to in footnote 14 of said paper.}
\begin{eqnarray}
    S=0\,,\qquad e^\mu_a b_\mu=0\,,
\end{eqnarray}
where $e^\mu_a b_\mu$ transforms as
\begin{equation}
    \delta\left(e^\mu_a b_\mu\right)=\mathcal{L}_\chi\left(e^\mu_a b_\mu\right)-\Lambda_D e^\mu_a b_\mu+\lambda^{ab}e^\mu_b b_\mu-\frac{1}{2}\lambda_a K+e^\mu_a\partial_\mu\Lambda_D+\sigma^a\,.
\end{equation}
This follows from \eqref{eq:trafoWeylcon} and \eqref{eq:vb}.
In this gauge we have $v^\rho\omega_\rho{}^a e^a_\mu = h^\rho_\mu\tilde b_\rho=a_\mu$.

In the following we will pick the gauge $S=0$ and $b_\mu=\tilde b_\mu$. Using \eqref{eq:vfa}--\eqref{eq:ef}, as well as the following results for the barred $B$ and $J$ curvatures \eqref{eq:vbarRBab}, \eqref{eq:spatialtracebarRB}, \eqref{eq:vbarRJ} and \eqref{eq:doubletracebarRJ}, we find that in this gauge
\begin{eqnarray}
    f_\mu & = & \overset{(0)}{P}_\mu+\frac{1}{2}\tau_\mu\left(\overset{(0)}{S}-a^2\right)\,,\label{eq:fmu}\\
    f_\mu{}^a & = & -\tau_\mu v^\rho e^\sigma_a\left(\partial_\rho\tilde b_\sigma-\partial_\sigma\tilde b_\rho\right)+\frac{1}{2}e^a_\mu\left(\overset{(0)}{S}-a^2\right)-\frac{1}{2}e^\nu_a N_{\mu\nu}\nonumber\\
    &&+\frac{1}{2}e^\nu_a\left(\mathcal{L}_v+\frac{1}{2}K\right)F_{\mu\nu}\,,\label{eq:fmua}
\end{eqnarray}
where $\overset{(0)}{P}_\mu$ is given in \eqref{eq:P0} and where $N_{\mu\nu}$ is the news tensor defined in \eqref{eq:newsdef}.

In the gauge \eqref{eq:gaugechoice} the $K$-curvatures are given by
\begin{eqnarray}
    v^\mu e^\nu_b R(K)_{\mu\nu}{}^a & = & -\frac{1}{2}e^\mu_a e^\nu_b\left[\mathcal{L}_v N_{\mu\nu}-2h^\rho_{\langle\mu}h^\sigma_{\nu\rangle}\left(\mathcal{D}_\rho+3a_\rho\right)G_\sigma\right]\,,\label{eq:veRK}\\
    v^\mu  R(K)_{\mu\nu} & = & \left(\mathcal{L}_v-\frac{1}{2}K\right)\overset{(0)}{P}_\nu+\frac{1}{2}h^\rho_\nu\left(\partial_\rho+2a_\rho\right)\left(\overset{(0)}{S}-a^2\right)\nonumber\\
    &&+\frac{1}{2}G_\sigma\left(C^\sigma{}_\nu-F^\sigma{}_\nu\right)\,,\\
    e^\mu_a e^\nu_b R(K)_{\mu\nu} & = & e^\mu_a e^\nu_b\left[\left(\partial_\mu+a_\mu\right)\overset{(0)}{P}_\nu-\left(\partial_\nu+a_\nu\right)\overset{(0)}{P}_\mu+\frac{1}{4}N_{\mu\rho}C^\rho{}_\nu-\frac{1}{4}N_{\nu\rho}C^\rho{}_\mu\right.\nonumber\\
    &&\left.+\left(\overset{(0)}{S}-a^2\right)F_{\mu\nu}\right]\,,\label{eq:eeRK}
\end{eqnarray}
where we defined
\begin{equation}\label{eq:G}
    G_\nu=-v^\mu \left(\partial_\mu\tilde b_\nu-\partial_\nu\tilde b_\mu\right)\,.
\end{equation}
As shown in \cite{Hartong:2026rbr} we can write $v^\mu  R(K)_{\mu\nu}$ also as
\begin{eqnarray}
     v^\mu  R(K)_{\mu\nu} & = & \frac{1}{2}h^{\rho\sigma}\mathcal{D}_\rho N_{\sigma\nu}+\frac{1}{2}h^{\rho}_\nu\left(\partial_\rho+2a_\rho\right)\left(\overset{(0)}{S}-a^2\right)\label{eq:PTWeylcov}\\
    &&-\frac{1}{2}G_\sigma F^{\sigma}{}_{\nu}+\frac{1}{2}h_{\rho\nu}\left(\mathcal{L}_v-\frac{3}{2}K\right)\mathcal{D}_\sigma F^{\sigma\rho}\,.\nonumber
\end{eqnarray}

In section 9.6 of \cite{Hartong:2026rbr} we showed that the variation of the Einstein--Hilbert action plus appropriate boundary terms placed on the cutoff hypersurface $r=\text{cst}$ when evaluated on shell is not Carroll boost invariant. This nonzero invariance of action under Carroll boosts was referred to as the Carroll boost anomaly and the expression for it is precisely the $K$-curvature $v^\mu  R(K)_{\mu\nu}$.

\section{Conformal Carroll algebra in the bulk}\label{sec:confcarbulk}

In this section we will connect the geometric approach of section \ref{sec:reviewCarcovBSgauge} with the algebraic approach of the previous section. 
An early version of these sorts of constructions can be found in \cite{Korovin:2017xqu}.

\subsection{Bulk gauge transformations and the conformal Carroll algebra}

In the gauge \eqref{eq:gaugechoice} we have $2e^a_{(\mu}\omega_{\nu)}{}^a=C_{\mu\nu}$. This follows from \eqref{eq:decompboostcon} and \eqref{eq:Sab}. We then observe that the near-boundary expansion of $g_{\mu\nu}$ as written in \eqref{eq:gexp} to order $r^0$ can be written as 
\begin{eqnarray}
    \overset{(-1)}{g}_{\mu\nu} & = & -\tau_\mu b_\nu-\tau_\nu b_\mu+\omega_\mu{}^a e^a_\nu+\omega_\nu{}^a e^a_\mu\,,\label{eq:g-1gaugecon}\\
    \overset{(0)}{g}_{\mu\nu} & = & -\tau_\mu f_\nu-\tau_\nu f_\mu+h^\rho_\mu h^\sigma_\nu\overset{(0)}{\Pi}_{\rho\sigma}\,,\label{eq:g0gaugecon}
\end{eqnarray}
where $f_\mu$ is given in \eqref{eq:fmu}.
Note that the connection $f_\mu{}^a$, given in \eqref{eq:fmua}, does not feature in this near-boundary expansion of the metric\footnote{We will see below that $f_\mu{}^a$ does feature in the Weyl tensor for this asymptotic metric.}.  
Also note that $h^\rho_\mu h^\sigma_\nu\os{0}{\Pi}_{\rho\sigma}$ is not captured by the gauging of the conformal Carroll algebra. The same is true for even lower order coefficients such as $\overset{(1)}{g}_{\mu\nu}$ which in 4D contains the boundary energy-momentum tensor.

The observation \eqref{eq:g-1gaugecon} and \eqref{eq:g0gaugecon} prompts us to take a closer look at the relation between the near-boundary expansion and the gauging of the conformal algebra. To this end we reconsider the near-boundary expansion of the metric. However, unlike what we did in section \ref{subsec:Carcovgauge} we will not want to assume any gauge choices in the bulk (with one exception that we explain shortly) in order not to prejudice the origin of the conformal Carroll gauge structure.

We decompose the metric in a null bein basis as 
\begin{equation}
    ds^2=-2UV+E^aE^a\,.
\end{equation}
The nonzero Lorentzian inner products between these vielbeine are as follows
\begin{equation}
    U\cdot V=-1\,,\qquad E^a\cdot E^b=\delta^{ab}\,.
\end{equation}
The bulk diffeomorphism and local Lorentz transformations act on the vielbeine as follows
\begin{subequations}
  \label{eq:gen-null-bulk-frame-diffeo-lor-action}
  \begin{align}
    \delta U_M
    &= \mathcal{L}_\xi U_M
    + \Lambda U_M
    + \Sigma_a E^a_M\,,
    \\
    \delta V_M
    &= \mathcal{L}_\xi V_M
    - \Lambda V_M
    + \Lambda_a E^a_M\,,
    \\
    \delta E^a_M
    &= \mathcal{L}_\xi E^a_M
    + \Lambda^a U_M
    + \Sigma^a V_M
    + \Lambda^a{}_b E^b_M\,.
  \end{align}
\end{subequations}
We split the bulk coordinates $x^M=(r,x^\mu)$ and we fix a gauge such that 
\begin{equation}\label{eq:gaugechoiceV}
    V^M\partial_M = -\partial_r\,.
\end{equation}
This implies that the following radial components are fixed to be
\begin{equation}
    U_r=1\,,\qquad V_r=0\,,\qquad E^a_r=0\,.
\end{equation}
The choice \eqref{eq:gaugechoiceV} is the only gauge choice we shall make in the bulk in this section. The reason behind this is explained in section 3 of \cite{Hartong:2026rbr}. As shown there and reconfirmed here, in this gauge the bulk vielbeine at leading order in the radial expansion induce a manifest Carroll geometry on the boundary.

We impose the boundary conditions
\begin{subequations}
  \label{eq:boundary-frame-with-M-falloff-mu-parametrization}
  \begin{eqnarray}
    V_\mu
    &=& \tau_\mu
    +r^{-1}\overset{(1)}{V}_\mu+ \mathcal{O}(r^{-2})\,,
    \\
    E^a_\mu
    &=& r e^a_\mu
    +\overset{(0)}{E}{}^a_\mu +\mathcal{O}(r^{-1})\,,\\
    U_\mu & = & rb_\mu+\overset{(0)}{U}_\mu+\mathcal{O}r^{-1})\,.
  \end{eqnarray}
\end{subequations}
The vacuum Einstein equations tells us that $v^\mu \overset{(1)}{V}_\mu=0$, so that we can write
\begin{equation}\label{eq:EOMconstraint}
    \overset{(1)}{V}_\mu=\overset{(1)}{V}{}^a e^a_\mu\,.
\end{equation}.

The gauge choice and boundary conditions are preserved by the bulk diffeomorphisms and local Lorentz transformations provided we have the following radial falloff conditions for the gauge parameters
\begin{subequations}
  \label{eq:boundary-frame-gauge-tr-expansions-diffeos}
  \begin{align}
    \xi^r
    &= r \Lambda_D
    + \os{0}{\xi}^r
    + \mathcal{O}(r^{-1})\,,
    \\
    \xi^\mu
    &= \chi^\mu
    + r^{-1} \overset{(1)}{\xi}{}^\mu
    + \mathcal{O}(r^{-2})\,,
  \end{align}
\end{subequations}
while the null boosts $\Lambda$,
null rotations $\Lambda^a$ and $\Sigma^a$
and spatial rotations $\Lambda^{ab}=-\Lambda^{ba}$ must satisfy
\begin{subequations}
  \label{eq:boundary-frame-gauge-tr-expansions-local-gauge}
  \begin{align}
    \Lambda
    &= -\Lambda_D
    + \mathcal{O}(r^{-1})\,,
    \\
    \Lambda^a
    &= r^{-1} \lambda^a
    + \mathcal{O}(r^{-2})\,,
    \\
    \Sigma^a
    &= \sigma^a+r^{-1}\overset{(1)}{\Sigma}{}^a
    + \mathcal{O}(r^{-2})\,,
    \\
    \Lambda^{ab}
    &= \lambda^{ab}+r^{-1}\overset{(1)}{\Lambda}{}^{ab}
    + \mathcal{O}(r^{-2})\,.
  \end{align}
\end{subequations}
The gauge choice \eqref{eq:gaugechoiceV} tells us that
\begin{equation}
    \overset{(1)}{\xi}{}^\mu=e^\mu_a\lambda^a\,.
\end{equation}
This induces the following transformation of $\tau_\mu$, $e_\mu^a$ and $b_\mu$
\begin{subequations}
  \label{eq:boundary-frame-gauge-tr-with-Vr-fixed}
  \begin{align}
    \delta \tau_\mu
    &= \mathcal{L}_\chi \tau_\mu
    + \Lambda_D \tau_\mu
    + \lambda_a e^a_\mu\,,
    \\
    \delta e^a_\mu
    &= \mathcal{L}_\chi e^a_\mu
    + \Lambda_D e^a_\mu
    + \lambda^a{}_b e^b_\mu\,,
    \\
    \delta b_\mu
    &= \mathcal{L}_\chi b_\mu
    + \partial_\mu \Lambda_D
    + \sigma_a e^a_\mu\,.
  \end{align}
\end{subequations}
Note that this agrees with \eqref{eq:trafotau}, \eqref{eq:trafoe} and \eqref{eq:trafoWeylcon} that we obtained from the gauging procedure.

The next-to-leading order connections in the radial expansion of $U_\mu$ and $E_\mu^a$ can be shown to transform as
\begin{eqnarray}
    \delta\os{0}{U}_\mu & = & \mathcal{L}_\chi \os{0}{U}_\mu-\Lambda_D \os{0}{U}_\mu+\partial_\mu\sigma+\sigma b_\mu+\sigma^a\os{0}{E}{}^a_\mu-\lambda^a f_\mu{}^a\nonumber\\
    &&+\os{1}{\xi}{}^\rho R(D)_{\rho\mu}+e^a_\mu\left(\os{1}{\xi}{}^\rho f_\rho{}^a+\os{1}{\Sigma}{}^a\right)\,,\\
    \delta \os{0}{E}{}^a_\mu & = & \mathcal{L}_\chi \os{0}{E}{}^a_\mu+\partial_\mu\lambda^a+\lambda^{ab}\os{0}{E}{}^b_\mu+\sigma e^a_\mu+\sigma^a\tau_\mu-\omega_\mu{}^{ab}\lambda^b\nonumber\\
    &&+\os{1}{\xi}{}^\rho R(P)_{\rho\mu}{}^a+\left(\os{1}{\xi}{}^\rho\omega_\rho{}^{ab}+\os{1}{\Lambda}{}^{ab}\right)e^b_\mu\,,
\end{eqnarray}
where we defined
\begin{equation}
\sigma=\os{0}{\xi}{}^r+\os{1}{\xi}{}^\rho b_\rho\,. 
\end{equation}
It can be shown that the $f_\mu{}^a$ and $\omega_\mu{}^{ab}$ connections drop out on the right-hand side. To see this one needs to use the expressions for the $P_a$ and $D$ curvatures.
We included them to make the connection with the gauging perspective more manifest. We now compare these results with the gauge transformations of $f_\mu$ and $\omega_\mu{}^a$ given in \eqref{eq:gaugetrafoboostcon} and \eqref{eq:deltaf}. We see that upon using the curvature constraints, which imply that $R(P)_{\rho\mu}{}^a=R(D)_{\mu\nu}=0$, we get almost perfect agreement if we identify $\os{0}{U}_\mu$ with $f_\mu$ and  $\os{0}{E}^a_\mu$ with $\omega_\mu{}^a$. The only terms that do not agree are the gauge transformations with parameters 
$\os{1}{\xi}{}^\rho f_\rho{}^a+\os{1}{\Sigma}{}^a$ and 
$\os{1}{\xi}{}^\rho\omega_\rho{}^{ab}+\os{1}{\Lambda}{}^{ab}$. These correspond to gauge transformations that are not included in the conformal Carroll algebra. 

The origin of these extra gauge transformations is easy to understand. If we write 
\begin{equation}
    g_{\mu\nu}=-U_\mu V_\nu-U_\nu V_\mu+E^a_\mu E^a_\nu\,,
\end{equation}
then we get
\begin{eqnarray}
    \overset{(-1)}{g}_{\mu\nu} & = & -b_\mu\tau_\nu-b_\nu\tau_\mu+e^a_\mu\overset{(0)}{E}{}^a_\nu+e^a_\nu\overset{(0)}{E}{}^a_\mu\,,\\
    \overset{(0)}{g}_{\mu\nu} & = & -\overset{(0)}{U}_\mu\tau_\nu-\overset{(0)}{U}_\nu\tau_\mu+\overset{(0)}{E}{}^a_\mu\overset{(0)}{E}{}^a_\nu\nonumber\\
    &&+e^a_\mu\left(\overset{(1)}{E}{}^a_\nu-\overset{(1)}{V}{}^a\tau_\nu\right)+e^a_\nu\left(\overset{(1)}{E}{}^a_\mu-\overset{(1)}{V}{}^a\tau_\mu\right)\,,
\end{eqnarray}
where we used \eqref{eq:EOMconstraint}.
The gauge transformation with the antisymmetric parameter $\os{1}{\xi}{}^\rho\omega_\rho{}^{ab}+\os{1}{\Lambda}{}^{ab}$ leaves the combination $e^a_\mu\overset{(0)}{E}{}^a_\nu+e^a_\nu\overset{(0)}{E}{}^a_\mu$ invariant. Hence, $\overset{(-1)}{g}_{\mu\nu}$ does not notice this gauge transformation. Likewise if we consider $\overset{(0)}{g}_{\mu\nu}$ then the part $-\overset{(0)}{U}_\mu\tau_\nu-\overset{(0)}{U}_\nu\tau_\mu+\overset{(0)}{E}{}^a_\mu\overset{(0)}{E}{}^a_\nu$ transforms under the gauge transformations with parameters $\os{1}{\xi}{}^\rho f_\rho{}^a+\os{1}{\Sigma}{}^a$ and 
$\os{1}{\xi}{}^\rho\omega_\rho{}^{ab}+\os{1}{\Lambda}{}^{ab}$ in such a way that the result is of the form $e^a_\mu \alpha^a_\nu+e^a_\nu \alpha^a_\mu$ for some $\alpha^a_\mu$. This can thus be cancelled by an appropriate transformation of 
$\overset{(1)}{E}{}^a_\nu-\overset{(1)}{V}{}^a\tau_\nu$.

Another way to phrase this is to say that the algebra of gauge transformations that acts on $\tau_\mu, b_\mu, \overset{(0)}{U}_\mu$ as well as $e^a_\mu, \overset{(0)}{E}{}^a_\mu$ is larger than what we get from the gauging of the conformal Carroll algebra, but that the latter can be obtained as a quotient of this larger bulk gauge algebra where we quotient out the transformations with parameters $\os{1}{\xi}{}^\rho f_\rho{}^a+\os{1}{\Sigma}{}^a$ and 
$\os{1}{\xi}{}^\rho\omega_\rho{}^{ab}+\os{1}{\Lambda}{}^{ab}$.

It would be interesting to see if there exists a larger gauge algebra that can account for more subleading terms in the $1/r$ expansion than the conformal Carroll algebra can account for.

\subsection{$K$-curvatures in the bulk}

For solutions to the vacuum Einstein equations the Riemann tensor is equal to the Weyl tensor. The latter has 10 independent components. In appendix \ref{app:Weyltensor} we use techniques developed in \cite{Hartong:2026rbr} to compute the Riemann tensor for an asymptotic geometry of the form as detailed in section \ref{subsec:Carcovgauge}.

In appendix \ref{app:Weyltensor} we show that at leading order the independent components of the Weyl tensor $C_{MNPQ}$ are
\begin{eqnarray}
    U^\rho U^\sigma C_{r\rho r\sigma} & = & r^{-3}\left(-v^\rho v^\sigma \overset{(1)}{g}_{\rho\sigma}-\frac{1}{16}K\left(F^2-C^2\right)\right.\nonumber\\
    &&\left.-\frac{1}{8}\left(\mathcal{L}_v-K\right)F^2-\frac{1}{4}C\cdot N\right)+\cdots\,,\label{eq:Weylmass}\\
    U^\rho\Pi^\sigma_\alpha C_{r\rho r\sigma} & = & r^{-3}\left(-\frac{3}{2}v^\rho g^\sigma_\alpha \overset{(1)}{g}_{\rho\sigma}+a_\rho D^\rho{}_\alpha+\frac{3}{32}a_\alpha\left(F^2-C^2\right)\right.\nonumber\\
    &&\left.-\frac{3}{32}h^\rho_\alpha\left(\partial_\rho+2a_\rho\right)\left(F^2-C^2\right)\right)+\cdots\,,\label{eq:Weylmomentum}\\
    \Pi^\rho_{\langle\alpha}\Pi^\sigma_{\beta\rangle} C_{r\rho r\sigma} & = & -r^{-2}D_{\alpha\beta}+\cdots\,,\label{eq:Dterm}\\
    U^\nu \Pi_\alpha^\rho\Pi^\sigma_\beta C_{r\nu\rho\sigma} & = & r^{-1}h^\rho_\alpha h^\sigma_\beta R(K)_{\rho\sigma}+\cdots\,,\label{eq:hhRK}\\
    U^\nu U^\rho\Pi^\sigma_\alpha C_{r\nu\rho\sigma} & = & r^{-1}v^\mu  R(K)_{\mu\alpha}+\cdots\,,\label{eq:anomaly}\\
    U^\mu U^\rho\Pi^\nu_{\langle\alpha}\Pi^\sigma_{\beta\rangle}C_{\mu\nu\rho\sigma} & = & -r v^\mu h^\nu_{\langle\alpha}e^a_{\beta\rangle}R(K)_{\mu\nu}{}^a+\cdots\,.\label{eq:timedernews}
\end{eqnarray}
In section \ref{sec:KcurvBondi} we will relate $U^\rho U^\sigma C_{r\rho r\sigma}$ and $U^\rho\Pi^\sigma_\alpha C_{r\rho r\sigma}$ at order $r^{-3}$ to the Bondi mass and angular momentum aspects.

We see that \eqref{eq:Dterm} leads to a violation of peeling whenever $D_{\alpha\beta}$ is nonzero. This can happen either when we allow for certain $\log r$ terms in the near-boundary expansion or when the boundary topology allows for a nonzero $D_{\alpha\beta}$. In the absence of $\log r$ terms $D_{\alpha\beta}$ is constrained by
\begin{equation}
    \mathcal{L}_v D_{\mu\nu}=0\,,\qquad h^{\rho\mu}\mathcal{D}_\rho D_{\mu\nu}=0\,,
\end{equation}
and depending on the topology of the boundary this either has or has no globally well-defined regular solutions. It is well-known for example that on the 2-sphere there are no such globally well-defined tensors. When we add the $\log r$ terms described by \eqref{eq:g1,1} then $D_{\mu\nu}$ is only constrained by $\mathcal{L}_v D_{\mu\nu}=0$. 

The remaining 5 components of the Weyl tensor, \eqref{eq:hhRK}--\eqref{eq:timedernews}, at leading order provide us with the $K$-curvatures. There is thus a clear relation between the $K$-curvatures and some of the Newman--Penrose scalars. By using the results of the next subsection, where we compute the $K$-curvatures in standard Bondi coordinates, we can see that $v^\mu h^\nu_{\langle\alpha}e^a_{\beta\rangle}R(K)_{\mu\nu}{}^a$ corresponds to the Newman--Penrose scalar $\psi_4$, that $v^\mu  R(K)_{\mu\nu}$ corresponds to $\psi_3$ and finally that $h^\mu_\alpha h^\nu_\beta R(K)_{\mu\nu}$ corresponds to the imaginary part of $\psi_2$.

We conclude from our analysis of the Weyl tensor that the $K$-curvatures capture the radiative properties of the spacetime in a boundary Carroll-covariant manner. The four types of spacetimes defined in \eqref{eq:hierarchyofrad} can thus be viewed as a hierarchy of (non-)radiativeness. Referring to the four types of $K$-curvature constraints in \eqref{eq:hierarchyofrad} we will distinguish the following four types:
\begin{equation}
\begin{array}{ccl}
    \text{i).} & &\text{vacuum} \\
    \text{ii).} & &\text{strongly non-radiative}\\
    \text{iii).} && \text{weakly non-radiative}\\
    \text{iv).} & &\text{radiative}
\end{array} 
\end{equation}
This is similar to various notions of non-radiative spacetimes discussed in \cite{Freidel:2021qpz}.

\subsection{$K$-curvatures in standard Bondi coordinates}\label{subsec:standardBScoord}

One of the aims of this paper is to provide an algebraic and boundary covariant notion of concepts such as vacuum and radiative shear as well as to understand the role of the $K$-curvatures in the Bondi loss equations (see section \ref{subsec:newBondiform} below). However, for familiarity we provide the expressions for the $K$-curvature in what we refer to as standard Bondi coordinates whose definition is given below. Before we gauge fix our boundary geometry we first discuss a useful ADM-like parametrisation of the Carroll geometry.

Rather than covariantising with respect to the full Carroll structure we can also split our boundary coordinates as $x^\mu=(u, x^i)$ and covariantise with respect to the $x^i$ coordinates. It is useful to parameterise the Carroll data as follows
\begin{eqnarray}
    h_{\mu\nu}dx^\mu dx^\nu & = & \sigma_{ij}\left(dx^i+N^i du\right)\left(dx^j+N^j du\right)\,,\\
    \label{eq:carroll-adm-tau-decomp}
    \tau_\mu dx^\mu & = & Ndu +\tau_i \left(dx^i+N^i du\right)\,,\\
    v^\mu\partial_\mu & = & -\frac{1}{N}\left(\partial_u-N^i\partial_i\right)\,,\label{eq:vparam}
\end{eqnarray}
where $\sigma_{ij}$ is a 2-dimensional Riemannian metric. We see that $h_{\mu\nu}$ contains 5=3+2 independent components, namely $\sigma_{ij}$ and $N^i$.

In this parametrisation we have
\begin{eqnarray}
K_{ij} & = & \frac{1}{2N}\left(\partial_u \sigma_{ij}-D_i N_j-D_j N_i\right)\,,\label{eq:Kij}\\
a_i & = & N^{-1}\partial_i N-N^{-1}\left(\partial_u\tau_i-\mathcal{L}_N\tau_i\right)\,,\\
F_{ij} & = & \partial_i\tau_j-\partial_j\tau_i\,.
\end{eqnarray}
The $u$ components of $K_{\mu\nu}$, $a_\mu$ and $F_{\mu\nu}$ follow from the fact that these are all spatial tensors, so that their contraction with $v^\mu$ in \eqref{eq:vparam} gives zero. In the expression for $K_{ij}$ we denote by $D_i$ the covariant derivative with respect to the Levi-Civita connection for the $2$-dimensional Riemannian manifold whose coordinates are $x^i$ and whose metric is $h_{ij}=\sigma_{ij}$ with inverse $\sigma^{jk}$. Note that the latter is not equal to $h^{jk}$. In fact we have
\begin{eqnarray}
    h^{ij} & = & \sigma^{ij}+N^{-1}\left(N^i\tau^j+N^j\tau^i\right)+N^{-2}\tau^2 N^i N^j\,,\\
    h^{ui} & = & -N^{-1}\tau^i-N^{-2}\tau^2 N^i\,,\\
    h^{uu} & = & N^{-2}\tau^2\,,
\end{eqnarray}
where we defined $\tau^i=\sigma^{ij}\tau_j$ and $\tau^2=\sigma^{kl}\tau_k\tau_l$.
It follows that 
\begin{equation}
    K=h^{\mu\nu}K_{\mu\nu}=\sigma^{ij}K_{ij}\,.
\end{equation}

Equation \eqref{eq:Kij} tells us that $K_{ij}$ is the extrinsic curvature of a $u=\text{cst}$ hypersurface. The constraint \eqref{eq:Kconstraint} states that these hypersurfaces are totally umbilical.

If we decompose $\sigma_{ij}$ in terms of Riemannian vielbeine as follows
\begin{equation}
    \sigma_{ij}=\delta_{ab}f^a_i f^b_j\,,
\end{equation}
then writing $h_{\mu\nu}dx^\mu dx^\nu=\delta_{ab}e^a_\mu e^b_\nu dx^\mu dx^\nu$ we have
\begin{equation}
    e^a_\mu dx^\mu=f^a_i\left(dx^i+N^i du\right)\,.
\end{equation}
For the integration measure we have
\begin{equation}
    e=N\sqrt{\sigma}\,.
\end{equation}

A Carroll boost acts on $\tau_\mu$ as $\tau'_\mu=\tau_\mu+\lambda_a e^a_\mu$. Hence, this gives in components
\begin{equation}
    N'=N\,,\qquad \tau'_i=\tau_i+\lambda_a  f^a_i\,.
\end{equation}
We can thus always completely fix the Carroll boosts by setting $\tau_i=0$, so that $\tau=Ndu$. We can fix the Weyl transformation by setting $N=1$. We will not put any restrictions on the topology of $\mathcal{I}^+$ other than that it is of the form $\mathbb{R}\times\Sigma$ where $\Sigma$ can be either compact or non-compact. After setting $\tau=du$ by fixing Carroll boosts and Weyl transformations we still have 3 boundary diffeomorphisms that we can use to fix the Carroll metric $h_{\mu\nu}$. We will use these to fix $N^i=0$ and $K=0$. The residual gauge transformations are those for which
\begin{eqnarray}
    \lambda_i & = & -\partial_i\chi^u\,,\\
    \Lambda_D & = & -\partial_u\chi^u\,,\\
    \partial_u\chi^i & = & 0\,,\\
    \partial_u^2\chi^u & = & 0\,.
\end{eqnarray}
We are now in a gauge in which
\begin{eqnarray}
    \tau & = & du\,,\label{eq:taustandardBS}\\
    h & = & \sigma_{ij}dx^i dx^j\,,\label{eq:hstandardBS}
\end{eqnarray}
where $\sigma_{ij}$ does not depend on $u$ but is otherwise arbitrary. We will refer to a boundary coordinate system for which $\tau_\mu$ and $h_{\mu\nu}$ are given by \eqref{eq:taustandardBS} and \eqref{eq:hstandardBS} as standard Bondi--Sachs coordinates.

The residual gauge transformations of the standard Bondi--Sachs coordinates are
\begin{equation}
    \delta\sigma_{ij}=\mathcal{L}_\xi \sigma_{ij}-2\partial_u\chi^u \sigma_{ij}\,,
\end{equation}
where $\mathcal{L}_\xi$ denotes the Lie derivative along $\xi^i$ which is an arbitrary 2-dimensional vector that is independent of $u$ and $\partial_u\chi^u$ is any 2-dimensional function independent of $u$. There is enough residual gauge freedom to fix $\sigma_{ij}$ fully (locally). Once we do this the residual gauge transformations are those for which 
\begin{equation}
    \delta\sigma_{ij}=\mathcal{L}_\xi \sigma_{ij}-2\partial_u\chi^u \sigma_{ij}=0\,.
\end{equation}
This implies
\begin{eqnarray}
    \partial_u\chi^u & = & \frac{1}{2}D_i\chi^i\,,\\
    D_i\chi_j+D_j\chi_i & = & \sigma_{ij} D_k\chi^k\,,
\end{eqnarray}
where $D_i$ is a covariant derivative with respect to the 2-dimensional Levi-Civita connection. Hence, we get
\begin{eqnarray}
    \chi^u & = & a(x)+\frac{1}{2}u D_k\chi^k\,,\\
    \chi^i & = & \chi^i(x)\,,
\end{eqnarray}
where $\chi^i$ is a conformal Killing vector of $\sigma_{ij}$. These are the well-known generators of the extended BMS$_4$ algebra with $a(x)$ describing supertranslations and $\chi^i(x)$ superrotations.

If we evaluate the $K$-curvatures using these standard BS coordinates we obtain
\begin{eqnarray}
R(K)_{ui}{}^a e^a_j & = & -\frac{1}{2} \partial_u N_{ij}\,,\\
     R(K)_{ui} & = & \partial_u\overset{(0)}{P}_i-\frac{1}{2}\partial_i\overset{(0)}{S}\,,\\
    R(K)_{ij} & = & \partial_i\overset{(0)}{P}_j-\partial_j\overset{(0)}{P}_i+\frac{1}{4}N_{ik}C^k{}_j-\frac{1}{4}N_{jk}C^k{}_i\,,    
\end{eqnarray}
where
\begin{eqnarray}
    \overset{(0)}{S} & = & \frac{1}{2}R[\sigma]\,,\\
    \overset{(0)}{P}_i & = & -\frac{1}{2}D_k C^k{}_i\,,\\
        N_{ij} & = & \partial_u C_{ij}\,.
\end{eqnarray}

We will solve for the vacuum case when all three $K$-curvatures vanish. We will solve this following the hierarchy \eqref{eq:CarrollboostKcurv}. Consider first the case $R(K)_{ui}{}^a e^a_j=0$. Solving this gives
\begin{equation}\label{eq:weaknonradC}
    C_{ij}=S_{ij}+u T_{ij}\,,
\end{equation}
where $S_{ij}$ and $T_{ij}$ are independent of $u$ and STF with respect to $\sigma^{ij}$. Let us next add to this the second curvature condition and $R(K)_{ui}=0$. This now  
reads
\begin{equation}\label{eq:divergenceT}
    D_k T^{k}{}_i=-\frac{1}{2}\partial_i R[\sigma]\,.
\end{equation}
The remaining curvature is then
\begin{equation}\label{eq:RKstandbdry}
    R(K)_{ij}=-\frac{1}{2}\left(\partial_i D_k S^{k}{}_j-\frac{1}{2}T_{ik}S^{k}{}_j-\left(i\leftrightarrow j\right)\right)\,.
\end{equation}
Note that this is independent of $u$ and an equation for $S_{ij}$.
Setting $R(K)_{ij}=0$ leads to the following expression for $S_{ij}$,
\begin{equation}
    S_{ij}=-2D_{\langle i}\partial_{j\rangle} C+CT_{ij}\,.
\end{equation}

We conclude that in standard BS coordinates the vacuum shear, i.e. the solution for setting all the $K$-curvatures to zero, is given by
\begin{equation}\label{eq:CvacBScoord}
C^{\text{vac}}_{ij}=-2D_{\langle i}\partial_{j\rangle} C+\left(C+u\right)T_{ij}\,,
\end{equation}
where $T_{ij}$ obeys \eqref{eq:divergenceT}. This agrees with the literature (see e.g. \cite{Donnay:2023mrd}). In the next section we will be more precise about the form of $T_{ij}$ (see footnote \ref{fn:Tij}).

\section{Vacuum shear}\label{sec:vacshear}

In this section we will show that the vanishing of all the $K$-curvatures can be used to construct a boundary covariant expression for the vacuum (or soft) shear. The final expression for the vacuum shear is given in \eqref{eq:vacshearfinal} and is written in terms of boundary Carroll geometric objects as well as two Carrollian scalar fields denoted by $F$ and $\Phi$.

\subsection{K-curvatures measure deviation from vacuum shear}

The $K$-curvatures lead to the notion of vacuum shear and the corresponding vacuum news. These are the solutions for when all the $K$-curvatures vanish. This can be made manifest by rewriting the $K$-curvatures as we will do in this section. This will lead to a Carroll-covariant definition of the vacuum shear in terms of two boundary scalar fields.

The $K$-curvature hierarchy under the Carroll boosts \eqref{eq:CarrollboostKcurv} suggests that setting the curvatures to zero should be done in accordance with the gauge structure. So we start with the weakest condition which is setting $v^\mu e^\nu_b R(K)_{\mu\nu}{}^a$ equal to zero. Before solving $v^\mu e^\nu_b R(K)_{\mu\nu}{}^a=0$ we shall first rewrite \eqref{eq:veRK}, which we repeat here for convenience
\begin{equation}\label{eq:cuvr1}
    v^\mu e^\nu_b R(K)_{\mu\nu}{}^a = -\frac{1}{2}e^\mu_a e^\nu_b\left[\mathcal{L}_v N_{\mu\nu}-2h^\rho_{\langle\mu}h^\sigma_{\nu\rangle}\left(\mathcal{D}_\rho+3a_\rho\right)G_\sigma\right]\,.
\end{equation}
We remind the reader that 
\begin{equation}\label{eq:defG}
    G_\nu=-v^\mu \left(\partial_\mu\tilde b_\nu-\partial_\nu\tilde b_\mu\right)\,.
\end{equation}

In order not to digress into a lot of technical details we have decided to move the calculational details to appendix \ref{app:rewritingcurv1}. There we derive the following result
\begin{equation}\label{eq:defeqvacNews}
    2h^\rho_{\langle\mu}h^\sigma_{\nu\rangle}\left(\mathcal{D}_\rho+3a_\rho\right)G_\sigma=\mathcal{L}_v N^{\text{vac}}_{\mu\nu}\,.
\end{equation}
In here we defined the vacuum news tensor $N^{\text{vac}}_{\mu\nu}$ as
\begin{equation}\label{eq:vacnews}
    N^{\text{vac}}_{\mu\nu}=2h^\rho_{\langle\mu}h^\sigma_{\nu\rangle}\left(\mathcal{D}_\rho\partial_\sigma\varphi+\partial_\rho\varphi\partial_\sigma\varphi-\mathcal{D}_\rho a_\sigma-a_\rho a_\sigma\right)+\mathcal{T}_{\mu\nu}\,,
\end{equation}
where at this stage $\mathcal{T}_{\mu\nu}$ is any spatial STF tensor that obeys\footnote{Later we will also require that $h^{\rho\sigma}\mathcal{D}_\rho \mathcal{T}_{\sigma\nu}=0$ for reasons that will become clear below. } 
\begin{equation}
    \mathcal{L}_v \mathcal{T}_{\mu\nu}=0\,.
\end{equation}
With the definition of the vacuum news tensor we can thus write
\begin{equation}\label{eq:cuvr1-v2}
    v^\mu e^\nu_b R(K)_{\mu\nu}{}^a = -\frac{1}{2}e^\mu_a e^\nu_b\mathcal{L}_v\left(N_{\mu\nu}-N^{\text{vac}}_{\mu\nu}\right)\,.
\end{equation}

The next curvature in the hierarchy is \eqref{eq:PTWeylcov} which we repeat here for convenience
\begin{eqnarray}
     v^\mu  R(K)_{\mu\nu} & = & \frac{1}{2}h^{\rho\sigma}\mathcal{D}_\rho N_{\sigma\nu}+\frac{1}{2}h^{\rho}_\nu\left(\partial_\rho+2a_\rho\right)\left(\overset{(0)}{S}-a^2\right)\label{eq:cuvr2}\\
    &&-\frac{1}{2}G_\sigma F^{\sigma}{}_{\nu}+\frac{1}{2}h_{\rho\nu}\left(\mathcal{L}_v-\frac{3}{2}K\right)\mathcal{D}_\sigma F^{\sigma\rho}\,.\nonumber
\end{eqnarray}
This can likewise be written very succinctly in terms of the vacuum news tensor. This follows from an identity that is derived in appendix \ref{app:rewritingcurv2} and given in equation \eqref{eq:divNvac} which we repeat here for convenience
\begin{eqnarray}
    h^{\rho\sigma}\mathcal{D}_\rho N^{\text{vac}}_{\sigma\nu} & = & -h^\sigma_\nu\left(\partial_\sigma+2a_\sigma\right)\left(\overset{(0)}{S}-a^2\right)+G_\rho F^\rho{}_\nu \nonumber\\
    &&-h_{\nu\rho}\left(\mathcal{L}_v-\frac{3}{2}K\right)\mathcal{D}_\sigma F^{\sigma\rho}\,,
\end{eqnarray}
where we now also assume that $\mathcal{T}_{\mu\nu}$ additionally obeys
\begin{equation}\label{eq:divX}
    h^{\rho\sigma}\mathcal{D}_\rho \mathcal{T}_{\sigma\nu}=0\,.
\end{equation}
With the help of this result equation \eqref{eq:cuvr2} becomes
\begin{equation}\label{eq:cuvr2-v2}
     v^\mu  R(K)_{\mu\nu}
    =\frac{1}{2}h^{\rho\sigma}\mathcal{D}_\rho\left(N_{\sigma\nu}-N^{\text{vac}}_{\sigma\nu}\right)\,.
\end{equation}

In order to do something similar with the last $K$-curvature, $e^\mu_a e^\nu_b R(K)_{\mu\nu}$, which we remind the reader is given by
\begin{eqnarray}
    e^\mu_a e^\nu_b R(K)_{\mu\nu} & = & e^\mu_a e^\nu_b\left[\left(\partial_\mu+a_\mu\right)\overset{(0)}{P}_\nu-\left(\partial_\nu+a_\nu\right)\overset{(0)}{P}_\mu+\frac{1}{4}N_{\mu\rho}C^\rho{}_\nu-\frac{1}{4}N_{\nu\rho}C^\rho{}_\mu\right.\nonumber\\
    &&\left.+\left(\overset{(0)}{S}-a^2\right)F_{\mu\nu}\right]\,,\label{eq:3rdKcurv}
\end{eqnarray}
we first need to find the vacuum shear tensor $C^{\text{vac}}_{\mu\nu}$ such that
\begin{equation}\label{eq:vacN}
    N^{\text{vac}}_{\mu\nu}=-\left(\mathcal{L}_v+\frac{1}{2}K\right)C^{\text{vac}}_{\mu\nu}\,.
\end{equation}
In appendix \ref{app:vacNtovacC} we show that the most general solution to this equation is given by
\begin{equation}\label{eq:Cvac}
    C^{\text{vac}}_{\mu\nu}=2P^\rho_{\langle\mu}P^\sigma_{\nu\rangle}\left(-\mathcal{D}_\rho\partial_\sigma F+F\left(\mathcal{D}_\rho a_\sigma+a_\rho a_\sigma\right)\right)+FN^{\text{vac}}_{\mu\nu}+Y_{\mu\nu}\,,
\end{equation}
where $F$ is any function that obeys
\begin{equation}
    \left(\mathcal{L}_v+\frac{1}{2}K\right)F=-1\,,
\end{equation}
and $Y_{\mu\nu}$ is any spatial STF tensor that for now obeys
\begin{equation}\label{eq:conY1}
    \left(\mathcal{L}_v+\frac{1}{2}K\right)Y_{\mu\nu}=0\,.
\end{equation}
Further below we will impose one extra condition on it.
We are now going to express $e^\mu_a e^\nu_b R(K)_{\mu\nu}$ in terms of the vacuum news and shear tensors.

The expression for $e^\mu_a e^\nu_b R(K)_{\mu\nu}$ is given in \eqref{eq:eeRK} which contains $\overset{(0)}{P}_\mu$ that is defined in \eqref{eq:P0}. 
In order to understand the solutions to $e^\mu_a e^\nu_b R(K)_{\mu\nu}=0$
we need to know what $\overset{(0)}{P}_\mu$ equates to when the shear is given by the vacuum shear. We denote this expression by $\overset{(0)}{P}{}^{\text{vac}}_\mu$. In appendix \ref{app:P0vac} we show that
\begin{eqnarray}
    \overset{(0)}{P}{}^{\text{vac}}_\mu & = & \left[-\frac{1}{2}N^{\text{vac}}{}^\rho{}_\mu+\frac{1}{2}\left(S^{(0)}-a^2\right)h^\rho_\mu-\frac{1}{2}h^{\kappa\rho}\left(\mathcal{L}_v+\frac{1}{2}K\right)F_{\kappa\mu}\right]\left(\partial_\rho-a_\rho\right)F\nonumber\\
    &&+\frac{1}{2}h^\rho_\mu\left(\partial_\rho+a_\rho\right)\left[h^{\kappa\lambda}\left(\mathcal{D}_\kappa-a_\kappa\right)\left(\partial_\lambda-a_\lambda\right)F-Fh^{\kappa\lambda}\mathcal{D}_\kappa\left(\partial_\lambda\varphi-a_\lambda\right)\right]\,.\nonumber\\
    &&
\end{eqnarray}

The goal is to express \eqref{eq:3rdKcurv} entirely in terms of $C_{\mu\nu}$ and $C^{\text{vac}}_{\mu\nu}$ and the corresponding news tensors. To achieve this we will use the following identity
\begin{equation}\label{eq:idcurv3}
    h^\mu_\alpha h^\nu_\beta\left(\left(\partial_\mu+a_\mu\right)\overset{(0)}{P}{}^{\text{vac}}_\nu+\frac{1}{4}N^{\text{vac}}{}^\rho{}_\mu C^{\text{vac}}_{\nu\rho}-\left(\mu\leftrightarrow\nu\right)\right)=-\left(\overset{(0)}{S}-a^2\right)F_{\alpha\beta}\,,
\end{equation}
which is derived in appendix \ref{app:rewritingcurv3}.
We use this to eliminate $\left(\overset{(0)}{S}-a^2\right)F_{\mu\nu}$ from the right-hand side of \eqref{eq:3rdKcurv}.
Doing so it follows that we can finally write the third curvature \eqref{eq:eeRK} as
\begin{eqnarray}
    e^\mu_a e^\nu_b R(K)_{\mu\nu} & = & e^\mu_a e^\nu_b\left[\left(\partial_\mu+a_\mu\right)\left(\overset{(0)}{P}_\nu-\overset{(0)}{P}{}^{\text{vac}}_\nu\right)+\frac{1}{4}h^{\rho\sigma}\left(N_{\mu\rho}C_{\nu\sigma}-N^{\text{vac}}_{\mu\rho}C^{\text{vac}}_{\nu\sigma}\right)-\left(\mu\leftrightarrow\nu\right)\right]\,.\nonumber\\
    &&\label{eq:curv3}
\end{eqnarray}

Recall that in the definition of $C^{\text{vac}}_{\mu\nu}$ we allowed for a tensor $Y_{\mu\nu}$ that was such that it does not appear in the vacuum news. The vacuum shear is going to be that shear tensor for which all the $K$-curvatures vanish. This means that $e^\mu_a e^\nu_b R(K)_{\mu\nu}$ should be expressible in terms of only the shear and vacuum shear and derivatives thereof. This means that the $Y_{\mu\nu}$ tensor in the vacuum shear has to be such that \eqref{eq:curv3} remains unaffected by including it in the vacuum shear. This is the same as demanding that \eqref{eq:idcurv3} is unaffected by the $Y_{\mu\nu}$ tensor. This means that we will demand that it obeys
\begin{equation}\label{eq:conY2}
    h^\mu_\alpha h^\nu_\beta\left(-\frac{1}{2}\left(\partial_\mu+a_\mu\right)\left(h^{\rho\sigma}\left(\mathcal{D}_\rho-a_\rho\right)Y_{\sigma\nu}\right)+\frac{1}{4}N^{\text{vac}}{}^\rho{}_\mu Y_{\nu\rho}-\left(\mu\leftrightarrow\nu\right)\right)=0\,.
\end{equation}

\subsubsection*{Summary}

To summarise, we defined
\begin{eqnarray}
    N^{\text{vac}}_{\mu\nu} & = & 2h^\rho_{\langle\mu}h^\sigma_{\nu\rangle}\left(\mathcal{D}_\rho\partial_\sigma\varphi+\partial_\rho\varphi\partial_\sigma\varphi-\mathcal{D}_\rho a_\sigma-a_\rho a_\sigma\right)+\mathcal{T}_{\mu\nu}\,,\label{eq:vacnews2}\\
    C^{\text{vac}}_{\mu\nu} & = & 2h^\rho_{\langle\mu}h^\sigma_{\nu\rangle}\left(-\mathcal{D}_\rho\partial_\sigma F+F\left(\mathcal{D}_\rho a_\sigma+a_\rho a_\sigma\right)\right)+FN^{\text{vac}}_{\mu\nu}+Y_{\mu\nu}\,,\label{eq:vacshear2}
\end{eqnarray}
where $\mathcal{T}_{\mu\nu}$ and $Y_{\mu\nu}$ are spatial STF tensors that obey $\mathcal{L}_v \mathcal{T}_{\mu\nu}=0$, and $h^{\rho\sigma}\mathcal{D}_\rho \mathcal{T}_{\sigma\nu}=0$ as well as $\left(\mathcal{L}_v+\frac{1}{2}K\right)Y_{\mu\nu}=0$ and equation \eqref{eq:conY2}. Furthermore $F$ is any scalar that obeys $\left(\mathcal{L}_v+\frac{1}{2}K\right)F=-1$. In terms of these vacuum objects we can write the $K$-curvatures as
\begin{eqnarray}
    v^\mu e^\nu_b R(K)_{\mu\nu}{}^a & = &  -\frac{1}{2}e^\mu_a e^\nu_b\mathcal{L}_v\left(N_{\mu\nu}-N^{\text{vac}}_{\mu\nu}\right)\,,\\
     v^\mu  R(K)_{\mu\nu} & = & \frac{1}{2}h^{\rho\sigma}\mathcal{D}_\rho\left(N_{\sigma\nu}-N^{\text{vac}}_{\sigma\nu}\right)\,,\\
     e^\mu_a e^\nu_b R(K)_{\mu\nu} & = & e^\mu_a e^\nu_b\left[\left(\partial_\mu+a_\mu\right)\left(\overset{(0)}{P}_\nu-\overset{(0)}{P}{}^{\text{vac}}_\nu\right)\right.\nonumber\\
     &&\left.+\frac{1}{4}h^{\rho\sigma}\left(N_{\mu\rho}C_{\nu\sigma}-N^{\text{vac}}_{\mu\rho}C^{\text{vac}}_{\nu\sigma}\right)-\left(\mu\leftrightarrow\nu\right)\right]\,.
\end{eqnarray}
It is now obvious that if we set the shear equal to the vacuum shear that all these curvatures vanish. Conversely setting all the $K$-curvatures to zero leads to the solution $C_{\mu\nu}=C^{\text{vac}}_{\mu\nu}+\cdots$ where the dots are terms that we can absorb into $\mathcal{T}_{\mu\nu}$ and $Y_{\mu\nu}$, so that without loss of generality the shear equals the vacuum shear if and only if all the $K$-curvatures vanish. This then implies that the $K$-curvatures measure the deviation from the vacuum shear.

\subsection{Carroll-covariant vacuum shear}

The vacuum news follows from the vacuum shear so we will focus now on the latter and study its properties. Combining \eqref{eq:vacnews2} and \eqref{eq:vacshear2} we obtain
\begin{equation}\label{eq:vacshear100}
    C^{\text{vac}}_{\mu\nu} = 2h^\rho_{\langle\mu}h^\sigma_{\nu\rangle}\left(-\mathcal{D}_\rho\partial_\sigma F+F\left(\mathcal{D}_\rho \partial_\sigma\varphi+\partial_\rho\varphi \partial_\sigma\varphi\right)\right)+F\mathcal{T}_{\mu\nu}+Y_{\mu\nu}\,,
\end{equation}
where $F$ obeys $\left(\mathcal{L}_v+\frac{1}{2}K\right)F=-1$.

In this section we take a closer look at the roles of $\mathcal{T}_{\mu\nu}$ and $Y_{\mu\nu}$. We will see that without loss of generality we can always set $Y_{\mu\nu}=0$ and that $\mathcal{T}_{\mu\nu}$ must take a specific form which is dictated by 
invariance of the vacuum shear with respect to the conformal isometry acting on $\varphi$, see equations \eqref{eq:confisom1} and \eqref{eq:confisom2}.

We remind the reader that $\mathcal{T}_{\mu\nu}$ and $Y_{\mu\nu}$ are spatial STF tensors that obey $\mathcal{L}_v \mathcal{T}_{\mu\nu}=0$, and $h^{\rho\sigma}\mathcal{D}_\rho \mathcal{T}_{\sigma\nu}=0$ as well as $\left(\mathcal{L}_v+\frac{1}{2}K\right)Y_{\mu\nu}=0$ and equation \eqref{eq:conY2}. These are all first order PDEs on a 3-dimensional Carroll manifold with $h_{\mu\nu}dx^\mu dx^\nu=e^{2\varphi}dX^a dX^a$.

The most general solutions for $\mathcal{T}_{\mu\nu}$ and $Y_{\mu\nu}$ are constructed in appendix \ref{app:SimplePDEs} with the solutions given in equations 
\eqref{eq:STFX} and \eqref{eq:STFY} which we repeat here for convenience:
\begin{eqnarray}
    \mathcal{T}_{\mu\nu} & = & \partial_\mu X^a\partial_\nu X^b\tilde{\mathcal{T}}_{ab}(X)\,,\qquad\text{where $\partial_a\tilde{\mathcal{T}}_{ab}=0$}\,,\\
    Y_{\mu\nu} & = & e^\varphi \partial_\mu X^a\partial_\nu X^b \left(\partial_{\langle a}\partial_{b\rangle} -\frac{1}{2}\tilde{\mathcal{T}}_{ab}\right)\tilde Y(X)\,,
\end{eqnarray}
where $\tilde{\mathcal{T}}_{ab}$ is STF with respect to the Kronecker delta. 

Using equation \eqref{eq:DdXa} from appendix \ref{app:SimplePDEs} we can write $Y_{\mu\nu}$ also as
\begin{equation}\label{eq:covformY}
    Y_{\mu\nu}=e^\varphi h^\rho_{\langle\mu}h^\sigma_{\nu\rangle}\left(\mathcal{D}_\rho\partial_\sigma Y+2\partial_\rho\varphi\partial_\sigma Y-\frac{1}{2}\mathcal{T}_{\rho\sigma}Y\right)\,,
\end{equation}
where $Y(x)=\tilde Y(X)$.

The condition on the function $F$ in the vacuum shear which is $\left(\mathcal{L}_v+\frac{1}{2}K\right)F=-1$ can be solved as follows. First we define $f$ as 
\begin{equation}
 F=e^\varphi f\,,
\end{equation}
so that we find 
\begin{equation}\label{eq:conf}
    v^\mu\partial_\mu f=-e^{-\varphi}\,.
\end{equation}
Next defining 
\begin{equation}
    v^\mu = e^{-\varphi}\tilde v^\mu\,,
\end{equation}
we find 
\begin{equation}
    \tilde v^\mu\partial_\mu f=-1\,.
\end{equation}
The most general solution is thus
\begin{equation}\label{eq:STGoldstone}
    f=C(X)+U\,,
\end{equation}
where $C(X)$ is any real function of $X^a$, that we will call the supertranslation field, and $U$ is a new scalar field (on par with $X^a$) which is defined by
\begin{equation}\label{eq:tildevdf}
    \tilde v^\mu\partial_\mu U=-1\,.
\end{equation}
Recall that we also have 
\begin{equation}
    \tilde v^\mu\partial_\mu X^a=0\,,
\end{equation}
since $\mathcal{L}_v X^a=0$. We can thus solve for $v^\mu$ leading to 
\begin{equation}
    v^\mu\partial_\mu=-e^{-\varphi}\partial_U\,.
\end{equation}

Equation \eqref{eq:tildevdf} can be read as providing a different way of expressing $\tau_\mu dx^\mu$, namely as
\begin{equation}
    \tau_\mu dx^\mu=e^\varphi\left( dU+B^a dX^a\right)\,,
\end{equation}
where $B^a$ is any 2-dimensional vector that depends on $(U,X^a)$. Combining this observation with the expression for $h_{\mu\nu}dx^\mu dx^\nu$ in \eqref{eq:metrich} we see that we can obtain $\tau_\mu$ and $h_{\mu\nu}$ as follows. We start with a flat Carroll target spacetime whose coordinates are $(U,X^a)$ and where we define the Carroll metric data $dU$ and $\delta_{ab}dX^a dX^b$. We then pull these fields back to $\mathcal{I}^+$ and Weyl rescale with appropriate powers of $e^{\varphi}$. Finally we apply a Carroll boost to $\tau_\mu$.
We can think of $f=\text{cst}$ as an equation on the target space which defines a cut by solving for $U$. After using the pullback construction this can then be seen as providing a Carroll-covariant notion of a cut of $\mathcal{I}^+$.

We can thus associate with $\mathcal{I}^+$ a flat Carrollian target spacetime by mapping points $x^\mu$ on $\mathcal{I}^+$ to $(U,X^a)$. The differential equations for $\mathcal{T}_{\mu\nu}$, $Y_{\mu\nu}$ and $F$ discussed in this section and in appendix \ref{app:SimplePDEs} where rewritten as differential equations on this target spacetime.

If we substitute $F=e^\varphi f$ into the vacuum shear \eqref{eq:vacshear100} we get
\begin{equation}
    C^{\text{vac}}_{\mu\nu} = 2e^\varphi h^\rho_{\langle\mu}h^\sigma_{\nu\rangle}\left(-\mathcal{D}_\rho\partial_\sigma f-2\partial_\rho\varphi\partial_\sigma f+\frac{1}{2}f\mathcal{T}_{\rho\sigma}+\frac{1}{2}\mathcal{D}_\rho\partial_\sigma Y+\partial_\rho\varphi\partial_\sigma Y-\frac{1}{4}\mathcal{T}_{\rho\sigma}Y\right)\,,
\end{equation}
where we used \eqref{eq:covformY}. We thus see that we can always absorb $Y(x)=\tilde Y(X)$ into $f=C(X)+U$. Since $\mathcal{L}_v Y=0$ this is compatible with \eqref{eq:conf}. Hence, without loss of generality we can continue with $Y_{\mu\nu}=0$.

Going back to \eqref{eq:vacshear100} we now have
\begin{equation}\label{eq:vacCY=0}
    C^{\text{vac}}_{\mu\nu} = 2h^\rho_{\langle\mu}h^\sigma_{\nu\rangle}\left(-\mathcal{D}_\rho\partial_\sigma F+F\left(\mathcal{D}_\rho \partial_\sigma\varphi+\partial_\rho\varphi \partial_\sigma\varphi+\frac{1}{2}\mathcal{T}_{\rho\sigma}\right)\right)\,.
\end{equation}
The field $\varphi$ suffers from the ambiguity \eqref{eq:confisom2}. This transformation leaves $h_{\mu\nu}$ and the connection $\mathcal{C}^\rho_{\mu\nu}$ invariant. We should demand that the vacuum shear is likewise invariant under this ambiguity in $\varphi$. This can be achieved by a nonzero $\mathcal{T}_{\mu\nu}$ that transforms appropriately under the ambiguity \eqref{eq:confisom1} and \eqref{eq:confisom2}. 

This form for $\mathcal{T}_{\mu\nu}$ can be obtained by St\"uckelberging the shift transformation. This means that we replace $\varphi$ by $\varphi+\zeta$ where $\zeta$ has to be such that 
\begin{equation}
    e^{2\varphi}dX^adX^a=e^{2\Phi}e^{-2\zeta}dX^a dX^a=e^{2\Phi}dY^a dY^a\,,
\end{equation}
where $\Phi=\varphi+\zeta$ is invariant under the ambiguity. This means that if we use the complex combination $Z=\frac{1}{\sqrt{2}}\left(X^1+iX^2\right)$ that we need
\begin{equation}
    2e^{-2\zeta}dZd\bar Z=2dWd\bar W\,,
\end{equation}
where $W=\frac{1}{\sqrt{2}}\left(Y^1+iY^2\right)$.
This implies that 
\begin{equation}\label{eq:zetatof(Z)}
    \zeta(x)=\tilde\zeta(X)=-\frac{1}{2}\log f'(Z)+\text{c.c.}\,,
\end{equation}
where $W=f(Z)$.

We demand that this $\zeta$ comes from $\mathcal{T}_{\rho\sigma}$ in such a manner that we have
\begin{equation}
    h^\rho_{\langle\mu}h^\sigma_{\nu\rangle}\left(\mathcal{D}_\rho \partial_\sigma\varphi+\partial_\rho\varphi \partial_\sigma\varphi+\frac{1}{2}\mathcal{T}_{\rho\sigma}\right)=h^\rho_{\langle\mu}h^\sigma_{\nu\rangle}\left(\mathcal{D}_\rho \partial_\sigma(\varphi+\zeta)+\partial_\rho(\varphi+\zeta) \partial_\sigma(\varphi+\zeta)\right)\,.
\end{equation}
We then read off that $\mathcal{T}_{\mu\nu}$ is given by
\begin{equation}\label{eq:choiceforcalT}
    \mathcal{T}_{\mu\nu}=2h^\rho_{\langle\mu}h^\sigma_{\nu\rangle}\left(\mathcal{D}_\rho \partial_\sigma \zeta+2\partial_\rho\varphi \partial_\sigma \zeta+\partial_\rho \zeta \partial_\sigma \zeta\right)\,.
\end{equation}
If we now use \eqref{eq:DdXa} we can write
\begin{equation}
    \mathcal{T}_{\mu\nu}=2\partial_\mu X^a\partial_\nu X^b\left(\partial_{\langle a}\partial_{b\rangle}\tilde\zeta+\partial_{\langle a}\tilde\zeta\partial_{b\rangle}\tilde\zeta\right)\,.
\end{equation}
This means that we have
\begin{equation}
    \tilde{\mathcal{T}}_{ab}=2\left(\partial_{\langle a}\partial_{b\rangle}\tilde\zeta+\partial_{\langle a}\tilde\zeta\partial_{b\rangle}\tilde\zeta\right)\,.
\end{equation}

Using the complex combination $Z=\frac{1}{\sqrt{2}}\left(X^1+iX^2\right)$ with corresponding derivative $\partial$ we get
\begin{equation}
    \tilde{\mathcal{T}}_{ZZ}=2\left(\partial^2\tilde\zeta+(\partial\tilde\zeta)^2\right)\,.
\end{equation}
Substituting \eqref{eq:zetatof(Z)} we then get
\begin{equation}\label{eq:Schwarzianf}
    \tilde{\mathcal{T}}_{ZZ}=-\left(\frac{f'''}{f'}-\frac{3}{2}\left(\frac{f''}{f'}\right)^2\right)\,.
\end{equation}
In other words $\tilde{\mathcal{T}}_{ZZ}$ is (minus) the Schwarzian derivative of the field $f$.

If we consider \eqref{eq:vacCY=0} with $\mathcal{T}_{\mu\nu}$ as given in \eqref{eq:choiceforcalT} we see that in the end we obtain\footnote{If we express this in standard BS coordinates and compare with \eqref{eq:CvacBScoord}
we see that $F=u+C$ and that $T_{ij}=2\left(D_{\langle i}\partial_{j\rangle}\Phi+\partial_{\langle i}\Phi\partial_{j\rangle}\Phi\right)$.\label{fn:Tij}}
\begin{eqnarray}
    C^{\text{vac}}_{\mu\nu} & = &  2h^\rho_{\langle\mu}h^\sigma_{\nu\rangle}\left(-\mathcal{D}_\rho\partial_\sigma F+F\left(\mathcal{D}_\rho \partial_\sigma\Phi+\partial_\rho\Phi\partial_\sigma\Phi\right)\right)\,,\label{eq:vacshearfinal}
\end{eqnarray}
where we defined
\begin{equation}\label{eq:DefPhi}
    \Phi=\varphi+\zeta\,,
\end{equation}
with $\zeta$ given in \eqref{eq:zetatof(Z)}.
The fields $F$ and $\Phi$ transform under boundary diffeomorphisms and Weyl transformations as
\begin{eqnarray}
    \delta\Phi & = & \mathcal{L}_\chi\Phi+\Lambda_D\,,\label{eq:trafoPhi}\\
    \delta F & = & \mathcal{L}_\chi F+\Lambda_D F\,.\label{eq:trafoF}
\end{eqnarray}
The transformation of $\Phi$ follows from the total gauge transformation of $\varphi$ in \eqref{eq:totgaugetrafovarphi} and the transformation of $\zeta$. The transformation of $F$ follows from the fact that the vacuum shear is a tensor with Weyl weight $+1$. When transforming the vacuum shear \eqref{eq:vacshearfinal} under Carroll boosts, one has to transform $\tau_\mu$ and $h^{\mu\nu}$ in $h^\rho_\mu$ and the connection $\mathcal{C}^\rho_{\mu\nu}$. We then reproduce the transformation \eqref{eq:trafoshear} when we assume that $\Phi$ and $F$ are inert under Carroll-boosts. In other words using 
\eqref{eq:trafoPhi} and \eqref{eq:trafoF} as well as the transformations \eqref{eq:tottrafotau} and \eqref{eq:tottrafoh} we obtain the following transformation of the vacuum shear \eqref{eq:vacshearfinal},
\begin{equation}
    \delta C^{\text{vac}}_{\mu\nu}=\mathcal{L}_\chi C^{\text{vac}}_{\mu\nu}+\Lambda_D C^{\text{vac}}_{\mu\nu}+2h^\rho_{\langle\mu}h^\sigma_{\nu\rangle}\left(\mathcal{D}_\rho+a_\rho\right)\lambda_\sigma\,.
\end{equation}
Equation \eqref{eq:vacshearfinal} is our final expression for the vacuum shear. In appendix \ref{app:vacsheardiffeos} we will derive this expression for the vacuum shear by acting with certain diffeomorphisms on Minkowski spacetime. This expression for the vacuum shear is fully boundary covariant and its associated vacuum news is given by \eqref{eq:vacnews2} where we choose $\mathcal{T}_{\mu\nu}$ to be given by \eqref{eq:choiceforcalT} leading to
\begin{equation}
    N^{\text{vac}}_{\mu\nu} = 2h^\rho_{\langle\mu}h^\sigma_{\nu\rangle}\left(\mathcal{D}_\rho\partial_\sigma \Phi+\partial_\rho \Phi\partial_\sigma \Phi-\mathcal{D}_\rho a_\sigma-a_\rho a_\sigma\right)\,.
\end{equation}
This vacuum news is the trace-free part of the Geroch tensor \cite{Campiglia:2020qvc,Nguyen:2020hot}. 

The vacuum shear transforms exactly like the total shear in \eqref{eq:trafoshear} and \eqref{eq:Delta}. In particular it transforms with the same inhomogenous shift term that we denoted by $\Delta$ in \eqref{eq:trafoshear}. This means that the non-vacuum shear $C_{\mu\nu}-C^{\text{vac}}_{\mu\nu}$ transforms homogeneously as
\begin{equation}
    \delta\left(C_{\mu\nu}-C^{\text{vac}}_{\mu\nu}\right)=\mathcal{L}_\chi\left(C_{\mu\nu}-C^{\text{vac}}_{\mu\nu}\right)+\Lambda_D\left(C_{\mu\nu}-C^{\text{vac}}_{\mu\nu}\right)\,.
\end{equation}
In particular $C_{\mu\nu}-C^{\text{vac}}_{\mu\nu}$ is invariant under Carroll boosts.
If we compute the $K$-curvatures for this difference and they go to zero at the corners of $\mathcal{I}^+$ then $C_{\mu\nu}-C^{\text{vac}}_{\mu\nu}$ is called the hard or radiative shear.

The fields $\Phi$ and $F$ are Carroll scalar fields that are defined on $\mathcal{I}^+$. Vacuum solutions thus correspond to Carroll geometries at $\mathcal{I}^+$ that contain $\tau_\mu$ and $h_{\mu\nu}$ (subject to the constraint \eqref{eq:Kconstraint}) as well as two scalar fields $\Phi$ and $F$ where the latter has to obey $\left(\mathcal{L}_v+\frac{1}{2}K\right)F=-1$. 

The field $F$ contains the supertranslation field $C$ via $F=e^\varphi(C(X)+U)$ (see equation \eqref{eq:STGoldstone}). The field $\Phi$ is sometimes called the Liouville or superboost field \cite{Compere:2018ylh}.

\section{$K$-curvatures and the Bondi loss equations}\label{sec:KcurvBondi}

We will start with a review of the definition of the boundary energy-momentum-news complex introduced in \cite{Hartong:2025jpp,Hartong:2026rbr}.
In these works it was shown that this object obeys a set of Ward identities associated with bulk diffeomorphisms that act on the boundary objects $\tau_\mu, h_{\mu\nu}, C_{\mu\nu}$ as in \eqref{eq:tottrafotau}-\eqref{eq:Delta}. The Bondi loss equations can be obtained from these Ward identities. In other words there is a boundary energy-momentum tensor whose diffeomorphism Ward identity takes the form of the Bondi loss equations which express the non-conservation of the energy-momentum tensor due to the presence of shear\footnote{%
    In~\cite{Fiorucci:2025twa} general symmetry properties of a putative Carrollian theory at $\mathcal{I}^+$ were used to construct a set of Ward identities for energy and momentum currents. It was then shown that the Bondi loss equations fit with the form of these Ward identities.
}. This boundary energy momentum tensor is defined as the response to varying $\tau_\mu$ and $h_{\mu\nu}$ keeping the shear fixed. This boundary energy-momentum tensor is not Carrollian in the sense that its energy flux is nonzero\footnote{It was observed in \cite{Ciambelli:2018wre,Mittal:2022ywl,Campoleoni:2023fug} that the presence of radiation leads to a nonzero energy flux.}. 

We will show in this section that we can rewrite the Bondi loss equations in such a manner that they are written in terms of another (Carrollian) energy-momentum tensor (i.e. with vanishing energy flux \cite{Hartong:2015usd,deBoer:2017ing}) and where furthermore the non-conservation is not due to the presence of a nonzero shear but due to the presence of a nonzero $C_{\mu\nu}-C_{\mu\nu}^{\text{vac}}$. More specifically the non-conservation is described by flux terms that are expressible in terms of the $K$-curvatures. We will furthermore see that the BMS currents are defined in terms of this Carrollian energy-momentum tensor and that their non-conservation is described by nonzero $K$-curvature flux terms.

\subsection{Review of the boundary energy-momentum-news complex}\label{subsec:reviewEMTnews}

In \cite{Hartong:2026rbr,Hartong:2025jpp} a boundary energy-momentum-news complex has been defined at $\mathcal{I}^+$. In order to define this object we consider the near-boundary expansion of section \ref{subsec:Carcovgauge} and choose a constant $r$ hypersurface. On this hypersurface we define a number of boundary terms that we add to the Einstein--Hilbert action. The total action is
\begin{equation}
S_{\text{EH}}+S_{\text{ext}}+S_{\text{norm}}+S_{\text{int}}\,.
\end{equation}
In here we defined 
\begin{align}
\begin{split}
    S_{\text{ext}} & =  2\int_\Sigma d^{3}\xi \sqrt{-g}\left(\delta^M_P+V^M N_P\right)\nabla_M N^P\nonumber\\
    &=  2\int d^3 x \sqrt{-g}\left[e^{-1}\partial_\mu\left(e U^\mu\right)+U^\mu\partial_\mu\beta+2r^{-1}S+\frac{1}{2}\partial_r S+S\partial_r\beta\right]\,,\\
        S_{\text{norm}} & =  \int_\Sigma d^3\xi\sqrt{-g}N^2\nabla_M V^M=-\int_{r=\Lambda} d^{3}x \sqrt{-g}\left(2r^{-1} S +S\partial_r\beta\right)\,,\\
    S_\text{int}
  & =  -\int_{r = \Lambda} d^{3}x\,\sqrt{-g}\, r\left(Q+\frac{1}{4}S\mathcal{F}^2\right)\,,
\end{split}
\end{align}
where $N_M=\partial_M r$ is the normal to the cutoff hypersurface, $V^M=-\delta^M_r$ is part of our Carroll-covariant gauge choice. In this gauge we can write
\begin{equation}
    \sqrt{-g}=r^2 e^\beta e\,,
\end{equation}
where $e=\text{det}\,(\tau_\mu\,, e^a_\mu)$.
The fields $S$ and $\beta$ are defined in section \ref{subsec:Carcovgauge} and $\nabla_M$ is the bulk Levi-Civita connection. 
In the intrinsic counterterm $S_\text{int}$ we defined $\mathcal{F}^2=\Pi^{MP}\Pi^{NQ}\mathcal{F}_{MN}\mathcal{F}_{PQ}$ where $\mathcal{F}_{MN}$ is defined in 
\eqref{eq:defcalF}. Finally, $Q$ is defined as
\begin{equation}
    Q=\Pi^{\mu\rho}\hat R_{\mu\rho}+\hat D_\mu \mathcal{A}^\mu\,,
\end{equation}
where $\hat R_{\mu\rho}$ is the Ricci tensor associated with the constant $r$ hypersurface connection
\eqref{eq:concstr}.

The extrinsic counterterm\footnote{An extrinsic counterterm is one that depends on the normal direction.} $S_{\text{ext}}$ is the generalisation of the Gibbons--Hawking--York extrinisc counterterm to the case of a general hypersurface such as $r=\text{cst}$ which is neither timelike nor spacelike. It is also not null. The appropriate extrinsic counterterm was worked out in \cite{Parattu:2016trq}. The second counterterm $S_{\text{norm}}$ is a combination of a term proportional to the norm of the normal $N_M$, i.e. $S=g^{rr}=N^2$ and a finite counterterm involving $S\partial_r\beta$. Finally the last boundary term $S_{\text{int}}$ depends on the intrinsic curvature of the constant $r$ hypersurfaces which is captured by $Q$ as well as a second finite counterterm containing $S\mathcal{F}^2$. These counterterms first of all ensure that the variation of the total action evaluated on shell remains finite in the large $r$ limit. Secondly they ensure that the result is of the form 
\begin{align} &\delta\left(S_{\text{EH}}+S_{\text{ext}}+S_{\text{norm}}+S_{\text{int}}\right)\big\vert_{\text{os}}
  \nonumber
  \\
  &{}\qquad
  =\cdots+2\int_{\mathcal{I}^+}d^3 xe\left(T^\mu\delta\tau_\mu+\frac{1}{2}T^{\mu\nu}\delta h_{\mu\nu}+\frac{1}{2}S^{\mu\nu}\delta C_{\mu\nu}\right)\,,
  \label{eq:variationosd=2}
\end{align}
where the dots denote terms that pertain to any other boundaries that are not $\mathcal{I}^+$.
The role of the finite counterterms was to ensure that the objects $T^\mu$, $T^{\mu\nu}$ and $S^{\mu\nu}$ all have a definite Weyl weight. 
The response to the variation of the shear is 
\begin{equation}
    S^{\mu\nu}=\frac{1}{2}h^{\mu\rho}h^{\nu\sigma}N_{\rho\sigma}\,,
\end{equation}
where 
\begin{equation}
    N_{\rho\sigma}=-\left(\mathcal{L}_v+\frac{1}{2}K\right)C_{\rho\sigma}\,,
\end{equation}
is the Weyl invariant news tensor. This fact is closely related to the Ashtekar--Streubel Poisson bracket between the shear and the news \cite{Ashtekar:1981bq}.

The bulk diffeomorphisms that respect the Carroll-covariant BS gauge act on the boundary objects $\tau_\mu, h_{\mu\nu}$ and $C_{\mu\nu}$ as in \eqref{eq:tottrafotau}--\eqref{eq:Delta}. In \cite{Hartong:2026rbr,Hartong:2025jpp} we have shown that the on shell variation of the action as defined in 
\eqref{eq:variationosd=2} is invariant under boundary diffeomorphisms and Weyl transformations. On the other hand with respect to Carroll boosts the variation of \eqref{eq:variationosd=2} goes into the Carroll boost anomaly. Concretely we found that 
\begin{equation}
\delta_\lambda\left(S_{\text{EH}}+S_{\text{ext}}+S_{\text{norm}}+S_{\text{int}}\right)\big\vert_{\text{os}}=\cdots+\int_{\mathcal{I}^+}d^3 xe\lambda_\mu\mathcal{A}_B^\mu\,,
\end{equation}
where the spatial Carroll boost parameter is $\lambda_\mu$. In \cite{Hartong:2026rbr} we showed that the anomaly is given by the right hand side of \eqref{eq:PTWeylcov}.
We thus see that we can now recognise the boost anomaly as given by the $K$-curvature
\begin{eqnarray}    
    \mathcal{A}^{\text{B}}_\mu & = & v^\rho  R(K)_{\rho\mu} \,.
\end{eqnarray}

The Weyl and anomalous Carroll boost Ward identities are
\begin{equation}\label{eq:WeylWI}
 \tau_\mu T^\mu+h_{\mu\nu}T^{\mu\nu}+\frac{1}{4}C_{\mu\nu}N^{\mu\nu}=0\,,
\end{equation}
and
\begin{equation}\label{eq:boostanom}
  h_{\mu\rho} T^\rho-\frac{1}{2}h^{\rho\sigma}\mathcal{D}_\rho N_{\sigma\mu} = \mathcal{A}_\mu^{\text{B}}\,.
\end{equation}
We will use these relations to determine $h_{\mu\nu}T^{\mu\nu}$ and $h_{\mu\rho}T^\rho$. The remaining components of $T^\mu$ and $T^{\mu\nu}$ are given by\footnote{See section 8.4 of \cite{Hartong:2026rbr}.}
\begin{eqnarray}
\tau_\mu T^\mu & = &  -v^\rho v^\sigma \overset{(1)}{g}_{\rho\sigma}-\frac{1}{16}K\left(F^2-C^2\right)+\mathcal{D}_\rho\overset{(0)}{P}{}^\rho+\frac{1}{8}\left(\mathcal{L}_v-K\right)F^2\,,\label{eq:d=2-energy-density-Weyl-cov}\\
      h_{\rho\mu}\tau_\nu T^{\mu\nu} & = & \frac{3}{2}\left(h^\mu_\rho v^\nu\os{1}{g}_{\mu\nu}-\frac{1}{16}\left(F^2-C^2\right)a_\rho\right)-a_\sigma D^\sigma{}_{\rho}\\
    &&+\frac{1}{32}h^\sigma_\rho\left(\partial_\sigma+2a_\sigma\right)\left(F^2-C^2\right)-\frac{1}{2}\overset{(0)}{P}_\sigma C^\sigma{}_\rho+\frac{5}{2}\overset{(0)}{P}_\sigma F^\sigma{}_\rho+\frac{1}{2}h^{\sigma\nu}\mathcal{D}_\sigma\chi_{\nu\rho}\,,\nonumber\\
    h^{\langle\rho}_\mu h^{\sigma\rangle}_\nu T^{\mu\nu} & = & \frac{1}{2}\left(\overset{(0)}{S}-a^2\right)C^{\rho\sigma} -\frac{1}{2}C^{\alpha(\rho}h^{\sigma)\beta}\left(\partial_\alpha\tilde b_\beta-\partial_\beta\tilde b_\alpha\right)+h^{\alpha\langle\rho}h^{\sigma\rangle\beta}\left(\mathcal{D}_\alpha+3a_\alpha\right)\overset{(0)}{P}_\beta\nonumber\\
    &&-\frac{1}{4}h^{\mu\rho}h^{\nu\sigma}\mathcal{L}_v\left(F_{\mu\kappa}C^\kappa{}_\nu\right)+\frac{1}{2}K D^{\rho\sigma}+\frac{1}{2}h^{\mu\rho}h^{\nu\sigma}\mathcal{L}_v\chi_{\mu\nu}\,.
\end{eqnarray}
We can also write $h^{\langle\rho}_\mu h^{\sigma\rangle}_\nu T^{\mu\nu}$ as
\begin{eqnarray}
    h^{\langle\rho}_\mu h^{\sigma\rangle}_\nu T^{\mu\nu} & = & \frac{1}{2}\left(\overset{(0)}{S}-a^2\right)C^{\rho\sigma} +h^{\alpha\langle\rho}h^{\sigma\rangle\beta}\left(\mathcal{D}_\alpha+3a_\alpha\right)\overset{(0)}{P}_\beta+\frac{1}{2}F^{\rho\alpha}N_{\alpha}{}^\sigma\nonumber\\
    &&+\frac{1}{2}K D^{\rho\sigma}-\frac{1}{4}h^{\mu\rho}h^{\nu\sigma}\mathcal{L}_v\left(C_\mu{}^\alpha F_{\alpha\nu}\right)+\frac{1}{2}h^{\mu\rho}h^{\nu\sigma}\mathcal{L}_v\chi_{\mu\nu}\label{eq:tildeT}\,,
\end{eqnarray}
which is how we will use it in this paper. This rewriting follows from the identity
\begin{align}
    &-\frac{1}{2}C^{\alpha(\rho}h^{\sigma)\beta}\left(\partial_\alpha\tilde b_\beta-\partial_\beta\tilde b_\alpha\right)-\frac{1}{4}h^{\mu\rho}h^{\nu\sigma}\mathcal{L}_v\left(F_{\mu\kappa}C^\kappa{}_\nu\right)\nonumber\\
    =&\frac{1}{2}F^{\rho\alpha}N_{\alpha}{}^\sigma-\frac{1}{4}h^{\mu\rho}h^{\nu\sigma}\mathcal{L}_v\left(C_\mu{}^\alpha F_{\alpha\nu}\right)\,,
\end{align}
which can be derived using \eqref{eq:ASSTF}.

The object $\chi_{\mu\nu}$ is any spatial STF tensor. As explained in \cite{Hartong:2026rbr} the terms involving $\chi_{\mu\nu}$ encode a fundamental ambiguity due to the constraint $K_{\langle\mu\nu\rangle}=0$. This means that $\delta h_{\mu\nu}$ should be viewed as a variation in the solution space of this constraint. As a result there are aspects of the response $T^{\mu\nu}$ that are not fixed. One can view this ambiguity as an improvement transformation. The diffeomorphism Ward identity which we will discuss shortly does not depend on $\chi_{\mu\nu}$.

The components $\tau_\mu T^\mu$ and $h_{\rho\mu}\tau_\nu T^{\mu\nu}$ are the energy and angular momentum densities. Note that $h^{\langle\rho}_\mu h^{\sigma\rangle}_\nu T^{\mu\nu}$ is not related to a particular component of the metric in the way that $\tau_\mu T^\mu$ and $h_{\rho\mu}\tau_\nu T^{\mu\nu}$ are. In other words, one cannot solve for a particular component of the $1/r$ expansion of $g_{\mu\nu}$ in terms of $h^{\langle\rho}_\mu h^{\sigma\rangle}_\nu T^{\mu\nu}$.

In \cite{Hartong:2026rbr} it was shown that the Bondi loss equations (appropriately generalised to account for a Carroll-covariant boundary description) can be written as the diffeomorphism Ward identity for this EMT-news complex. This is expressed through the following equations
\begin{eqnarray}
    0 & = &  -\left(\mathcal{L}_v-\frac{3}{2}K\right)\left(\tau_\mu T^\mu\right)-\frac{1}{4}N^{\rho\sigma}N_{\rho\sigma}+\left(\mathcal{D}_\mu+a_\mu\right)\left(T^\rho h^\mu_\rho\right)\,,\label{eq:diffeoWtimeproj}\\
    0 & = & -\left(\mathcal{L}_v-K\right) P_\kappa+ h_{\kappa\sigma}\mathcal{D}_\mu\tilde {T}^{\mu\sigma}+\frac{1}{2}h^\mu_\kappa\left(\partial_\mu+3a_\mu\right)\left(T^{\rho\sigma}h_{\rho\sigma}+\frac{1}{2}N^{\rho\sigma}C_{\rho\sigma}\right) \nonumber\\
    &&+\frac{1}{4}h_{\kappa\sigma}\mathcal{D}_\mu\left(N^{\mu\lambda}C_\lambda{}^\sigma-N^{\sigma\lambda}C_\lambda{}^\mu\right)-\frac{1}{4}N^{\mu\sigma}h^\nu_\kappa\left(\mathcal{D}_\nu+a_\nu\right) C_{\mu\sigma}+T^\sigma h^\mu_\sigma F_{\mu\kappa}\,,\nonumber\\
    &&\label{eq:diffeoWIspatialproj}
\end{eqnarray}
where we defined
\begin{equation}
    P_\kappa=T^{\mu\nu}\tau_\mu h_{\nu\kappa}\,,\qquad\tilde{T}^{\rho\sigma}=T^{\mu\nu}h^{\langle\rho}_\mu h^{\sigma\rangle}_\nu\,.
\end{equation}

\subsection{Another energy-momentum tensor, Bondi loss and $K$-curvatures}\label{subsec:newBondiform}

We have so far treated the triplet $(\tau_\mu, h_{\mu\nu}, C_{\mu\nu})$ on an equal footing. However, this democracy gets broken by the boundary conditions. In the variation \eqref{eq:variationosd=2} we will impose Dirichlet boundary conditions on the variations of $\tau_\mu$ and $h_{\mu\nu}$ but not for the shear $C_{\mu\nu}$. This makes sense physically because gravitational radiation enters $\mathcal{I}^+$ and so from the perspective of $\mathcal{I}^+$ we are dealing with an open system. To properly discuss the boundary conditions for the variation of the shear we need to split it into its vacuum (soft) and radiative (hard) components and include the boundary at past null infinity. We will not discuss these issues here as for our purposes we only need the general form \eqref{eq:variationosd=2} and not all the details of a well-posed variational principle. 

We will perform a rewriting of the Bondi loss equations where all aspects that pertain to radiation, i.e. the $K$-curvatures feature as source terms that render a certain boundary energy-momentum tensor not conserved. This will be a different boundary energy-momentum tensor than the one we introduced in the previous section.

The boundary energy-momentum tensor from the previous section has a number of properties that are unusual from a Carroll perspective. Ordinarily if we have Carroll boost symmetry the energy flux, i.e. $T^\mu h_\mu^\rho$, vanishes \cite{Hartong:2015usd,deBoer:2017ing}. What we found is that $T^\mu h_\mu^\rho$ is nonzero and that this is due to two effects: a boost anomaly and the presence of another source, the shear, that transforms under Carroll boosts. See \cite{Ciambelli:2018wre,Mittal:2022ywl,Campoleoni:2023fug} for similar observations. We will show that there is a way of writing the Bondi loss equations where the diffeomorphism Ward identity is the one for an energy-momentum tensor of an ordinary Carroll theory, i.e. with vanishing energy flux. In this way of writing there are now flux terms on the right-hand side that describe the non-conservation of this new energy-momentum tensor. We will see that these flux terms are described by the $K$-curvatures\footnote{Because the Carroll boost anomaly is one of the $K$-curvatures we can choose to put it either on the right-hand side, so that the energy flux is zero or we can move it to the left-hand side so that there are fewer flux terms but this leads to an energy flux that is purely anomalous. From the perspective of the Bondi loss equations both perspectives are possible.}. It is this new Carroll energy-momentum tensor in terms of which the BMS charges are most readily defined.

As the actual rewriting of the Bondi loss equations is quite an involved technical exercise we decided to move all aspects of this calculation to 
appendix \ref{app;rewritingBondiloss}. There it is shown that the Bondi loss equations can be rewritten as
\begin{eqnarray}
    0 & = & -\left(\mathcal{L}_v-\frac{3}{2}K\right)\tau\cdot T'+\frac{1}{2}C^{\rho\sigma}v^\mu R(K)_{\mu\rho}{}^a e^a_\sigma+\left(\mathcal{D}_\mu+a_\mu\right)\mathcal{A}^\mu_B\,,\label{eq:BLeq1}\\
    0 & = & -\left(\mathcal{L}_v-K\right)P'_\kappa+h_{\kappa\sigma}\mathcal{D}_\mu\left(\tilde {T}'^{\mu\sigma}\right)-\frac{1}{2}h^\mu_\kappa\left(\partial_\mu+3a_\mu\right)\left(\tau\cdot T'\right)\nonumber\\
    &&+\frac{1}{2}h_{\kappa\sigma}\left(\mathcal{D}_\mu+2a_\mu\right)R(K)^{\mu\sigma}+C_{\sigma\kappa}\mathcal{A}^\sigma_B\,,\label{eq:BLeq2}
\end{eqnarray}
where we defined a new energy density $\tau\cdot T'$ and a new momentum density $P'_\kappa$ which are given by
\begin{eqnarray}
    \tau\cdot T' & = & U^\rho U^\sigma C_{r\rho r\sigma}\big\vert_{\mathcal{O}(r^{-3})}\nonumber\\
    &=&-v^\rho v^\sigma g^{(1)}_{\rho\sigma}-\frac{1}{16}K\left(F^2-C^2\right)-\frac{1}{8}\left(\mathcal{L}_v-K\right)F^2-\frac{1}{4}C\cdot N\nonumber\\
    &=&\tau\cdot T-\frac{1}{4}C\cdot N-\frac{1}{4}\left(\mathcal{L}_v-K\right)F^2-\mathcal{D}_\rho\overset{(0)}{P}{}^\rho\,,\label{eq:redefenergyden}\\
    P'_\kappa & = &  -U^\rho\Pi^\sigma_\kappa C_{r\rho r\sigma}\big\vert_{\mathcal{O}(r^{-3})}+\frac{1}{2}h^{\sigma\nu}\mathcal{D}_\sigma\left(\chi_{\nu\kappa}-\frac{1}{2}C_\nu{}^{\rho}F_{\rho\kappa}\right)\nonumber\\
    & = & \frac{3}{2}v^\mu h^\nu_\kappa g^{(1)}_{\mu\nu}-\frac{3}{32}a_\kappa\left(F^2-C^2\right)+\frac{3}{32}h^\rho_\kappa\left(\partial_\rho+2a_\rho\right)\left(F^2-C^2\right)\nonumber\\
    &&-a_\sigma D^\sigma{}_\kappa+\frac{1}{2}h^{\sigma\nu}\mathcal{D}_\sigma\left(\chi_{\nu\kappa}-\frac{1}{2}C_\nu{}^{\rho}F_{\rho\kappa}\right)\nonumber\\
    & = & P_\kappa+\frac{1}{2}C^\rho{}_\kappa\overset{(0)}{P}_\rho-\frac{5}{2}F^\rho{}_\kappa\overset{(0)}{P}_\rho+\frac{1}{16}h^\rho_\kappa\left(\partial_\rho+2a_\rho\right)\left(F^2-C^2\right)\nonumber\\
    &&-\frac{1}{4}h^{\sigma\nu}\mathcal{D}_\sigma\left(C_\nu{}^\rho F_{\rho\kappa}\right)\,.
\end{eqnarray}
In the first equality for $\tau\cdot T'$ and $P'_\kappa$ we see that these are equal to specific components of the Weyl tensor at leading order in $1/r$. These components were computed in 
 \eqref{eq:Weylmass} and \eqref{eq:Weylmomentum} and so we see that the new energy density $\tau\cdot T'$ and momentum density $P'_\kappa$ can be expressed in terms of Newman--Penrose scalar (up to the ambiguity in $P'_\kappa$ described by $\chi_{\mu\nu}$). 
 In the second equality we expressed the energy density $\tau\cdot T'$ and momentum density $P'_\kappa$ in terms of quantities that appear in the near-boundary expansion of the metric. Finally, in the third equality we express the new $\tau\cdot T'$ and $P'_\kappa$ in terms of the unprimed quantities that we defined in the previous subsection.
The object $\tilde {T}'^{\mu\sigma}$ in \eqref{eq:BLeq2} is 
\begin{eqnarray}
    \tilde {T}'^{\mu\sigma} & = & \frac{1}{2}K D^{\mu\sigma}+\frac{1}{2}h^{\mu\alpha}h^{\sigma\beta}\mathcal{L}_v\left(\chi_{\alpha\beta}-\frac{1}{2}C_\alpha{}^\rho F_{\rho\beta}\right)\,.
\end{eqnarray}
Hence we see that except for the term containing $D^{\mu\sigma}$ almost all of $\tilde T^{\mu\sigma}$ has disappeared (of course apart from the ambiguity).

The Bondi loss equations \eqref{eq:BLeq1} and \eqref{eq:BLeq2} can be combined into the following equation\footnote{The $v^\mu$ projection of \eqref{eq:diffeoWInew} gives \eqref{eq:BLeq1} and the spatial projection gives \eqref{eq:BLeq2}.}
\begin{equation}\label{eq:diffeoWInew}
    -e^{-1}\partial_\mu \left(e T'^\mu{}_\nu\right)+T'^\mu\partial_\nu\tau_\mu+\frac{1}{2}T'^{\mu\rho}\partial_\nu h_{\mu\rho}=\mathcal{F}_\nu\,,
\end{equation}
where we defined the primed energy-momentum tensor 
\begin{equation}
T'^\mu{}_\nu=T'^\mu\tau_\nu+T'^{\mu\rho}h_{\rho\nu}\,,
\end{equation}
with
\begin{eqnarray}
    T'^\mu & = & -\tau\cdot T' v^\mu\,,\label{eq:T'mu}\\
    T'^{\mu\nu} & = & -2v^{(\mu}h^{\nu)\kappa}P'_\kappa+\frac{1}{2}K D^{\mu\nu}-\frac{1}{2}h^{\mu\nu}\tau\cdot T'\,,\label{eq:T'munu}
\end{eqnarray}
where we left out the ambiguity with parameter $\chi_{\mu\nu}-\frac{1}{2}C_\mu{}^\rho F_{\rho\nu}$.
In this way of writing $T'^\mu{}_\nu$ is a `standard' traceless Carroll energy-momentum tensor with vanishing energy flux. This agrees with observations in \cite{Donnay:2022wvx}. The non-conservation of this Carrollian energy-momentum tensor is due to the  flux $\mathcal{F}_\nu$. This can be expressed  in terms of the $K$-curvatures as
\begin{eqnarray}
    v^\nu\mathcal{F}_\nu & = & -\left(\mathcal{D}_\mu+a_\mu\right)\mathcal{A}^\mu_B-\frac{1}{2}v^\nu C^{\rho\sigma} R(K)_{\nu\rho}{}^a e^a_\sigma\,,\\
    h^\nu_\kappa\mathcal{F}_\nu & = & \frac{1}{2}h_{\kappa\sigma}\left(\mathcal{D}_\mu+2a_\mu\right)R(K)^{\mu\sigma}+C^\sigma{}_\kappa\mathcal{A}_\sigma^B\,.
\end{eqnarray}

If we contract \eqref{eq:diffeoWInew} with any vector $K^\nu$ we find
\begin{equation}
    -e^{-1}\partial_\mu\left(e T'^\mu{}_\nu K^\nu\right)+T'^\mu\mathcal{L}_K\tau_\mu+\frac{1}{2}T'^{\mu\nu}\mathcal{L}_K h_{\mu\nu}=K^\nu\mathcal{F}_\nu\,.
\end{equation}
The vector $K^\mu$ corresponds to a Carroll conformal Killing vector if we can solve
\begin{eqnarray}
    0=\delta \tau_\mu & = & \mathcal{L}_\chi\tau_\mu+\Lambda_D\tau_\mu+\lambda_\mu\,,\label{eq:asym1}\\
    0=\delta h_{\mu\nu} & = & \mathcal{L}_\chi h_{\mu\nu}+2\Lambda_D h_{\mu\nu}\,,\label{eq:asym2}
\end{eqnarray}
for $\chi^\mu=K^\mu$. 
For every such $K^\mu$ we define the BMS currents $J^\mu_{\text{BMS}}$ as
\begin{equation}
J^\mu_{\text{BMS}}=e T'^\mu{}_\nu K^\nu\,.
\end{equation}
Using that $T'^\mu$ has vanishing flux \eqref{eq:T'mu} and that the trace $T'^\mu{}_\mu=0$ vanishes \eqref{eq:T'munu} we see that 
\begin{equation}
    \partial_\mu J^\mu_{\text{BMS}} =-e K^\nu\mathcal{F}_\nu\,.
\end{equation}
This makes manifest that the non-conservation of the BMS currents $J^\mu_{\text{BMS}}$ is controlled by the $K$-curvatures.
Whenever we can write $K^\nu\mathcal{F}_\nu$ as the divergence of some current we obtain a conserved current at $\mathcal{I}^+$.
We expect this way of expressing the BMS currents agrees with what is called the `good prescription' in the review paper \cite{Donnay:2023mrd} (see also \cite{Compere:2018ylh,Donnay:2020lur,Freidel:2021qpz,Donnay:2021wrk}).

One may wonder what the solution space for these conformal Carroll Killing vectors is like. We can always by a combination of a Carroll boost, Weyl rescaling and boundary diffeomorphism bring the boundary geometry (locally) to the case where we have a round 2-sphere and an exact $\tau=du$ with $u$ retarded time. Hence we expect that we can always find $K^\mu$ that form the extended BMS algebra. More covariantly we can decompose $K^\mu$ as follows
\begin{equation}\label{eq:CCKV1}
    K^\mu=\phi v^\mu+h^{\mu\nu}\psi_\nu\,,
\end{equation}
where we take $\psi_\nu$ to be spatial.
This leads to the following three equations
\begin{subequations}
\begin{eqnarray}
    \left(\mathcal{L}_v+K\right)\psi_\mu & = & 0\,,\\
    h^\rho_{\langle\mu}h^\sigma_{\nu\rangle}\mathcal{D}_\rho\psi_\sigma & = & 0\,,\\
    \left(\mathcal{L}_v+\frac{1}{2}K\right)\phi-\frac{1}{2}h^{\rho\sigma}\left(\mathcal{D}_\rho-2a_\rho\right)\psi_\sigma & = & 0\,,
\end{eqnarray}
\end{subequations}
as well as
\begin{equation}
    \begin{split}
        \Lambda_D & = -\frac{1}{2}h^{\rho\sigma}\left(\mathcal{D}_\rho-2a_\rho\right) \psi_\sigma-\tilde b_\mu K^\mu\,,\\
    \lambda_\mu & =  h^\rho_\mu\left(\partial_\rho-a_\rho\right)\phi-\psi_\rho F^\rho{}_\mu\,.\label{eq:CCKV6}
    \end{split}
\end{equation}
These can in principle be solved in the order in which they are presented. It basically says that every time one has a conformal Killing vector on the 2-dimensional base (first two equations) you can extend it to a conformal Killing vector on the 3-dimensional Carroll manifold by solving the latter equation. The homogeneous solution to the third equation are supertranslations. The former are superrotations. The role of $a_\mu$ and $K$ is to make the equations Weyl covariant. The weights\footnote{The Weyl weights are defined such that $h_{\mu\nu}$ has weight $+2$. } follow from the fact that $K^\mu$ has Weyl weight zero. For example $\lambda_\mu$ has weight $+1$, $\phi$ has $+1$ and $\psi_\mu$ has weight $+2$.

The fact the new primed (Carrollian)
energy-momentum tensor is non-conserved due to the presence of $C_{\mu\nu}-C^{\text{vac}}_{\mu\nu}$ suggests that we should consider the responses to varying $\tau_\mu$ and $h_{\mu\nu}$ keeping $C_{\mu\nu}-C^{\text{vac}}_{\mu\nu}$ fixed (instead of $C_{\mu\nu}$) \cite{Hartong:2026WIP}.

\subsection*{Acknowledgements}
I would like to thank Tim Adamo, Glenn Barnich, Jan de Boer, Emil Have, James Lucietti, Lionel Mason, Vijay Nenmeli, Niels Obers, Gerben Oling and Stefan Vandoren for useful discussions. I also wish to thank Gerben Oling for reading and commenting on earlier versions of this manuscript.
The work of JH was supported by the 
Royal Society URF Renewal grant URF\textbackslash R\textbackslash 221038.

\appendix

\section{Useful $d=2$ identities}

In this appendix we collect a number of useful identities that are special for $d=2$ and that are frequently used in calculations. 

In the following let $A_{\mu\nu}$  and $\tilde A_{\mu\nu}$ both be a spatial antisymmetric tensor. Furthermore, let $S_{\mu\nu}$ and $\tilde S_{\mu\nu}$ both be spatial STF tensors. Then we have the following 2D identities:
\begin{eqnarray}
    h^{\rho\sigma}A_{\rho\mu}\tilde A_{\sigma\nu}-h^{\rho\sigma}A_{\rho\nu}\tilde A_{\sigma\mu} & = & 0\,,\label{eq:AtildeA}\\
    h^{\rho\sigma}A_{\rho\mu}S_{\sigma\nu}-h^{\rho\sigma}A_{\rho\nu}S_{\sigma\mu} & = & 0\,,\label{eq:ASSTF}\\
    h^{\rho\sigma}S_{\rho\mu}\tilde S_{\sigma\nu}+h^{\rho\sigma}S_{\rho\nu}\tilde S_{\sigma\mu} & = & h_{\mu\nu}S\cdot \tilde S\,.\label{eq:STFSTF}
\end{eqnarray}
These identities follow from properties of 2D tensors and they can be proven by using spatial vielbeine. Since they are spatial tensors we can always write 
\begin{equation}
    A_{\mu\nu}=e^a_\mu e^b_\nu \varepsilon_{ab}A\,,\qquad \tilde A_{\mu\nu}=e^a_\mu e^b_\nu \varepsilon_{ab}\tilde A\,,\qquad S_{\mu\nu}=e^a_\mu e^b_\nu S_{ab}\,,
\end{equation}
where $S_{ab}$ is STF with respect to $\delta^{ab}$. The first identity then follows from the familiar statement $\varepsilon_{ac}\varepsilon_{bc}=\delta_{ab}$ whereas the second identity follows from 
\begin{equation}
    \varepsilon_{ac}S_{bc}+\varepsilon_{bc}S_{ac}=0\,,
\end{equation}
which can be readily verified. 

The next set of identities follows from the following 2D identity involving a triple product of Kronecker deltas:
\begin{equation}\label{eq:tripledeltas}
    \left(\delta^d_a\delta^e_b-\delta^d_b\delta^e_a\right)\delta^f_c-\delta_{ac}\left(\delta^{df}\delta^e_b-\delta^{ef}\delta^d_b\right)+\delta_{bc}\left(\delta^{df}\delta^e_a-\delta^{ef}\delta^d_a\right)=0\,.
\end{equation}
By using spatial vielbeine this can be turned into an identity involving $h_{\mu\nu}$. Appropriately contracting three of the indices with either $\mathcal{D}_\mu A_{\nu\rho}$ or $\mathcal{D}_\mu S_{\sigma\nu}$ leads to the following identities
\begin{equation}\label{eq:DASA}
    h^\mu_\alpha h^\nu_\beta h^\rho_\gamma\mathcal{D}_\mu A_{\nu\rho}=\left(h_{\alpha\beta}h^\rho_\gamma-h_{\alpha\gamma}h^\rho_\beta\right)h^{\kappa\lambda}\mathcal{D}_\kappa A_{\lambda\rho}\,,
\end{equation}
for any spatial antisymmetric tensor $A_{\mu\nu}$, and
\begin{equation}\label{eq:DSTFid}
    h^\mu_\alpha h^\nu_\beta h^{\sigma}_\gamma\Big(\mathcal{D}_\mu S_{\sigma\nu}-\mathcal{D}_\nu S_{\sigma\mu}\Big)=-2h^\mu_{[\alpha}h_{\beta]\gamma}h^{\rho\sigma}\mathcal{D}_\rho S_{\sigma\mu}\,,
\end{equation}
for any spatial STF tensor $S_{\mu\nu}$.
In a similar way we can show the following result
\begin{equation}\label{eq:XYA}
    X_\alpha Y^\rho A_{\rho\beta}-X_\beta Y^\rho A_{\rho\alpha}=A_{\alpha\beta}Y^\rho X_\rho\,,
\end{equation}
for any spatial $X_\rho, Y^\rho$ and antisymmetric $A_{\rho\sigma}$.

\section{Boundary Riemann tensor}\label{app:bdryRiem}

The purpose of this appendix is to give details of the boundary curvature tensor associated with the connection \eqref{eq:conn}. One of the main results is the expression for the Ricci scalar in \eqref{eq:Ricscalar}.

Given the connection \eqref{eq:conn} we can define a boundary Riemann tensor in the usual manner as follows
\begin{eqnarray}
    \left[\mathcal{D}_\mu\,,\mathcal{D}_\nu\right]X_\rho & = & \mathcal{R}_{\mu\nu\rho}{}^\sigma X_\sigma\,,\\
    \left[\mathcal{D}_\mu\,,\mathcal{D}_\nu\right]X^\sigma & = & -\mathcal{R}_{\mu\nu\rho}{}^\sigma X^\rho\,.
\end{eqnarray}
This leads to the familiar result
\begin{equation}
    \mathcal{R}_{\mu\nu\rho}{}^\sigma=-\partial_\mu\mathcal{C}_{\nu\rho}^\sigma-\mathcal{C}_{\mu\lambda}^\sigma\mathcal{C}^\lambda_{\nu\rho}-\left(\mu\leftrightarrow\nu\right)\,.
\end{equation}
We define the Ricci tensor as follows
\begin{equation}
    \mathcal{R}_{\mu\rho}=\mathcal{R}_{\mu\nu\rho}{}^\nu\,.
\end{equation}
The Ricci tensor is symmetric as a result of the properties \eqref{eq:propconn}. Because the connection is symmetric the algebraic and differential Bianchi identities are simply
\begin{eqnarray}
    \mathcal{R}_{[\mu\nu\rho]}{}^\sigma & = & 0\,,\\
    \mathcal{D}_{[\mu}\mathcal{R}_{\nu\rho]\lambda}{}^\sigma & = & 0\,.
\end{eqnarray}

We next define
\begin{eqnarray}
    \mathcal{Q}_{\mu\nu\rho\sigma} & = & \mathcal{R}_{\mu\nu\rho}{}^\lambda h_{\lambda\sigma}+\frac{1}{2}\partial_\mu K\tau_\sigma h_{\nu\rho}-\frac{1}{2}\partial_\nu K\tau_\sigma h_{\mu\rho}+\frac{1}{4}K^2\tau_\mu\tau_\sigma h_{\nu\rho}-\frac{1}{4}K^2\tau_\nu\tau_\sigma h_{\mu\rho}\nonumber\\
    &&+\frac{1}{4}KF_{\mu\sigma}h_{\nu\rho}-\frac{1}{4}KF_{\nu\sigma}h_{\mu\rho}-\frac{1}{2}K\tau_\mu a_\sigma h_{\nu\rho}+\frac{1}{2}K\tau_\nu a_\sigma h_{\mu\rho}\,.\label{eq:defS}
\end{eqnarray}
This object has been constructed to possess the following symmetry properties
\begin{eqnarray}
    \mathcal{Q}_{(\mu\nu)\rho\sigma} & = & 0\,,\label{eq:prop1}\\
    \mathcal{Q}_{[\mu\nu\rho]\sigma} & = & 0\,,\label{eq:prop2}\\
    \mathcal{Q}_{\mu\nu(\rho\sigma)} & = & 0\,.\label{eq:prop3}
\end{eqnarray}
To prove the latter of these we can use the following identity
\begin{eqnarray}
    \left[\mathcal{D}_\mu\,,\mathcal{D}_\nu\right]h_{\rho\sigma} & = & \mathcal{R}_{\mu\nu\rho}{}^\lambda h_{\sigma\lambda}+\mathcal{R}_{\mu\nu\sigma}{}^\lambda h_{\rho\lambda}\nonumber\\
    & = & -\frac{1}{2}\partial_\mu K\tau_\rho h_{\nu\sigma}-\frac{1}{4}KF_{\mu\rho}h_{\nu\sigma}+\frac{1}{2}K\tau_\mu a_\rho h_{\nu\sigma}-\frac{1}{4}K^2\tau_\mu\tau_\rho h_{\nu\sigma}\nonumber\\
    &&+\left(\rho\leftrightarrow\sigma\right)-\left(\mu\leftrightarrow\nu\right)\,,
\end{eqnarray}
where we used \eqref{eq:covDh}.

Using standard argument it then follows that $\mathcal{Q}_{\mu\nu\rho\sigma}$ also obeys
\begin{eqnarray}
    \mathcal{Q}_{\mu\nu\rho\sigma} & = & \mathcal{Q}_{\rho\sigma\mu\nu}\,,\label{eq:prop4}\\
    \mathcal{Q}_{[\mu\nu\rho\sigma]} & = & 0\,,\label{eq:prop5}
\end{eqnarray}
and furthermore that properties \eqref{eq:prop1}, \eqref{eq:prop3}, \eqref{eq:prop4}, and \eqref{eq:prop5} are independent. Property \eqref{eq:prop2} can be seen to be a consequence of the other properties. A $(0,4)$ tensor obeying \eqref{eq:prop1}, \eqref{eq:prop3}, \eqref{eq:prop4}, and \eqref{eq:prop5} is known as a Riemann-like tensor as it has the same symmetry properties as the Riemann tensor associated with the Levi-Civita connection of a (pseudo-)Riemannian geometry. 

If we contract $\mathcal{Q}_{\mu\nu\rho\sigma}$  with a single $v^\lambda$ it is enough to know e.g. $\mathcal{Q}_{\mu\nu\rho\sigma}v^\sigma$ as any other contraction follows from the symmetry properties. Now $\mathcal{Q}_{\mu\nu\rho\sigma}v^\sigma$ can be readily computed from the definition \eqref{eq:defS}. The same is true for any $v^\kappa$ contraction of $\mathcal{R}_{\mu\nu\rho}{}^\lambda h_{\lambda\sigma}$. For example we obtain
\begin{eqnarray}
    \mathcal{R}_{\mu\nu\rho}{}^\lambda h_{\lambda\sigma} v^\rho & = & \frac{1}{2}\partial_\mu K h_{\nu\sigma}-\frac{1}{4}K^2\tau_\nu h_{\mu\sigma}-\left(\mu\leftrightarrow\nu\right)\,,\\
    \mathcal{R}_{\mu\nu\rho}{}^\lambda h_{\lambda\sigma} v^\nu & = & \frac{1}{2}Ka_\sigma h_{\mu\rho}-\frac{1}{2}\partial_\rho K h_{\sigma\mu}+\frac{1}{2}h_\sigma^\alpha\partial_\alpha K h_{\rho\mu}-\frac{1}{4}K^2\tau_\rho h_{\sigma\mu}\,.\label{eq:vRiem2nd}
\end{eqnarray}
The first of these two identities can also be computed from the fact that $\mathcal{R}_{\mu\nu\rho}{}^\lambda  v^\rho=-\left[\mathcal{D}_\mu\,,\mathcal{D}_\nu\right]v^\lambda$.

If we solve \eqref{eq:defS} for $\mathcal{R}_{\mu\nu\rho}{}^\lambda$ we obtain
\begin{eqnarray}
    \mathcal{R}_{\mu\nu\rho}{}^\kappa & = & \mathcal{Q}_{\mu\nu\rho\sigma}h^{\sigma\kappa}-v^\kappa \mathcal{R}_{\mu\nu\rho}{}^\lambda\tau_\lambda-\frac{1}{4}K F_\mu{}^\kappa h_{\nu\rho}+\frac{1}{4}K F_\nu{}^\kappa h_{\mu\rho}\nonumber\\
    &&+\frac{1}{2}K\tau_\mu a^\kappa h_{\nu\rho}-\frac{1}{2}K\tau_\nu a^\kappa h_{\mu\rho}\,,\label{eq:Riem}
\end{eqnarray}
where $\mathcal{R}_{\mu\nu\rho}{}^\lambda\tau_\lambda=\left[\mathcal{D}_\mu\,,\mathcal{D}_\nu\right]\tau_\rho$ and so is a known quantity. Explicitly, we have
\begin{eqnarray}
    \mathcal{R}_{\mu\nu\rho}{}^\lambda\tau_\lambda & = & \frac{1}{2}\mathcal{D}_\mu F_{\nu\rho}-\frac{1}{2}\mathcal{D}_\nu F_{\mu\rho}-a_\rho F_{\mu\nu}+\tau_\mu\left(\mathcal{D}_\nu a_\rho+a_\nu a_\rho\right)-\tau_\nu\left(\mathcal{D}_\mu a_\rho+a_\mu a_\rho\right)\,,\nonumber\\
    &&\label{eq:Riemtau}
\end{eqnarray}
where we used \eqref{eq:covDtau}.

We conclude that all contractions of $\mathcal{R}_{\mu\nu\rho}{}^\kappa$ with $v^\sigma$ can be explicitly computed. This leaves us with the fully spatial part, i.e. $h^\mu_\alpha h^\nu_\beta h^\rho_\gamma h_{\kappa\delta}\mathcal{R}_{\mu\nu\rho}{}^\kappa$ which is equal to 
\begin{equation}
   h^\mu_\alpha h^\nu_\beta h^\rho_\gamma h_{\kappa\delta}\mathcal{R}_{\mu\nu\rho}{}^\kappa= h^\mu_\alpha h^\nu_\beta h^\rho_\gamma h^\sigma_\delta \mathcal{Q}_{\mu\nu\rho\sigma}-\frac{1}{4}KF_{\alpha\delta}h_{\beta\gamma}+\frac{1}{4}KF_{\beta\delta}h_{\alpha\gamma}\,.
\end{equation}
Because $\mathcal{Q}_{\mu\nu\rho\sigma}$ is a Riemann-like tensor so is $h^\mu_\alpha h^\nu_\beta h^\rho_\gamma h^\sigma_\delta \mathcal{Q}_{\mu\nu\rho\sigma}$ and furthermore because it is spatial we can write it as $e^a_\alpha e^b_\beta e^c_\gamma e^d_\delta \mathcal{Q}_{abcd}$ where $\mathcal{Q}_{abcd}$ is a 2-dimensional Riemann-like tensor, so that it only has one independent component and we can write 
\begin{equation}
    \mathcal{Q}_{abcd}=\frac{1}{2}\mathcal{Q}\left(\delta_{ac}\delta_{bd}-\delta_{ad}\delta_{bc}\right)\,.
\end{equation}
Hence, we have
\begin{equation}\label{eq:2DRiem}
    h^\mu_\alpha h^\nu_\beta h^\rho_\gamma h_{\kappa\delta}\mathcal{R}_{\mu\nu\rho}{}^\kappa= \frac{1}{2}\mathcal{Q}\left(h_{\alpha\gamma}h_{\beta\delta}-h_{\alpha\delta}h_{\beta\gamma}\right)-\frac{1}{4}KF_{\alpha\delta}h_{\beta\gamma}+\frac{1}{4}KF_{\beta\delta}h_{\alpha\gamma}\,.
\end{equation}
It remains to determine $\mathcal{Q}$.

However, before we determine $\mathcal{Q}$ let us first compute the Ricci tensor. Contracting $\kappa$ and $\nu$ in \eqref{eq:Riem} leads to
\begin{equation}
    \mathcal{R}_{\mu\rho} = \mathcal{Q}_{\mu\nu\rho\sigma}h^{\nu\sigma}-\frac{1}{2}\tau_\mu\mathcal{L}_v a_\rho-\frac{1}{2}\tau_\rho\mathcal{L}_v a_\mu-\mathcal{D}_{(\mu}a_{\rho)}-a_\mu a_\rho\,. 
\end{equation}
If we trace this with $h^{\mu\rho}$ we obtain
\begin{equation}\label{eq:StoRicscalar}
\mathcal{R}:=h^{\mu\rho}\mathcal{R}_{\mu\rho}=\mathcal{Q}-e^{-1}\partial_\mu\left(e a^\mu\right)\,.
\end{equation}
This determines $\mathcal{Q}$ in terms of $\mathcal{R}$. We can compute $\mathcal{R}$ directly from its definitions and the form of the metric \eqref{eq:metrich}. This leads to\footnote{One way to do this calculation is to define $h_{\mu\nu}=e^{2\varphi}\tilde h_{\mu\nu}$ where $\tilde h_{\mu\nu}=\delta_{ab}\partial_\mu X^a\partial_\nu X^b$ and split the connection as 
\begin{equation}
    \mathcal{C}^\rho_{\mu\nu}=-\frac{1}{2}v^\rho\left(\partial_\mu\tau_\nu+\partial_\nu\tau_\mu+a_\mu\tau_\nu+a_\nu\tau_\mu\right)+2h^\rho_{\langle\mu}\partial_{\nu\rangle}\varphi+\tilde{\mathcal{C}}^\rho_{\mu\nu}\,,
\end{equation}
where 
\begin{equation}
    \tilde{\mathcal{C}}^\rho_{\mu\nu}=\frac{1}{2}\tilde h^{\rho\sigma}\left(\partial_\mu \tilde h_{\nu\sigma}+\partial_\nu \tilde h_{\mu\sigma}-\partial_\sigma \tilde h_{\mu\nu}\right)\,.
\end{equation}
The object $\tilde h^{\rho\sigma}$ obeys
\begin{align}
    & \tilde h^{\rho\sigma}\partial_\rho X^a\partial_\sigma X^b=\delta^{ab}\,,\\
    &\tilde h^{\rho\sigma}\tau_\rho=0\,.
\end{align}
Using this split of the connection we can compute 
\begin{equation}
\mathcal{R}=-2e^{-1}\partial_\mu\left(e h^{\mu\nu}\partial_\nu\varphi\right)-e^{-1}\partial_\mu\left(e a^\mu\right)+2a^\mu\partial_\mu\varphi+Y\,,
\end{equation}
where we defined
\begin{equation}
    Y=h^{\mu\rho}\left(-v^\lambda\left(\partial_\lambda\tau_\nu\right)\tilde {\mathcal{C}}^\nu_{\mu\rho}-h^\kappa_\lambda\left(\partial_\mu v^\lambda \right)\partial_\rho\tau_\kappa-\partial_\mu\tilde{\mathcal{C}}^\nu_{\nu\rho}+\partial_\nu\tilde{\mathcal{C}}^\nu_{\mu\rho}-\tilde{\mathcal{C}}_{\mu\lambda}^\nu\tilde{\mathcal{C}}_{\nu\rho}^\lambda+\tilde{\mathcal{C}}_{\nu\lambda}^\nu\tilde{\mathcal{C}}_{\mu\rho}^\lambda\right)\,.
\end{equation}
Using the special form of $\tilde h_{\mu\nu}$ it can be shown that $Y=0$.} 
\begin{eqnarray}\label{eq:Ricscalar}
    \mathcal{R} & = & -2e^{-1}\partial_\mu\left(e h^{\mu\nu}\partial_\nu\varphi\right)-e^{-1}\partial_\mu\left(e a^\mu\right)+2a^\mu\partial_\mu\varphi\,,\nonumber\\
    & = & -2h^{\mu\nu}\mathcal{D}_\mu\partial_\nu\varphi-e^{-1}\partial_\mu\left(e a^\mu\right)\,.
\end{eqnarray}

\section{Bulk Weyl tensor}\label{app:Weyltensor}

On shell the Weyl tensor is equal to the Riemann tensor. In this appendix we will first compute $R_{MNPQ}$ in terms of the variables $\beta, S, \Pi_{\mu\nu}$ and then we will compute the LO non-vanishing component of the Weyl tensor. 

We start with the definition of the Riemann tensor in terms of the LC connection
\begin{equation}
    R_{MNPQ}=\left(-\partial_M\Gamma^S_{NP}+\partial_N\Gamma^S_{MP}-\Gamma^S_{MR}\Gamma^R_{NP}+\Gamma^S_{NR}\Gamma^R_{MP}\right)g_{SQ}\,.
\end{equation}
We will use the Carroll-covariant BS gauge choice \eqref{eq:CarcovBSgauge}.

There are three sets of Riemann tensor components depending on how many indices are equal to $r$, i.e. $R_{r\rho r\sigma}$, $R_{r\nu\rho\sigma}$ and $R_{\mu\nu\rho\sigma}$. On shell there are 10 independent components. Let us first discuss what the Einstein equations tell us about the various Riemann tensor components and then we calculate below the ones that are left undetermined. The notation used here for the bulk metric components is introduced in equations \eqref{eq:gtoPi}--\eqref{eq:propsUVPiinvPi}.

From $R_{rr}=0$ we learn that
\begin{equation}
    \Pi^{\mu\nu}R_{r\mu r\nu}=0\,.
\end{equation}
Using $R_{\rho r}=0$ we learn that
\begin{equation}
    -U^\sigma R_{r\rho r\sigma}+\Pi^{\mu\nu}R_{r\mu\rho\nu}=0\,.
\end{equation}
This tells us that
\begin{eqnarray}
    U^\rho \Pi^{\mu\nu}R_{r\mu\rho\nu} & = & U^\rho U^\sigma R_{r\rho r\sigma}\,,\label{eq:traceUPi}\\
    \Pi^{\mu\nu}\Pi^\rho_\alpha R_{r\mu\rho\nu} & = & \Pi^\rho_\alpha U^\sigma R_{r\rho r\sigma}\,.\label{eq:UPiR}
\end{eqnarray}
Next if we use the Bianchi identity
\begin{equation}
    R_{r\rho\mu\nu}+R_{r\mu\nu\rho}+R_{r\nu\rho\mu}=0\,.
\end{equation}
By contracting this with $U^\rho\Pi^\mu_\alpha\Pi^\nu_\beta$ we obtain
\begin{equation}\label{eq:ASUPi}
    2U^\rho\Pi^\mu_{[\alpha}\Pi^\nu_{\beta]} R_{r\mu\rho\nu}=U^\rho\Pi^\mu_\alpha\Pi^\nu_\beta R_{r\rho\mu\nu}\,.
\end{equation}
In $d=2$ we have the following identity
\begin{eqnarray}
    \Pi^\rho_\gamma\Pi^\mu_\alpha\Pi^\nu_\beta R_{r\rho\mu\nu} & = & E^a_\gamma E^b_\alpha E^c_\beta R_{rabc}=E^a_\gamma E^b_\alpha E^c_\beta\left(\delta_{ab}R_{rddc}-\delta_{ac}R_{rddb}\right)\nonumber\\
    & = & 2\Pi_{\gamma[\alpha}\Pi^\rho_{\beta]}\Pi^{\mu\nu}R_{r\mu\nu\rho}=-2\Pi_{\gamma[\alpha}\Pi^\rho_{\beta]}U^\sigma R_{r\rho r\sigma}\,,
\end{eqnarray}
where in the first equality we used $R_{rabc}=E^\mu_a E^\nu_b E^\rho_c R_{r\mu\nu\rho}$. The second equality is where we used an identity that is special for $d=2$. In the last equality we used \eqref{eq:UPiR}.

Next, we consider the $\rho\sigma$ components of $R_{MN}=0$ which reads
\begin{equation}\label{eq:Rrhosigma}
    SR_{r\rho r\sigma}+U^\mu R_{r\sigma\mu\rho}+U^\mu R_{r\rho\mu\sigma}+\Pi^{\mu\nu}R_{\rho\mu\sigma\nu}=0\,.
\end{equation}
We have the following two $d=2$ identities
\begin{eqnarray}
    \Pi^\mu_\alpha \Pi^\nu_\beta\Pi^\rho_\gamma\Pi^\sigma_\delta R_{\mu\nu\rho\sigma} & = & \frac{1}{2}\left(\Pi_{\alpha\gamma}\Pi_{\beta\delta}-\Pi_{\beta\gamma}\Pi_{\alpha\delta}\right)\Pi^{\kappa\lambda}\Pi^{\rho\sigma}R_{\kappa\rho\lambda\sigma}\,,\\
    U^\mu \Pi^\nu_\alpha\Pi^\rho_\beta\Pi^\sigma_\gamma R_{\mu\nu\rho\sigma} & = & 2\Pi_{\alpha[\beta}\Pi_{\gamma]}^\sigma\Pi^{\nu\rho} U^\mu R_{\mu\nu\rho\sigma}\,.
\end{eqnarray}
These can be proven by using vielbein projections of the Riemann tensor and its symmetry properties. Using \eqref{eq:Rrhosigma} we next deduce that
\begin{eqnarray}
    \Pi^{\kappa\lambda}\Pi^{\rho\sigma}R_{\kappa\rho\lambda\sigma} & = & -2 U^\rho U^\sigma R_{r\rho r\sigma}\,,\\
    \Pi_{\gamma}^\sigma\Pi^{\nu\rho} U^\mu R_{\mu\nu\rho\sigma} & = & \Pi^\sigma_\gamma U^\mu U^\nu R_{r\mu\nu\sigma}+SU^\rho \Pi^\sigma_\gamma R_{r\rho r\sigma}\,.
\end{eqnarray}
The other $U$ triple $\Pi$ projections of $R_{\mu\nu\rho\sigma}$ follow from the symmetry properties of the Riemann tensor. Equation \eqref{eq:Rrhosigma} finally also tells us that
\begin{eqnarray}
    2U^\mu\Pi^\rho_{\langle\alpha}\Pi^\sigma_{\beta\rangle}R_{r\rho\mu\sigma} & = & -S\Pi^\rho_\alpha\Pi^\sigma_\beta R_{r\rho r\sigma}\,,\label{eq:STFUPi}\\
    U^\mu U^\rho\Pi^{\nu\sigma}R_{\mu\nu\rho\sigma} & = & -SU^\rho U^\sigma R_{r\rho r\sigma}\,.
\end{eqnarray}

We can summarise our findings as follows
\begin{eqnarray}
    \Pi^{\mu\nu}R_{r\mu r\nu} & = & 0\,,\label{eq:tracelessrr}\\
    \Pi^\mu_\alpha \Pi^\nu_\beta\Pi^\rho_\gamma\Pi^\sigma_\delta R_{\mu\nu\rho\sigma} & = & -\left(\Pi_{\alpha\gamma}\Pi_{\beta\delta}-\Pi_{\beta\gamma}\Pi_{\alpha\delta}\right)U^\rho U^\sigma R_{r\rho r\sigma}\,,\\
    U^\mu \Pi^\nu_\alpha\Pi^\rho_\beta\Pi^\sigma_\gamma R_{\mu\nu\rho\sigma} & = & 2\Pi_{\alpha[\beta}\Pi_{\gamma]}^\sigma\left(U^\mu U^\nu R_{r\mu\nu\sigma}+SU^\rho R_{r\rho r\sigma}\right)\,,\\
    U^\mu U^\rho\Pi^{\nu\sigma}R_{\mu\nu\rho\sigma} & = & -SU^\rho U^\sigma R_{r\rho r\sigma}\,,\\
    \Pi^\rho_\gamma\Pi^\mu_\alpha\Pi^\nu_\beta R_{r\rho\mu\nu} & = & -2\Pi_{\gamma[\alpha}\Pi^\rho_{\beta]}U^\sigma R_{r\rho r\sigma}\,,\\
    U^\mu\Pi^\rho_{\alpha}\Pi^\sigma_{\beta}R_{r\rho\mu\sigma} & = & -\frac{1}{2}S\Pi^\rho_\alpha\Pi^\sigma_\beta R_{r\rho r\sigma}+\frac{1}{2}\Pi_{\alpha\beta}U^\rho U^\sigma R_{r\rho r\sigma}+\frac{1}{2}U^\rho\Pi^\mu_\alpha\Pi^\nu_\beta R_{r\rho\mu\nu}\,,\nonumber\\
    &&
\end{eqnarray}
where in the last expression we used \eqref{eq:traceUPi}, \eqref{eq:ASUPi} and \eqref{eq:STFUPi}.
This means that the objects left to compute are $R_{r\rho r\sigma}$ (5 components due to \eqref{eq:tracelessrr}), $U^\rho R_{r\rho\mu\nu}$ (3 components) and $U^\mu U^\rho\Pi^\nu_{\langle\alpha}\Pi^\sigma_{\beta\rangle}R_{\mu\nu\rho\sigma}$ (2 components).

In order to compute the 10 undetermined components listed above we adopt the same strategy as developed in section 4 of \cite{Hartong:2026rbr}.
As explained there it will prove convenient to  define the following objects
\begin{eqnarray}
    \mathcal{A}_\mu & = & \mathcal{L}_V U_\mu\,,\\
    \mathcal{Z}_\mu & = & \Pi_{\mu\nu}\partial_r U^\nu-\mathcal{L}_V U_\mu\,,\\
    \mathcal{G}_{\mu\nu} & = & \Pi^\rho_\mu\Pi^\sigma_\nu\left(\partial_r\Pi_{\rho\sigma}-2r^{-1}\Pi_{\rho\sigma}\right)\,,\\
    \mathcal{K}_{\mu\nu} & = & -\frac{1}{2}\mathcal{L}_U \Pi_{\mu\nu}\,,\\
    \mathcal{K} & = & \Pi^{\mu\nu}\mathcal{K}_{\mu\nu}\,,\\
    \mathcal{K}^T_{\mu\nu} & = & \mathcal{K}_{\mu\nu}-\frac{1}{2}\mathcal{K}\Pi_{\mu\nu}\,,\\
    \bar{\mathcal{K}} & = & \mathcal{K}-r^{-1}S\,,\\
    \mathcal{F}_{\mu\nu} & = & \Pi^\rho_\mu\Pi^\sigma_\nu\left(\partial_\rho V_\sigma-\partial_\sigma V_\rho\right)\,.\label{eq:defcalF}
\end{eqnarray}
Once the equations are written in terms of these objects it is easier to expand the result in $1/r$. It will be useful to express the Riemann tensor components using the following connection
\begin{equation}\label{eq:concstr}
    \hat C^\rho_{\mu\nu}=-U^\rho\partial_{(\mu}V_{\nu)}+\frac{1}{2}\Pi^{\rho\sigma}\left(\partial_\mu\Pi_{\nu\sigma}+\partial_\nu\Pi_{\mu\sigma}-\partial_\sigma\Pi_{\mu\nu}\right)-U^\rho\mathcal{A}_{(\mu}V_{\nu)}\,.
\end{equation}
At leading order in the radial expansion this becomes equal to \eqref{eq:conn}.
The associated covariant derivarive will be denoted by $\hat D_\mu$.

A bulk tensor is spatial if it is orthogonal to both $V_\mu$ and $U^\nu$. For spatial tensors we will raise and lower indices using $\Pi_{\mu\nu}$ and $\Pi^{\mu\nu}$. Finally, we will say that a bulk spatial tensor is STF if it is symmetric and traceless with respect to $\Pi^{\mu\nu}$.

We start with $R_{r\rho r\sigma}$. The various components can be written as
\begin{eqnarray}
    U^\rho U^\sigma R_{r\rho r\sigma} & = & \frac{1}{2}\partial_r^2 S-\frac{1}{2}U^\rho U^\sigma \partial_r^2\Pi_{\rho\sigma}+\frac{1}{4}\Pi_{\rho\sigma}\left(\partial_r U^\rho+\Pi^{\rho\kappa}\mathcal{A}_\kappa\right)\left(\partial_r U^\sigma+\Pi^{\sigma\lambda}\mathcal{A}_\lambda\right)\nonumber\\
    &&+\frac{3}{2}\partial_r S\partial_r\beta+S\partial_r^2\beta+S\left(\partial_r\beta\right)^2+\mathcal{L}_U\partial_r\beta\,,\\
    U^\rho\Pi^\sigma_\alpha R_{r\rho r\sigma} & = & \Pi^\rho_\alpha\partial_\rho\partial_r\beta+\frac{1}{4}\mathcal{G}_{\alpha\rho}\mathcal{Z}^\rho-\frac{1}{4}\mathcal{F}_{\alpha\rho}\mathcal{Z}^\rho+\frac{1}{2}\Pi^\rho_\alpha\left(\partial_r+r^{-1}\right)\mathcal{Z}_\rho\,,\\
\Pi^\rho_{\langle\alpha}\Pi^\sigma_{\beta\rangle} R_{r\rho r\sigma} & = & -\frac{1}{2}\Pi^\rho_\alpha\Pi^\sigma_\beta\partial_r\mathcal{G}_{\rho\sigma}+\frac{1}{4}\Pi_{\alpha\beta}\mathcal{G}^2+\frac{1}{2}\mathcal{G}_{\alpha\beta}\partial_r\beta-\frac{1}{4}\mathcal{F}_\alpha{}^\kappa\mathcal{G}_{\kappa\beta}-\frac{1}{4}\mathcal{F}_\beta{}^\kappa\mathcal{G}_{\kappa\alpha}\,.\nonumber\\
&&
\end{eqnarray}

\begin{eqnarray}
    U^\nu U^\rho\Pi^\sigma_\alpha R_{r\nu\rho\sigma} & = & \frac{1}{2}\mathcal{L}_U\mathcal{Z}_\alpha-\frac{1}{2}\mathcal{K}^T_{\alpha\rho}\left(\partial_r U^\rho+\mathcal{A}^\rho\right)-\frac{1}{4}\mathcal{K}\mathcal{Z}_\alpha-\frac{1}{2}\bar{\mathcal{K}}\mathcal{A}_\alpha-S\Pi^\rho_\alpha\partial_r\mathcal{A}_\rho\nonumber\\
    &&-\frac{1}{2}\Pi^\rho_\alpha\partial_\rho\left(\partial_r S-r^{-1}S\right)+\frac{3}{4}S\mathcal{Z}_\rho\mathcal{F}^\rho{}_\alpha+S\mathcal{A}_\rho\mathcal{F}^\rho{}_\alpha+\frac{1}{2}S\mathcal{A}_\rho\mathcal{G}^\rho{}_\alpha\nonumber\\
    &&+\frac{1}{4}\mathcal{F}^\rho{}_\alpha\partial_\rho S+\frac{1}{4}\mathcal{G}^\rho{}_\alpha\partial_\rho S-\Pi^\rho_\alpha\partial_\rho S\partial_r\beta\nonumber\\
    &&-\mathcal{A}_\alpha\left(\frac{1}{2}\partial_r S-\frac{1}{2}r^{-1}S+S\partial_r\beta\right)\,,\\
    U^\nu \Pi_\alpha^\rho\Pi^\sigma_\beta R_{r\nu\rho\sigma} & = & -\frac{1}{2}S\Pi^\rho_\alpha\Pi^\sigma_\beta\partial_r\mathcal{F}_{\rho\sigma}-\frac{1}{2}\left(\partial_r S-r^{-1}S+S\partial_r\beta+\bar{\mathcal{K}}\right)\mathcal{F}_{\alpha\beta}\nonumber\\
    &&+\frac{1}{2}\Pi^\rho_\alpha\Pi^\sigma_\beta\left(\partial_\rho\mathcal{Z}_\sigma-\partial_\sigma\mathcal{Z}_\rho\right)-\frac{1}{2}\mathcal{K}^T_{\alpha\rho}\mathcal{G}^\rho{}_\beta+\frac{1}{2}\mathcal{K}^T_{\beta\rho}\mathcal{G}^\rho{}_\alpha\,.
\end{eqnarray}

\begin{eqnarray}
    U^\mu U^\rho\Pi^\nu_{\langle\alpha}\Pi^\sigma_{\beta\rangle}R_{\mu\nu\rho\sigma} & = & \Pi^\rho_{\langle\alpha}\Pi^\sigma_{\beta\rangle}\left[\mathcal{L}_U\mathcal{K}^T_{\rho\sigma}-\frac{1}{2}\partial_rS\mathcal{K}^T_{\rho\sigma}-S\partial_r\beta\mathcal{K}^T_{\rho\sigma}+S\mathcal{K}^T_{\rho\kappa}\mathcal{F}^\kappa{}_\sigma\right.\nonumber\\
    &&\left.+S\left(\hat D_\rho \mathcal{A}_\sigma+\mathcal{A}_\rho\mathcal{A}_\sigma\right)+\frac{1}{2}\hat D_\rho\partial_\sigma S+\mathcal{A}_\rho\Pi^\kappa_\sigma\partial_\kappa S+\frac{1}{4}S\mathcal{Z}_\rho\mathcal{Z}_\sigma\nonumber\right.\\
    &&\left.-\frac{1}{2}\mathcal{Z}_\rho\Pi^\kappa_\sigma\partial_\kappa S+\frac{1}{4}S\partial_r S\mathcal{G}_{\rho\sigma}+\frac{1}{2}S^2\partial_r\beta\mathcal{G}_{\rho\sigma}-\frac{1}{4}\mathcal{G}_{\rho\sigma}\mathcal{L}_U S\right]\,.\nonumber\\
    &&
\end{eqnarray}

A non-vanishing $D_{\alpha\beta}$ violates peeling. It is possible for $D_{\alpha\beta}$ to be nonzero even when there are no logs by choosing a boundary topology $\mathbb{R}\times\Sigma$ where $\Sigma$ is not the 2-sphere.

Using the above expression for the Riemann tensor and the fact that for solutions to the vacuum Einstein equations the Riemann tensor is equal to the Weyl tensor we obtain the following leading order coefficients in the near-boundary expansion of the Weyl tensor
\begin{eqnarray}
    U^\rho U^\sigma C_{r\rho r\sigma} & = & r^{-3}\left(-v^\rho v^\sigma \overset{(1)}{g}_{\rho\sigma}-\frac{1}{16}K\left(F^2-C^2\right)\right.\nonumber\\
    &&\left.-\frac{1}{8}\left(\mathcal{L}_v-K\right)F^2-\frac{1}{4}C\cdot N\right)+\cdots\,,\\
    U^\rho\Pi^\sigma_\alpha C_{r\rho r\sigma} & = & r^{-3}\left(-\frac{3}{2}v^\rho g^\sigma_\alpha \overset{(1)}{g}_{\rho\sigma}+a_\rho D^\rho{}_\alpha+\frac{3}{32}a_\alpha\left(F^2-C^2\right)\right.\nonumber\\
    &&\left.-\frac{3}{32}h^\rho_\alpha\left(\partial_\rho+2a_\rho\right)\left(F^2-C^2\right)\right)+\cdots\,,\\
    \Pi^\rho_{\langle\alpha}\Pi^\sigma_{\beta\rangle} C_{r\rho r\sigma} & = & -r^{-2}D_{\alpha\beta}+\cdots\,,\\
    U^\nu \Pi_\alpha^\rho\Pi^\sigma_\beta C_{r\nu\rho\sigma} & = & r^{-1}h^\rho_\alpha h^\sigma_\beta R(K)_{\rho\sigma}+\cdots\,,\\
    U^\nu U^\rho\Pi^\sigma_\alpha C_{r\nu\rho\sigma} & = & r^{-1}v^\mu h^\nu_\alpha R(K)_{\mu\nu}+\cdots\,,\\
    U^\mu U^\rho\Pi^\nu_{\langle\alpha}\Pi^\sigma_{\beta\rangle}C_{\mu\nu\rho\sigma} & = & -r v^\mu h^\nu_{\langle\alpha}e^a_{\beta\rangle}R(K)_{\mu\nu}{}^a+\cdots\,.
\end{eqnarray}

\section{From $K$-curvatures to vacuum shear}\label{app:vacshear}

This appendix provides calculation details for results discussed in section \ref{sec:vacshear}.

\subsection{Rewriting $v^\mu e^\nu_b R(K)_{\mu\nu}{}^a$}\label{app:rewritingcurv1}

We will now provide the details for the rewriting of $v^\mu e^\nu_b R(K)_{\mu\nu}{}^a$ from \eqref{eq:veRK} to 
\begin{equation}\label{eq:finalexpcurv1app}
    v^\mu e^\nu_b R(K)_{\mu\nu}{}^a = -\frac{1}{2}e^\mu_a e^\nu_b\mathcal{L}_v \left(N_{\mu\nu}-N^{\text{vac}}_{\mu\nu}\right)\,.
\end{equation}
We will start by rewriting \eqref{eq:veRK} which we repeat here for convenience
\begin{equation}\label{eq:startingpointKcurv1}
    v^\mu e^\nu_b R(K)_{\mu\nu}{}^a = -\frac{1}{2}e^\mu_a e^\nu_b\left[\mathcal{L}_v N_{\mu\nu}-2h^\rho_{\langle\mu}h^\sigma_{\nu\rangle}\left(\mathcal{D}_\rho+3a_\rho\right)G_\sigma\right]\,,
\end{equation}
where 
\begin{equation}\label{eq:defGapp}
    G_\nu=-v^\mu \left(\partial_\mu\tilde b_\nu-\partial_\nu\tilde b_\mu\right)=-\left(\mathcal{L}_v\tilde b_\nu+\frac{1}{2}\partial_\nu K\right)\,.
\end{equation}
In here we used $v^\alpha\tilde b_\alpha=-\frac{1}{2}K$.

In order to achieve our objective we will need to write $2h^\rho_{\langle\mu}h^\sigma_{\nu\rangle}\left(\mathcal{D}_\rho+3a_\rho\right)G_\sigma$ as the Lie derivative along $v^\mu$ of something. The first step will involve commuting $\mathcal{L}_v$ in \eqref{eq:defGapp} with $\mathcal{D}_\rho$ in \eqref{eq:startingpointKcurv1}. To this end we will use the following identity. For any spatial STF $X^{\rho\sigma}$ tensor and for any $Y_\sigma$ we have
\begin{equation}\label{eq:exchangeLiecovD}
    X^{\rho\sigma}\mathcal{L}_v\mathcal{D}_\rho Y_\sigma=X^{\rho\sigma}\left(\mathcal{D}_\rho\mathcal{L}_v Y_\sigma+v^\kappa Y_\kappa\left(\mathcal{D}_\rho a_\sigma+a_\rho a_\sigma\right)+Y_\sigma\partial_\rho K\right)\,.
\end{equation}
This results from commuting the Lie derivative along $v^\mu$ with the covariant derivative. This operation leads to curvature tensors contracted with $v^\mu$ and we have used the results of appendix \ref{app:bdryRiem}.

We can write
\begin{align}
    & 2h^\rho_{\langle\mu}h^\sigma_{\nu\rangle}\left(\mathcal{D}_\rho+3a_\rho\right)\left[v^\alpha\left(\partial_\alpha\tilde b_\sigma-\partial_\sigma\tilde b_\alpha\right)\right]\nonumber\\
    =&2h^\rho_{\langle\mu}h^\sigma_{\nu\rangle}\left(\mathcal{D}_\rho+3a_\rho\right)\left[\mathcal{L}_v\tilde b_\sigma+\frac{1}{2}\partial_\sigma K\right]\nonumber\\
    =&2h^\rho_{\langle\mu}h^\sigma_{\nu\rangle}\left[\mathcal{L}_v\mathcal{D}_\rho\tilde b_\sigma+\frac{1}{2}K\left(\mathcal{D}_\rho a_\sigma+a_\rho a_\sigma\right)+3a_\rho\mathcal{L}_v\tilde b_\sigma+\frac{1}{2}\mathcal{D}_\rho\partial_\sigma K+\frac{1}{2}a_\rho\partial_\sigma K\right]\label{eq:firstfewsteps}
\end{align}
where in the first equality we used that $v^\alpha\tilde b_\alpha=-\frac{1}{2}K$ and in the second equality we used the identity \eqref{eq:exchangeLiecovD}.
Next we use that for any tensor $X_{\mu\nu}$ we have
\begin{equation}\label{eq:interchangeLieSTFP}
    \mathcal{L}_v\left(2h^\rho_{\langle\mu}h^\sigma_{\nu\rangle}X_{\rho\sigma}\right)=2h^\rho_{\langle\mu}h^\sigma_{\nu\rangle}\left(\mathcal{L}_v X_{\rho\sigma}+a_\rho \left(v^\kappa h^\lambda_\sigma+v^\lambda h^\kappa_\sigma\right)X_{\kappa\lambda}\right)\,.
\end{equation}
Applying this to $X_{\rho\sigma}=\mathcal{D}_\rho\tilde b_\sigma$ we obtain
\begin{equation}
    \mathcal{L}_v\left(2h^\rho_{\langle\mu}h^\sigma_{\nu\rangle}\mathcal{D}_\rho\tilde b_\sigma\right)
    =2h^\rho_{\langle\mu}h^\sigma_{\nu\rangle}\left(\mathcal{L}_v \mathcal{D}_\rho\tilde b_\sigma+a_\rho\mathcal{L}_v\tilde b_\sigma+Ka_\rho a_\sigma-\frac{1}{2}a_\rho\partial_\sigma K\right)\,.
\end{equation}
With the help of this result we can rewrite \eqref{eq:firstfewsteps} as
\begin{align}
    & 2h^\rho_{\langle\mu}h^\sigma_{\nu\rangle}\left(\mathcal{D}_\rho+3a_\rho\right)\left[v^\alpha\left(\partial_\alpha\tilde b_\sigma-\partial_\sigma\tilde b_\alpha\right)\right]\nonumber\\
    =&\mathcal{L}_v\left(2h^\rho_{\langle\mu}h^\sigma_{\nu\rangle}\mathcal{D}_\rho\tilde b_\sigma\right)+2h^\rho_{\langle\mu}h^\sigma_{\nu\rangle}\left[\frac{1}{2}K\left(\mathcal{D}_\rho a_\sigma+a_\rho a_\sigma\right)+2a_\rho\mathcal{L}_v\tilde b_\sigma\right.\nonumber\\
    &\left.+\frac{1}{2}\mathcal{D}_\rho\partial_\sigma K+a_\rho\partial_\sigma K-Ka_\rho a_\sigma\right]\nonumber\\
    =&\mathcal{L}_v\left(2h^\rho_{\langle\mu}h^\sigma_{\nu\rangle}\mathcal{D}_\rho\tilde b_\sigma\right)+2h^\rho_{\langle\mu}h^\sigma_{\nu\rangle}\left[\frac{1}{2}K\left(\mathcal{D}_\rho a_\sigma+a_\rho a_\sigma\right)+2a_\rho\mathcal{L}_v a_\sigma\right.\nonumber\\
    &\left.+\frac{1}{2}\mathcal{D}_\rho\partial_\sigma K+a_\rho\partial_\sigma K\right]\nonumber\\
    =&\mathcal{L}_v\left(2h^\rho_{\langle\mu}h^\sigma_{\nu\rangle}\left(\mathcal{D}_\rho a_\sigma+a_\rho a_\sigma\right)\right)+2h^\rho_{\langle\mu}h^\sigma_{\nu\rangle}\left[\frac{1}{2}K\left(\mathcal{D}_\rho a_\sigma+a_\rho a_\sigma\right)\right.\nonumber\\
    &\left.+\frac{1}{2}\mathcal{D}_\rho\partial_\sigma K+a_\rho\partial_\sigma K\right]\,.\label{eq:idcurv1}
\end{align}
Armed with \eqref{eq:idcurv1} we can rewrite the terms in square brackets in \eqref{eq:startingpointKcurv1} as follows
\begin{align}
     &\mathcal{L}_v N_{\mu\nu}-2h^\rho_{\langle\mu}h^\sigma_{\nu\rangle}\left(\mathcal{D}_\rho+3a_\rho\right)G_\sigma=h^\rho_{\langle\mu}h^\sigma_{\nu\rangle}\left[K\left(\mathcal{D}_\rho a_\sigma+a_\rho a_\sigma\right)+\left(\mathcal{D}_\rho+2a_\rho\right)\partial_\sigma K\right]\nonumber\\
     &+\mathcal{L}_v \left[N_{\mu\nu}+2h^\rho_{\langle\mu}h^\sigma_{\nu\rangle}\left(\mathcal{D}_\rho a_\sigma+a_\rho a_\sigma\right)\right]\,.
\end{align}

As we said earlier the goal is to write \eqref{eq:idcurv1} as the $\mathcal{L}_v$ derivative of something. To make progress we use that $K=-2\mathcal{L}_v\varphi$ (see equation \eqref{eq:Kinvarphi}). 
We now derive an identity for the terms in square brackets on the last line in \eqref{eq:idcurv1}. This identity is
\begin{equation}\label{eq:id2curv1}
    h^\rho_{\langle\mu}h^\sigma_{\nu\rangle}\left(\mathcal{D}_\rho\partial_\sigma K+2a_\rho\partial_\sigma K+K\left(\mathcal{D}_\rho a_\sigma+a_\rho a_\sigma\right)\right)=-2\mathcal{L}_v\left(h^\rho_{\langle\mu}h^\sigma_{\nu\rangle}\left(\mathcal{D}_\rho\partial_\sigma\varphi+\partial_\rho\varphi\partial_\sigma\varphi\right)\right)\,,
\end{equation}
which we will prove next.

This goes along similar lines as the previous result. First of all we use $K=-2\mathcal{L}_v\varphi$ so that we have
\begin{align}
    & -\frac{1}{2}X^{\rho\sigma}\mathcal{D}_\rho\partial_\sigma K=X^{\rho\sigma}\mathcal{D}_\rho\mathcal{L}_v\partial_\sigma\varphi\nonumber\\    =&X^{\rho\sigma}\left(\mathcal{L}_v\mathcal{D}_\rho\partial_\sigma\varphi+\frac{1}{2}K\left(\mathcal{D}_\rho a_\sigma+a_\rho a_\sigma\right)-\partial_\sigma\varphi\partial_\rho K\right)\nonumber\\
=&X^{\rho\sigma}\left(\mathcal{L}_v\left(\mathcal{D}_\rho\partial_\sigma\varphi+\partial_\rho\varphi\partial_\sigma\varphi\right)+\frac{1}{2}K\left(\mathcal{D}_\rho a_\sigma+a_\rho a_\sigma\right)\right)\,,
\end{align}
for any spatial and STF $X^{\rho\sigma}$, where we used \eqref{eq:exchangeLiecovD} with $Y_\sigma=\partial_\sigma\varphi$.
Removing $X^{\rho\sigma}$ this means that we have
\begin{align}
    & -\frac{1}{2}h^\rho_{\langle\mu}h^\sigma_{\nu\rangle}\mathcal{D}_\rho\partial_\sigma K\nonumber\\    =&h^\rho_{\langle\mu}h^\sigma_{\nu\rangle}\left(\mathcal{L}_v\left(\mathcal{D}_\rho\partial_\sigma\varphi+\partial_\rho\varphi\partial_\sigma\varphi\right)+\frac{1}{2}K\left(\mathcal{D}_\rho a_\sigma+a_\rho a_\sigma\right)\right)\,.
\end{align}
Now we use \eqref{eq:interchangeLieSTFP} with $X_{\rho\sigma}=\mathcal{D}_\rho\partial_\sigma\varphi+\partial_\rho\varphi\partial_\sigma\varphi$ so that $\tilde X_\sigma=-h_\sigma^\rho\partial_\rho K$, which leads to
\begin{align}
    & -\frac{1}{2}h^\rho_{\langle\mu}h^\sigma_{\nu\rangle}\mathcal{D}_\rho\partial_\sigma K\nonumber\\    =&\mathcal{L}_v\left(h^\rho_{\langle\mu}h^\sigma_{\nu\rangle}\left(\mathcal{D}_\rho\partial_\sigma\varphi+\partial_\rho\varphi\partial_\sigma\varphi\right)\right)+h^\rho_{\langle\mu}h^\sigma_{\nu\rangle}\left(a_\rho\partial_\sigma K+\frac{1}{2}K\left(\mathcal{D}_\rho a_\sigma+a_\rho a_\sigma\right)\right)\,.
\end{align}
Rearranging gives \eqref{eq:id2curv1}.

Using \eqref{eq:id2curv1} we finally obtain
\begin{equation}
      v^\mu e^\nu_b R(K)_{\mu\nu}{}^a = -\frac{1}{2}e^\mu_a e^\nu_b \mathcal{L}_v \left[N_{\mu\nu}+2h^\rho_{\langle\mu}h^\sigma_{\nu\rangle}\left(\mathcal{D}_\rho a_\sigma+a_\rho a_\sigma-\mathcal{D}_\rho\partial_\sigma\varphi-\partial_\rho\varphi\partial_\sigma\varphi\right)\right]\,,
\end{equation}
which upon defining the vacuum news as in \eqref{eq:vacnews} leads to the desired result \eqref{eq:finalexpcurv1app}.

\subsection{Rewriting $v^\mu R(K)_{\mu\nu}$}\label{app:rewritingcurv2}

Just as in the previous subsection we will here provide the details for the rewriting of the curvature $v^\mu  R(K)_{\mu\nu}$ from equation \eqref{eq:PTWeylcov} to 
\begin{equation}\label{eq:cuvr2-v2app}
     v^\mu  R(K)_{\mu\nu}
    =\frac{1}{2}h^{\rho\sigma}\mathcal{D}_\rho\left(N_{\sigma\nu}-N^{\text{vac}}_{\sigma\nu}\right)\,.
\end{equation}
The starting point is thus
\begin{eqnarray}
     v^\mu  R(K)_{\mu\nu} & = & \frac{1}{2}h^{\rho\sigma}\mathcal{D}_\rho N_{\sigma\nu}+\frac{1}{2}h^{\rho}_\nu\left(\partial_\rho+2a_\rho\right)\left(\overset{(0)}{S}-a^2\right)\label{eq:PTWeylcovapp}\\
    &&-\frac{1}{2}G_\sigma F^{\sigma}{}_{\nu}+\frac{1}{2}h_{\rho\nu}\left(\mathcal{L}_v-\frac{3}{2}K\right)\mathcal{D}_\sigma F^{\sigma\rho}\,.\nonumber
\end{eqnarray}

The vacuum news \eqref{eq:vacnews} can be written as
\begin{equation}
    N^{\text{vac}}_{\mu\nu}=2h^\rho_{\langle\mu}h^\sigma_{\nu\rangle}M_{\rho\sigma}\,,
\end{equation}
where
\begin{equation}
    M_{\rho\sigma}=\mathcal{D}_\rho\partial_\sigma\varphi+\partial_\rho\varphi\partial_\sigma\varphi-\mathcal{D}_{(\rho} a_{\sigma)}-a_\rho a_\sigma\,.
\end{equation}
Referring to the definition of the vacuum shear in \eqref{eq:vacnews} we have dropped the $\mathcal{T}_{\mu\nu}$ term as it is assumed to obey \eqref{eq:divX} and so it plays no role in deriving the identity.
We first use that 
\begin{equation}\label{eq:divNvac2}
    h^{\alpha\mu}\mathcal{D}_\alpha N^{\text{vac}}_{\mu\nu}=2h^{\alpha\rho}h^\sigma_\nu\mathcal{D}_\alpha M_{\rho\sigma}-h^\alpha_\nu\partial_\alpha\left(h^{\rho\sigma}M_{\rho\sigma}\right)+v^\rho F^\sigma{}_\nu M_{\rho\sigma}\,.
\end{equation}
Using various results from section \ref{subsec:bdrygeom}, in particular equations \eqref{eq:covDh}--\eqref{eq:covDinvh}, and equations 
\eqref{eq:Riem}, \eqref{eq:Riemtau}, \eqref{eq:2DRiem}, 
as well as \eqref{eq:StoRicscalar} and \eqref{eq:Ricscalar}, we can derive
\begin{eqnarray}
    h^{\alpha\rho}h^\sigma_\nu\mathcal{D}_\alpha M_{\rho\sigma} & = & h^\sigma_\nu\mathcal{D}_\sigma\left(h^{\alpha\rho}\mathcal{D}_\alpha\partial_\rho\varphi+\frac{1}{2}h^{\alpha\rho}\partial_\alpha\varphi\partial_\rho\varphi\right)-\frac{1}{2}F^\rho{}_\nu\left(\partial_\rho K+K a_\rho\right)\nonumber\\
    &&-\frac{1}{4}Kh^{\alpha\rho}h^\sigma_\nu\mathcal{D}_\alpha F_{\rho\sigma}-h^{\alpha\rho}h^\sigma_\nu\mathcal{D}_\alpha \left(\mathcal{D}_{(\rho} a_{\sigma)}+a_\rho a_\sigma\right)\,,
\end{eqnarray}
Substituting this result into \eqref{eq:divNvac2} gives
\begin{eqnarray}
    h^{\alpha\mu}\mathcal{D}_\alpha N^{\text{vac}}_{\mu\nu} & = & h^\sigma_\nu\mathcal{D}_\sigma\left(h^{\alpha\rho}\mathcal{D}_\alpha\partial_\rho\varphi+e^{-1}\partial_\rho\left(e a^\rho\right)\right)-\frac{1}{2}Kh^\sigma_\nu\mathcal{D}_\rho F^\rho{}_\sigma\label{eq:divvacnewsintermed}\\
    &&+F^\rho{}_\nu\left(v^\sigma M_{\rho\sigma}-\partial_\rho K-\frac{1}{2}K a_\rho\right)-2h^{\alpha\rho}h^\sigma_\nu\mathcal{D}_\alpha \left(\mathcal{D}_{(\rho} a_{\sigma)}+a_\rho a_\sigma\right)\,.\nonumber
\end{eqnarray}
In here we have 
\begin{equation}
v^\sigma M_{\rho\sigma}-\frac{1}{2}\mathcal{L}_v a_\rho  =  G_\rho\,,
\end{equation}
where $G_\rho$ is given in \eqref{eq:G}.

We next compute $h^{\alpha\rho}h^\sigma_\nu\mathcal{D}_\alpha \left(\mathcal{D}_{(\rho} a_{\sigma)}+a_\rho a_\sigma\right)$.
We start with the following rewriting
\begin{equation}\label{eq:curv2idstep1}
    h^{\alpha\rho}h^\sigma_\nu\mathcal{D}_\alpha \left(\mathcal{D}_{(\rho} a_{\sigma)}+a_\rho a_\sigma\right) = h^{\alpha\rho}h^\sigma_\nu\mathcal{D}_\alpha \left(\mathcal{D}_\sigma a_\rho+\frac{1}{2}\mathcal{D}_\rho a_\sigma-\frac{1}{2}\mathcal{D}_\sigma a_\rho+a_\rho a_\sigma\right)\,.
\end{equation}
The first term on the right-hand side we rewrite as 
\begin{equation}
    h^{\alpha\rho}h^\sigma_\nu\mathcal{D}_\alpha \mathcal{D}_\sigma a_\rho=    h^{\alpha\rho}h^\sigma_\nu\left[\mathcal{D}_\alpha\,, \mathcal{D}_\sigma\right] a_\rho+    h^{\alpha\rho}h^\sigma_\nu\mathcal{D}_\sigma \mathcal{D}_\alpha a_\rho\,,
\end{equation}
where we will use the Riemann tensor \eqref{eq:Riem} for the commutator term. For the 2nd and 3rd terms on the right-hand side of \eqref{eq:curv2idstep1} we use
\begin{equation}
    \mathcal{D}_\rho a_\sigma-\mathcal{D}_\sigma a_\rho=\mathcal{L}_v F_{\rho\sigma}-\tau_\rho\mathcal{L}_v a_\sigma+\tau_\sigma\mathcal{L}_v a_\rho\,,
\end{equation}
which follows directly from taking the Lie derivative along $v^\mu$ of \eqref{eq:dtau}. After some rewriting where we use various results from appendix \ref{app:bdryRiem} we obtain
\begin{eqnarray}
    h^{\alpha\rho}h^\sigma_\nu\mathcal{D}_\alpha \left(\mathcal{D}_{(\rho} a_{\sigma)}+a_\rho a_\sigma\right) & = &  \left(\frac{1}{2}\mathcal{Q}+h^{\alpha\rho}\mathcal{D}_\alpha a_\rho\right)a_\nu+h^\sigma_\nu\mathcal{D}_\sigma\left(h^{\alpha\rho}\mathcal{D}_\alpha a_\rho+\frac{1}{2}a^2\right)\nonumber\\
    &&\hspace{-3cm}+\frac{1}{2}h^{\alpha\rho}h^\sigma_\nu\mathcal{D}_\alpha\mathcal{L}_v F_{\rho\sigma}+\frac{3}{4}F^\rho{}_\nu\mathcal{L}_v a_\rho+\frac{1}{4}Ka_\rho F^\rho{}_\nu+a^\rho\mathcal{L}_v F_{\rho\nu}\,,\label{eq:curv2idstep2}
\end{eqnarray}
where $\mathcal{Q}$ follows from \eqref{eq:StoRicscalar}. We note that the $\mathcal{Q}$ in the above expression can be replaced via
\begin{eqnarray}
    \frac{1}{2}\mathcal{Q}+h^{\alpha\rho}\mathcal{D}_\alpha a_\rho & = & \overset{(0)}{S}-a^2\,,\label{eq:calS}
\end{eqnarray}
where $\overset{(0)}{S}$ is given in \eqref{eq:S0}.
Next, we rewrite the first term on the second line of \eqref{eq:curv2idstep2} as
\begin{eqnarray}
    h^{\alpha\rho}h^\sigma_\nu\mathcal{D}_\alpha\mathcal{L}_v F_{\rho\sigma} & = & h_{\nu\lambda}\mathcal{D}_\alpha\mathcal{L}_v F^{\alpha\lambda}-2h_{\nu\lambda}a_\kappa\mathcal{L}_v F^{\kappa\lambda}-2K\mathcal{D}_\rho F^\rho{}_\nu\nonumber\\
    &&+F^\rho{}_\nu\left(-\mathcal{L}_v a_\rho+3Ka_\rho-2\partial_\rho K\right)\,,\label{eq:divLieF}
\end{eqnarray}
where we introduced $F^{\mu\nu}=h^{\mu\rho}h^{\nu\sigma}F_{\rho\sigma}$ and where we used the identity
\begin{equation}\label{eq:comcovLieF}
    \mathcal{D}_\rho\mathcal{L}_v F^{\rho\sigma}=\mathcal{L}_v\mathcal{D}_\rho F^{\rho\sigma}+F^{\rho\sigma}\partial_\rho K\,.
\end{equation}

Substituting \eqref{eq:curv2idstep2}
into \eqref{eq:divvacnewsintermed} and using \eqref{eq:divLieF} we finally obtain the result we were after, namely
\begin{eqnarray}
    h^{\alpha\mu}\mathcal{D}_\alpha N^{\text{vac}}_{\mu\nu} & = & -h^\sigma_\nu\left(\partial_\sigma+2a_\sigma\right)\left(\overset{(0)}{S}-a^2\right)+F^\rho{}_\nu G_\rho\nonumber\\
    &&-h_{\nu\lambda}\left(\mathcal{L}_v-\frac{3}{2}K\right)\mathcal{D}_\alpha F^{\alpha\lambda}\,.\label{eq:divNvac}
\end{eqnarray}
Using this we see that the expression for the curvature $v^\mu  R(K)_{\mu\nu}$ given in \eqref{eq:PTWeylcovapp} can indeed be written as \eqref{eq:cuvr2-v2app}.

\subsection{From vacuum news to vacuum shear}\label{app:vacNtovacC}

We will construct the most general shear tensor $C^{\text{vac}}_{\mu\nu}$
such that 
\begin{equation}\label{eq:vacNapp}
    N^{\text{vac}}_{\mu\nu}=-\left(\mathcal{L}_v+\frac{1}{2}K\right)C^{\text{vac}}_{\mu\nu}\,,
\end{equation}
holds with the vacuum news given by \eqref{eq:vacnews}. To find the most general such $C_{\mu\nu}^{\text{vac}}$ we make the following redefinition
\begin{equation}\label{eq:CtoZ}
    C^{\text{vac}}_{\mu\nu}=2h^\rho_{\langle\mu}h^\sigma_{\nu\rangle}Z_{\rho\sigma}+FN^{\text{vac}}_{\mu\nu}+Y_{\mu\nu}\,,
\end{equation}
where we take $Z_{\rho\sigma}$ to be symmetric and where $F$ is any function that obeys
\begin{equation}\label{eq:F}
    \left(\mathcal{L}_v+\frac{1}{2}K\right)F=-1\,,
\end{equation}
and $Y_{\mu\nu}$ is any spatial STF tensor that obeys
\begin{equation}
    \left(\mathcal{L}_v+\frac{1}{2}K\right)Y_{\mu\nu}=0\,.
\end{equation}
The $Y_{\mu\nu}$ tensor is a solution to the homogeneous version \eqref{eq:vacNapp} (i.e. with zero on the left-hand side).
Substituting \eqref{eq:CtoZ} into \eqref{eq:vacNapp} we obtain
\begin{align}
    &h^\rho_{\langle\mu}h^\sigma_{\nu\rangle}\left[\left(\mathcal{L}_v+\frac{1}{2}K\right)Z_{\rho\sigma}+a_\rho\left(v^\kappa h^\lambda_\sigma+v^\lambda h^\kappa_\sigma\right)Z_{\kappa\lambda}\right]\\
    =&h^\rho_{\langle\mu}h^\sigma_{\nu\rangle}F\left[\frac{1}{2}\mathcal{D}_\rho\partial_\sigma K+\left(\mathcal{L}_v+\frac{1}{2}K\right)\left(\mathcal{D}_{(\rho}a_{\sigma)}+a_\rho a_\sigma\right)+a_\rho\left(\mathcal{L}_v a_\sigma+Ka_\sigma+\partial_\sigma K\right)\right]\,,\nonumber
\end{align}
where we used \eqref{eq:vacnews} and \eqref{eq:defeqvacNews}.

We can simplify the latter equation a bit by redefining $Z_{\rho\sigma}$ as 
\begin{equation}\label{eq:ZtoZ'}
    Z_{\mu\nu}=F\left(\mathcal{D}_{(\rho}a_{\sigma)}+a_\rho a_\sigma\right)+Z'_{\mu\nu}\,,
\end{equation}
which leads to the following equation for $Z'_{\mu\nu}$
\begin{align}\label{eq:Zprime}
    &h^\rho_{\langle\mu}h^\sigma_{\nu\rangle}\left[\left(\mathcal{L}_v+\frac{1}{2}K\right)Z'_{\rho\sigma}+a_\rho\left(v^\kappa h^\lambda_\sigma+v^\lambda h^\kappa_\sigma\right)Z'_{\kappa\lambda}\right]\nonumber\\
    =&h^\rho_{\langle\mu}h^\sigma_{\nu\rangle}\left[\frac{1}{2}F\mathcal{D}_\rho\partial_\sigma K-\left(\mathcal{D}_{(\rho}a_{\sigma)}+a_\rho a_\sigma\right)\mathcal{L}_v F+a_\rho\partial_\sigma K\right]\,.
\end{align}

Since we already accounted for the most general homogeneous solution to \eqref{eq:vacN} (which is the $Y_{\mu\nu}$ term in \eqref{eq:CtoZ}) we just need to construct a given particular solution for $Z'_{\mu\nu}$.

We will derive an identity that can be used to find a particular solution for $Z'_{\mu\nu}$. To this end we consider
\begin{equation}
    h^\rho_{\langle\mu}h^\sigma_{\nu\rangle}\mathcal{L}_v\mathcal{D}_\rho\partial_\sigma F=h^\rho_{\langle\mu}h^\sigma_{\nu\rangle}\left(v^\alpha\mathcal{R}_{\alpha\rho\sigma}{}^\gamma\partial_\gamma F+\mathcal{D}_\rho\mathcal{L}_v\partial_\sigma F-\partial_\alpha F\mathcal{D}_\rho\mathcal{D}_\sigma v^\alpha\right)\,,
\end{equation}
which follows from commuting $\mathcal{L}_v$ and $\mathcal{D}_\rho$. Here we take $F$ to be any function obeying \eqref{eq:F}. Using equations \eqref{eq:vRiem2nd} and \eqref{eq:Riemtau} we find
\begin{equation}\label{eq:idsolCvac}
    h^\rho_{\langle\mu}h^\sigma_{\nu\rangle}\left(\mathcal{L}_v+\frac{1}{2}K\right)\mathcal{D}_\rho\partial_\sigma F=h^\rho_{\langle\mu}h^\sigma_{\nu\rangle}\left(-\frac{1}{2}F\mathcal{D}_\rho\partial_\sigma K+\left(\mathcal{D}_\rho a_\sigma+a_\rho a_\sigma\right)\mathcal{L}_v F\right)\,.
\end{equation}

Using the identity \eqref{eq:idsolCvac} we find that \eqref{eq:Zprime} is solved by 
\begin{equation}\label{eq:Z'=DDF}
    Z'_{\rho\sigma}=-\mathcal{D}_\rho\partial_\sigma F\,.
\end{equation}
We conclude that the most general solution to \eqref{eq:vacN} is given by
\begin{equation}\label{eq:Cvacapp}
    C^{\text{vac}}_{\mu\nu}=2h^\rho_{\langle\mu}h^\sigma_{\nu\rangle}\left(-\mathcal{D}_\rho\partial_\sigma F+F\left(\mathcal{D}_\rho a_\sigma+a_\rho a_\sigma\right)\right)+FN^{\text{vac}}_{\mu\nu}+Y_{\mu\nu}\,.
\end{equation}

\subsection{Computing $\os{0}{P}^{\text{vac}}_\mu$}\label{app:P0vac}

In the expression for $e^\mu_a e^\nu_b R(K)_{\mu\nu}$ we encounter the object $\os{0}{P}_\mu$. Here we will compute 
$\os{0}{P}_\mu$ for the case where the shear is given by the vacuum shear. We first define
\begin{equation}
    \overset{(0)}{P}{}^{\text{vac}}_\mu = -\frac{1}{2}h^{\rho\sigma}\left(\mathcal{D}_\rho-a_\rho\right)C^{\text{vac}}_{\sigma\mu}+\frac{1}{2}\mathcal{D}_\rho F^\rho{}_\mu\,,
\end{equation}
where we will for the moment set $Y_{\mu\nu}$ in the vacuum shear equal to zero. We will reintroduce it later. 

If we do this then using \eqref{eq:CtoZ} we obtain
\begin{eqnarray}
    h^{\rho\sigma}\left(\mathcal{D}_\rho-a_\rho\right)C^{\text{vac}}_{\sigma\mu} & = & h^{\rho\sigma}\left(\mathcal{D}_\rho-a_\rho\right)\left(2h^\kappa_{\langle\sigma}h^\lambda_{\mu\rangle}Z_{\kappa\lambda}\right)+Fh^{\rho\sigma}\mathcal{D}_\rho N^{\text{vac}}_{\sigma\mu}\nonumber\\
    &&+N^{\text{vac}}{}^\rho{}_\mu\left(\partial_\rho-a_\rho\right)F\,,
\end{eqnarray}
where the middle term on the right-hand side is given by \eqref{eq:divNvac}.
Next we use that 
\begin{eqnarray}
    h^{\rho\sigma}\left(\mathcal{D}_\rho-a_\rho\right)\left(2h^\kappa_{\langle\sigma}h^\lambda_{\mu\rangle}Z_{\kappa\lambda}\right) & = & 2h^{\rho\kappa}h^\lambda_\mu\left(\mathcal{D}_\rho-a_\rho\right)Z_{\kappa\lambda}-h^\rho_\mu\left(\partial_\rho-a_\rho\right)\left(h^{\kappa\lambda}Z_{\kappa\lambda}\right)\nonumber\\
    &&+F^\kappa{}_\mu v^\lambda Z_{\kappa\lambda}\,.
\end{eqnarray}
Both these results follow from \eqref{eq:covDh}--\eqref{eq:covDinvh}. Equations \eqref{eq:ZtoZ'} and \eqref{eq:Z'=DDF} tell us that
\begin{equation}\label{eq:Zmunu}
    Z_{\rho\sigma}=-\mathcal{D}_\rho\partial_\sigma F+F\left(\mathcal{D}_{(\rho}a_{\sigma)}+a_\rho a_\sigma\right)\,.
\end{equation}
Using this as well as the properties of the curvature tensor given in \eqref{eq:Riem}, \eqref{eq:Riemtau} and \eqref{eq:2DRiem} we find after commuting a few covariant derivatives
\begin{align}
    &h^{\rho\kappa}h^\lambda_\mu\left(\mathcal{D}_\rho-a_\rho\right)Z_{\kappa\lambda}\nonumber\\
    =&h^{\rho\kappa}h^\lambda_\mu\left(\mathcal{D}_{[\kappa}a_{\lambda]}+a_\kappa a_\lambda\right)\left(\partial_\rho-a_\rho\right)F+\frac{1}{2}a^2h^\lambda_\mu\partial_\lambda F\nonumber\\
    &-\frac{1}{2}\mathcal{L}_v F\mathcal{D}_\rho F^\rho{}_\mu+\frac{1}{2}F F^\kappa{}_\mu\partial_\kappa K+\frac{1}{4}K F^\kappa{}_\mu\partial_\kappa F-\frac{1}{2}\mathcal{Q}h^\lambda_\mu\partial_\lambda F\nonumber\\
    &-h^\lambda_\mu\partial_\lambda\left(h^{\rho\sigma}\mathcal{D}_\rho\partial_\sigma F-a^\rho\partial_\rho F+\frac{1}{2}a^2 F\right)+Fh^{\rho\kappa}h^\lambda_\mu\mathcal{D}_\rho \left(\mathcal{D}_{(\kappa}a_{\lambda)}+a_\kappa a_\lambda\right)\,,\label{eq:divZ}
\end{align}
where $\mathcal{Q}$ is given by \eqref{eq:calS}. Finally, we use \eqref{eq:curv2idstep2}--\eqref{eq:comcovLieF} to rewrite the last term $Fh^{\rho\kappa}h^\lambda_\mu\mathcal{D}_\rho\left(\mathcal{D}_{(\kappa}a_{\lambda)}+a_\kappa a_\lambda\right)$ in \eqref{eq:divZ}. Combining everything we obtain
\begin{eqnarray}
    \overset{(0)}{P}{}^{\text{vac}}_\mu & = & \left[-\frac{1}{2}N^{\text{vac}}{}^\rho{}_\mu+\frac{1}{2}\left(S^{(0)}-a^2\right)h^\rho_\mu-\frac{1}{2}h^{\kappa\rho}\left(\mathcal{L}_v+\frac{1}{2}K\right)F_{\kappa\mu}\right]\left(\partial_\rho-a_\rho\right)F\nonumber\\
    &&+\frac{1}{2}h^\rho_\mu\left(\partial_\rho+a_\rho\right)\left[h^{\kappa\lambda}\left(\mathcal{D}_\kappa-a_\kappa\right)\left(\partial_\lambda-a_\lambda\right)F-Fh^{\kappa\lambda}\mathcal{D}_\kappa\left(\partial_\lambda\varphi-a_\lambda\right)\right]\,.\nonumber\\
    &&\label{eq:P0vacapp}
\end{eqnarray}

\subsection{Rewriting $e^\mu_a e^\nu_b R(K)_{\mu\nu}$}\label{app:rewritingcurv3}

In writing the third curvature in terms of the vacuum news and shear we had to use the identity \eqref{eq:idcurv3}. Here we will derive this. For brevity we will first write \eqref{eq:P0vacapp} more compactly as follows
\begin{equation}
    \overset{(0)}{P}{}^{\text{vac}}_\mu = X^\rho{}_\mu\left(\partial_\rho-a_\rho\right)F+\frac{1}{2}h^\lambda_\mu\left(\partial_\lambda+a_\lambda\right)Z\,,
\end{equation}
where we defined 
\begin{eqnarray}
    X^\rho{}_\mu & = & h^{\rho\sigma}\left(-\frac{1}{2}N^{\text{vac}}_{\sigma\mu}+\frac{1}{2}\left(\overset{(0)}{S}-a^2\right)h_{\sigma\mu}-\frac{1}{2}\left(\mathcal{L}_v+\frac{1}{2}K\right)F_{\sigma\mu}\right)\,,\label{eq:Xidcurv3}\\
    Z & = & h^{\rho\sigma}\mathcal{D}_\rho\partial_\sigma F-F h^{\rho\sigma}\left(\mathcal{D}_\rho a_\sigma+a_\rho a_\sigma\right)-2a^\rho\left(\partial_\rho-a_\rho\right)F+F\left(\overset{(0)}{S}-a^2\right)\,.\nonumber\\
    &&\label{eq:Zidcurv3}
\end{eqnarray}
We can then derive
\begin{align}\label{eq:spatcurlP0}
    & h^\mu_\alpha h^\nu_\beta\left(\left(\partial_\mu+a_\mu\right)\overset{(0)}{P}{}^{\text{vac}}_\nu-\left(\partial_\nu+a_\nu\right)\overset{(0)}{P}{}^{\text{vac}}_\mu\right)\nonumber\\
    =& h^\mu_\alpha h^\nu_\beta\Big(\frac{1}{2}\mathcal{L}_v\left(Z F_{\mu\nu}\right)+\left(\mathcal{D}_\mu X^\rho{}_\nu-\mathcal{D}_\nu X^\rho{}_\mu\right)\left(\partial_\rho-a_\rho\right)F\\
    &+X^\rho{}_\mu \tilde Z_{\nu\rho}-X^\rho{}_\nu \tilde Z_{\mu\rho}-X^\rho{}_\mu\left(a_\nu\partial_\rho F-a_\rho\partial_\nu F\right)+X^\rho{}_\nu\left(a_\mu\partial_\rho F-a_\rho\partial_\mu F\right)\Big)\,,\nonumber
\end{align}
where $\tilde Z_{\mu\nu}$ is defined as
\begin{equation}
    \tilde Z_{\mu\nu}=h^\kappa_\mu h^\lambda_\nu\left(-\mathcal{D}_\kappa\partial_\lambda F+F\left(\mathcal{D}_\kappa a_\lambda+a_\kappa a_\lambda\right)\right)=h^\kappa_\mu h^\lambda_\nu Z_{\kappa\lambda}+\frac{1}{2}F\mathcal{L}_v F_{\kappa\lambda}\,,
\end{equation}
where $Z_{\kappa\lambda}$ is defined in \eqref{eq:Zmunu}.
It will be useful to split this in an STF part, a trace part, and an antisymmetric part leading to
\begin{equation}
    \tilde Z_{\mu\nu}=Z_{\langle\mu\nu\rangle}+\frac{1}{2}h_{\mu\nu}h\cdot Z+\frac{1}{2}F\mathcal{L}_v F_{\mu\nu}\,,
\end{equation}
where $Z_{\langle\mu\nu\rangle}=h^\kappa_{\langle\mu} h^\lambda_{\nu\rangle} Z_{\kappa\lambda}$ and $h\cdot Z=h^{\rho\sigma}Z_{\rho\sigma}$. 
Next we use the identities \eqref{eq:AtildeA} and 
\eqref{eq:ASSTF}.
 Using these results and \eqref{eq:Xidcurv3} we find for the terms on the second line of \eqref{eq:spatcurlP0},
\begin{align}
    &h^\mu_\alpha h^\nu_\beta\Big(X^\rho{}_\mu \tilde Z_{\nu\rho}-X^\rho{}_\nu \tilde Z_{\mu\rho}-X^\rho{}_\mu\left(a_\nu\partial_\rho F-a_\rho\partial_\nu F\right)+X^\rho{}_\nu\left(a_\mu\partial_\rho F-a_\rho\partial_\mu F\right)\Big)\nonumber\\
    =&h^\mu_\alpha h^\nu_\beta\Big(-\frac{1}{2}h^{\rho\sigma}N^{\text{vac}}_{\rho[\mu}C^{\text{vac}}_{\nu]\sigma}+\frac{1}{2}h\cdot Z\left(\mathcal{L}_v+\frac{1}{2}K\right)F_{\mu\nu}-\frac{1}{2}F\left(\overset{(0)}{S}-a^2\right)\mathcal{L}_v F_{\mu\nu}\nonumber\\
    &+\left(\overset{(0)}{S}-a^2\right)\left(a_\mu\partial_\nu F-a_\nu\partial_\mu F\right)\Big)\,,
\end{align}
where we used \eqref{eq:Cvacapp} (without the $Y_{\mu\nu}$-term) in the first term on the right-hand side. For the second term on the first line of \eqref{eq:spatcurlP0} we can write
\begin{equation}
    h^\mu_\alpha h^\nu_\beta\Big(\mathcal{D}_\mu X^\rho{}_\nu-\mathcal{D}_\nu X^\rho{}_\mu\Big)=\frac{1}{2}\left(\overset{(0)}{S}-a^2\right)v^\rho F_{\alpha\beta}+h^\mu_\alpha h^\nu_\beta h^{\rho\sigma}\Big(\mathcal{D}_\mu X_{\sigma\nu}-\mathcal{D}_\nu X_{\sigma\mu}\Big)\,.
\end{equation}
Combining these results and rearranging we can now write \eqref{eq:spatcurlP0} as
\begin{align}
    & h^\mu_\alpha h^\nu_\beta\left(\left(\partial_\mu+a_\mu\right)\overset{(0)}{P}{}^{\text{vac}}_\nu-\left(\partial_\nu+a_\nu\right)\overset{(0)}{P}{}^{\text{vac}}_\mu+\frac{1}{2}h^{\rho\sigma}N^{\text{vac}}_{\rho[\mu}C^{\text{vac}}_{\nu]\sigma}\right)\nonumber\\
    =& \frac{1}{2}F_{\alpha\beta}\left(\mathcal{L}_v-\frac{1}{2}K\right)Z-a^\rho\left(\partial_\rho-a_\rho\right)F \left(\mathcal{L}_v+\frac{1}{2}K\right)F_{\alpha\beta}-\frac{1}{2}\left(\overset{(0)}{S}-a^2\right)F_{\alpha\beta}\nonumber\\
    &+\left(\overset{(0)}{S}-a^2\right)\left(a_\alpha h_\beta^\rho\partial_\rho F-a_\beta h_\alpha^\rho\partial_\rho F\right)+h^\mu_\alpha h^\nu_\beta h^{\rho\sigma}\Big(\mathcal{D}_\mu X_{\sigma\nu}-\mathcal{D}_\nu X_{\sigma\mu}\Big)\left(\partial_\rho-a_\rho\right)F\,,\label{eq:spatcurlP0intermed}
\end{align}
where we collected terms proportional to $\left(\mathcal{L}_v+\frac{1}{2}K\right)F_{\alpha\beta}$ and used the relation
\begin{equation}
    Z+h\cdot Z-F\left(\overset{(0)}{S}-a^2\right)=-2a^\rho\left(\partial_\rho-a_\rho\right)F\,,
\end{equation}
where we remind the reader that $Z$ is defined in \eqref{eq:Zidcurv3}.

To continue we need to rewrite the last term in \eqref{eq:spatcurlP0intermed} which contains $X_{\mu\nu}=h_{\mu\rho}X^\rho{}_\nu$ where $X^\rho{}_\nu$ is defined in \eqref{eq:Xidcurv3}.
To this end we use the identities \eqref{eq:DASA} and \eqref{eq:DSTFid}.
In the notation of \eqref{eq:DASA} and \eqref{eq:DSTFid} we apply these identities to $A_{\nu\sigma}=\left(\mathcal{L}_v+\frac{1}{2}K\right)F_{\nu\sigma}$ and $S_{\sigma\nu}=N^{\text{vac}}_{\sigma\nu}$. In the case of $A_{\nu\sigma}=\left(\mathcal{L}_v+\frac{1}{2}K\right)F_{\nu\sigma}$ we obtain
\begin{align}
    &\frac{1}{2}h^\mu_\alpha h^\nu_\beta h^\sigma_\gamma\left(\mathcal{D}_\mu\left(\mathcal{L}_v+\frac{1}{2}K\right)F_{\nu\sigma}-\left(\mu\leftrightarrow\nu\right)\right)h^{\gamma\rho}\left(\partial_\rho-a_\rho\right)F\\
    =&\frac{1}{2}h^\mu_\alpha h^\rho_\beta\left[\left(h_{\mu\lambda}\left(\mathcal{L}_v-\frac{3}{2}K\right)\mathcal{D}_\kappa F^{\kappa\lambda}+F^\kappa{}_\mu G_\kappa\right)\left(\partial_\rho-a_\rho\right)F-\left(\mu\leftrightarrow\rho\right)\right]\nonumber\\
    &+a^\rho\left(\mathcal{L}_v+\frac{1}{2}K\right)F_{\rho\beta}h^\sigma_\alpha\left(\partial_\sigma-a_\sigma\right)F-a^\rho\left(\mathcal{L}_v+\frac{1}{2}K\right)F_{\rho\alpha}h^\sigma_\beta\left(\partial_\sigma-a_\sigma\right)F\,,\nonumber
\end{align}
where we used 
\eqref{eq:divLieF} and \eqref{eq:comcovLieF} in order to create the first term on the right hand side. Next applying the identity \eqref{eq:XYA} to the last line of the above equality we obtain
\begin{align}
    & h^\mu_\alpha h^\nu_\beta\left(\left(\partial_\mu+a_\mu\right)\overset{(0)}{P}{}^{\text{vac}}_\nu-\left(\partial_\nu+a_\nu\right)\overset{(0)}{P}{}^{\text{vac}}_\mu+\frac{1}{2}h^{\rho\sigma}N^{\text{vac}}_{\rho[\mu}C^{\text{vac}}_{\nu]\sigma}\right)\nonumber\\
    =& \left(\frac{1}{2}\left(\mathcal{L}_v-\frac{1}{2}K\right)Z-\frac{1}{2}\left(\overset{(0)}{S}-a^2\right)-G^\rho\left(\partial_\rho-a_\rho\right)F\right) F_{\alpha\beta}\,,
\end{align}
where we used \eqref{eq:divNvac} as well as \eqref{eq:XYA} to create the last term involving $G^\rho$. Using the identity
\begin{equation}
    \left(\mathcal{L}_v-\frac{1}{2}K\right)Z=-\left(\overset{(0)}{S}-a^2\right)+2G^\rho\left(\partial_\rho-a_\rho\right)F\,,
\end{equation}
which follows from a somewhat lengthy but straightforward calculation\footnote{Essentially we  need to commute a Lie derivative along $v^\mu$ with a covariant derivative and use that $\mathcal{L}_v\varphi=-\frac{1}{2}K$ as well as the results of appendix \ref{app:bdryRiem}. A useful identity is 
\begin{equation}
    \left(\mathcal{L}_v-K\right)\left(h^{\rho\sigma}\mathcal{D}_\rho\partial_\sigma\varphi\right)=-a^\rho\partial_\rho K-\frac{1}{2}h^{\rho\sigma}\mathcal{D}_\rho\partial_\sigma K-\frac{1}{2}Kh^{\rho\sigma}\left(\mathcal{D}_\rho a_\sigma+a_\rho a_\sigma\right)\,.
\end{equation}}. The final result is thus
\begin{equation}\label{eq:idcurv3-app}
    h^\mu_\alpha h^\nu_\beta\left(\left(\partial_\mu+a_\mu\right)\overset{(0)}{P}{}^{\text{vac}}_\nu+\frac{1}{4}N^{\text{vac}}{}^\rho{}_\mu C^{\text{vac}}_{\nu\rho}-\left(\mu\leftrightarrow\nu\right)\right)=-\left(\overset{(0)}{S}-a^2\right)F_{\alpha\beta}\,,
\end{equation}
which is what we used in equation \eqref{eq:idcurv3}.

\subsection{Solving for $\mathcal{T}_{\mu\nu}$ and $Y_{\mu\nu}$}\label{app:SimplePDEs}

In the construction of the vacuum shear \eqref{eq:Cvacapp} we encountered two integrating tensors $\mathcal{T}_{\mu\nu}$ and $Y_{\mu\nu}$ that each obey two differential equations. 
In this appendix we find the most general solutions to these PDEs.

Consider first the tensor $\mathcal{T}_{\mu\nu}$. For completeness we remind the reader that this object is spatial and STF, and which furthermore obeys the properties 
\begin{equation}
    \mathcal{L}_v \mathcal{T}_{\mu\nu}=0\,,\qquad h^{\rho\sigma}\mathcal{D}_\rho \mathcal{T}_{\sigma\nu}=0\,.
\end{equation}
These are first order PDEs on a 3-dimensional Carroll manifold with  metric $h_{\mu\nu}$ given by
\begin{equation}
    h_{\mu\nu}=e^{2\varphi}\delta_{ab}\partial_\mu X^a\partial_\nu X^b\,.
\end{equation}

Since $v^\mu h_{\mu\nu}=0$ it follows that $\mathcal{L}_v X^a=0$.
Moreover, since $h_{\mu\nu}$ has rank 2 we know that $\partial_\mu X^a$ describes two independent non-vanishing 1-forms. In fact we have
\begin{equation}\label{eq:invmetricpushforward}
    e^{2\varphi}h^{\mu\nu}\partial_\mu X^a\partial_\nu X^b=\delta^{ab}\,.
\end{equation}
Hence, any spatial STF tensor can be written as
\begin{equation}
    \mathcal{T}_{\mu\nu}=\partial_\mu X^a\partial_\nu X^b\tilde{\mathcal{T}}_{ab}\,,
\end{equation}
where $\tilde{\mathcal{T}}_{ab}$ is STF with respect to $\delta_{ab}$. Using that $\mathcal{L}_v X^a=0$ and the fact that $\partial_\mu X^a$ are two independent non-vanishing 1-forms it follows that $\mathcal{L}_v \mathcal{T}_{\mu\nu}=0$ is equivalent to demanding 
\begin{equation}
    \mathcal{L}_v \tilde{\mathcal{T}}_{ab}=v^\rho\partial_\rho \tilde{\mathcal{T}}_{ab}=0\,.
\end{equation}
The most general solution to $\mathcal{L}_v \tilde{\mathcal{T}}_{ab}=0$ is obtained by taking $\tilde{\mathcal{T}}_{ab}$ to be a function of the two scalars $X^c$ only.

It remains to implement $h^{\rho\sigma}\mathcal{D}_\rho \mathcal{T}_{\sigma\nu}=0$. Here it is useful to first observe that our connection $\mathcal{C}^\rho_{\mu\nu}$ is such that
\begin{equation}\label{eq:Xaharmonic}
    h^{\rho\sigma}\mathcal{D}_\rho\partial_\sigma X^a=0\,.
\end{equation}
We will show this below. Using this we see that
\begin{equation}
   0=h^{\rho\sigma}\mathcal{D}_\rho \mathcal{T}_{\sigma\nu}=h^{\rho\sigma}\partial_\sigma X^a\mathcal{D}_\rho\partial_\nu X^b\tilde{\mathcal{T}}_{ab}+h^{\rho\sigma}\partial_\sigma X^a\partial_\nu X^b\partial_\rho X^c\partial_c\tilde{\mathcal{T}}_{ab}\,,
\end{equation}
where $\partial_c$ denotes a derivative with respect to $X^c$.
If we use \eqref{eq:invmetricpushforward} we find
\begin{equation}\label{eq:hDX}
    0=h^{\rho\sigma}\mathcal{D}_\rho \mathcal{T}_{\sigma\nu}=h^{\rho\sigma}\mathcal{D}_\nu\left(\partial_\sigma X^a\partial_\rho X^b\right)\tilde{\mathcal{T}}_{ab}+e^{-2\varphi}\partial_\nu X^b\partial_a\tilde{\mathcal{T}}_{ab}\,.
\end{equation}
Next we observe that
\begin{equation}
    h^{\rho\sigma}\mathcal{D}_\nu\left(\partial_\sigma X^a\partial_\rho X^b\right)\tilde{\mathcal{T}}_{ab}=\mathcal{D}_\nu\left(h^{\rho\sigma}\partial_\sigma X^a\partial_\rho X^b\right)\tilde{\mathcal{T}}_{ab}-\left(\mathcal{D}_\nu h^{\rho\sigma}\right)\partial_\sigma X^a\partial_\rho X^b\tilde{\mathcal{T}}_{ab}=0\,.
\end{equation}
The first term on the right hand side vanishes on account of \eqref{eq:invmetricpushforward} and because $\tilde{\mathcal{T}}_{aa}=0$. The second term vanishes because $\mathcal{L}_v X^a=0$ (where we used \eqref{eq:covDinvh}). Therefore, \eqref{eq:hDX} is satisfied if and only if 
\begin{equation}
    \partial_a\tilde{\mathcal{T}}_{ab}=0\,.
\end{equation}

It remains to show that $h^{\rho\sigma}\mathcal{D}_\rho\partial_\sigma X^a=0$. To this end consider
\begin{eqnarray}
    \mathcal{D}_\rho\partial_\mu X^a & = & \partial_\rho\partial_\mu X^a-\mathcal{C}^\sigma_{\rho\mu}\partial_\sigma X^a\nonumber\\
    & = & \partial_\rho\partial_\mu X^a-\frac{1}{2}h^{\sigma\alpha}\left(\partial_\rho h_{\mu\alpha}+\partial_\mu h_{\rho\alpha}-\partial_\alpha h_{\rho\sigma}\right)\partial_\sigma X^a\nonumber\\
    & = & -2\partial_{\langle\rho}\varphi\partial_{\mu\rangle} X^a\,,\label{eq:DdXa}
\end{eqnarray}
where in the second equality we used the expression for the connection $\mathcal{C}^\sigma_{\mu\rho}$ and the fact that $\mathcal{L}_v X^a=0$ and in the third equality we used that $h_{\mu\nu}=e^{2\varphi}\partial_\mu X^a\partial_\nu X^a$. Since the term on the right hand side of the last line is traceless with respect to $h^{\mu\rho}$ the result follows. 

We conclude that the most general spatial STF tensor $\mathcal{T}_{\mu\nu}$ obeying $\mathcal{L}_v \mathcal{T}_{\mu\nu}=0$ and $h^{\rho\sigma}\mathcal{D}_\rho \mathcal{T}_{\sigma\nu}=0$ is given by
\begin{equation}\label{eq:STFX}
    \mathcal{T}_{\mu\nu}=\partial_\mu X^a\partial_\nu X^b\tilde{\mathcal{T}}_{ab}(X)\,,\qquad\text{where $\partial_a\tilde{\mathcal{T}}_{ab}=0$}\,,
\end{equation}
with $\tilde{\mathcal{T}}_{ab}(X)$ STF with respect to $\delta^{ab}$.

The next case we need to understand is a spatial STF tensor $Y_{\mu\nu}$ that obeys $\left(\mathcal{L}_v+\frac{1}{2}K\right)Y_{\mu\nu}=0$ as well as 
\begin{equation}\label{eq:diffconY}
    h^\mu_\alpha h^\nu_\beta\left(\left(\partial_\mu+a_\mu\right)\left(h^{\rho\sigma}\left(\mathcal{D}_\rho-a_\rho\right)Y_{\sigma\nu}\right)-\frac{1}{2}N^{\text{vac}}{}^\rho{}_\mu Y_{\nu\rho}-\left(\mu\leftrightarrow\nu\right)\right)=0\,.
\end{equation}
Using that $K=-2\mathcal{L}_v\varphi$ we see that $\left(\mathcal{L}_v+\frac{1}{2}K\right)Y_{\mu\nu}=0$ is equivalent to $\mathcal{L}_v\left(e^{-\varphi}Y_{\mu\nu}\right)=0$. Using the same arguments as before it follows that 
\begin{equation}
    Y_{\mu\nu}=e^\varphi \partial_\mu X^a\partial_\nu X^b \tilde Y_{ab}(X)\,,
\end{equation}
where $\tilde Y_{ab}(X)$ is STF with respect to $\delta^{ab}$.

We now need to impose \eqref{eq:diffconY}. First we rewrite it a bit. Using \eqref{eq:DSTFid} and \eqref{eq:ASSTF} we obtain
\begin{equation}\label{eq:diffconY2}
    h^\mu_\alpha h^\nu_\beta\left(\partial_\mu\left(h^{\rho\sigma}\mathcal{D}_\rho Y_{\sigma\nu}\right)-\frac{1}{2}h^{\rho\sigma}\left(N^{\text{vac}}_{\mu\sigma} +2\mathcal{D}_{(\mu}a_{\sigma)}+2a_\mu a_\sigma\right)Y_{\rho\nu}-\left(\mu\leftrightarrow\nu\right)\right)=0\,.
\end{equation}
Defining $Y_{\mu\nu}=e^\varphi\bar Y_{\mu\nu}$ and using again \eqref{eq:DSTFid} and \eqref{eq:ASSTF} we can derive
\begin{align}\label{eq:diffconY3}
   & h^\mu_\alpha h^\nu_\beta\left(\partial_\mu\left(h^{\rho\sigma}\mathcal{D}_\rho Y_{\sigma\nu}\right)-\left(\mu\leftrightarrow\nu\right)\right)=\\
   & h^\mu_\alpha h^\nu_\beta e^\varphi\left(\partial_\mu\left(h^{\rho\sigma}\mathcal{D}_\rho \bar Y_{\sigma\nu}\right)+2\partial_\mu\varphi h^{\rho\sigma}\mathcal{D}_\rho \bar Y_{\sigma\nu}+h^{\rho\sigma}\left(\mathcal{D}_\mu\partial_\sigma\varphi+\partial_\mu\varphi\partial_\sigma\varphi\right)\bar Y_{\rho\nu}-\left(\mu\leftrightarrow\nu\right)\right)\,.\nonumber
\end{align}
Next we use a similar set of steps as in the discussion below equation \eqref{eq:Xaharmonic} which for $\bar Y_{\mu\nu}=\partial_\mu X^a\partial_\nu X^b \tilde Y_{ab}(X)$ leads to
\begin{equation}\label{eq:divbarY}
    h^{\rho\sigma}\mathcal{D}_\rho \bar Y_{\sigma\nu}=e^{-2\varphi}\partial_c\tilde Y_{cb}\partial_\nu X^b\,.
\end{equation}
Substituting \eqref{eq:diffconY3} into \eqref{eq:diffconY2} and using 
\eqref{eq:vacnews} as well as 
\eqref{eq:STFX}, \eqref{eq:STFY} and \eqref{eq:divbarY} we finally find
\begin{equation}
    h^\mu_\alpha h^\nu_\beta \partial_\mu X^a\partial_\nu X^b\left[\partial_a\partial_c\tilde Y_{cb}-\frac{1}{2}\tilde{\mathcal{T}}_{ac}\tilde Y_{bc}-\left(a\leftrightarrow b\right)\right]=0\,.
\end{equation}
This will be zero if and only if\footnote{Using the complex combinations $Z=\frac{1}{\sqrt{2}}\left(X^1+iX^2\right)$ and $\bar Z=\frac{1}{\sqrt{2}}\left(X^1-iX^2\right)$ with corresponding derivatives $\partial$ and $\bar\partial$ we can write this as
\begin{equation}
    \left(\bar\partial^2-\frac{1}{2}\tilde{\mathcal{T}}_{\bar Z\bar Z}\right)\tilde Y_{ZZ}=    \left(\partial^2-\frac{1}{2}\tilde{\mathcal{T}}_{ZZ}\right)\tilde Y_{\bar Z\bar Z}\,,
\end{equation}
where $\tilde{\mathcal{T}}_{ZZ}$ is holomorphic. The fields $\tilde{\mathcal{T}}_{ZZ}$, $\tilde Y_{ZZ}$ etc are the components of $\tilde{\mathcal{T}}_{ab}$ and $\tilde Y_{ab}$ in a complex basis.}
\begin{equation}
\left(\partial_a\partial_c\tilde -\frac{1}{2}\tilde{\mathcal{T}}_{ac}\right)\tilde Y_{bc}-\left(a\leftrightarrow b\right)\,,
\end{equation}
for any $\tilde{\mathcal{T}}_{ab}$ that is STF and transverse, i.e. $\partial_a \tilde{\mathcal{T}}_{ab}=0$. The solution to this latter PDE is given by
\begin{equation}
    \tilde Y_{ab}=\left(\partial_{\langle a}\partial_{b\rangle}\tilde -\frac{1}{2}\tilde{\mathcal{T}}_{ab}\right)\tilde Y(X)\,,
\end{equation}
where $\tilde Y(X)$ is any real function of $X^a$. Hence, we conclude that the most general 
spatial STF tensor $Y_{\mu\nu}$ that obeys $\left(\mathcal{L}_v+\frac{1}{2}K\right)Y_{\mu\nu}=0$ and \eqref{eq:diffconY} is given by 
\begin{equation}\label{eq:STFY}
    Y_{\mu\nu}=e^\varphi \partial_\mu X^a\partial_\nu X^b \left(\partial_{\langle a}\partial_{b\rangle}\tilde -\frac{1}{2}\tilde{\mathcal{T}}_{ab}\right)\tilde Y(X)\,.
\end{equation}

\section{Vacuum shear from diffeomorphisms}\label{app:vacsheardiffeos}

In this appendix we will verify our  Carroll-covariant expression for the vacuum shear by deriving it from a class of diffeomorphisms that transform Minkowski spacetime into our Carroll-covariant BS gauge. The calculations presented here are a generalisation of what was done in section 3 of \cite{Compere:2018ylh}.

\subsection{Setting up the problem}

We start with Minkowski spacetime with a flat spatial metric at $\mathcal{I}^+$, i.e.
\begin{equation}\label{eq:Minkflatbdry}
    ds^2=-2d\bar rd\bar u+\bar r^2d\bar x^i d\bar x^i\,.
\end{equation}
The goal will be to construct a coordinate transformation that transforms this to
\begin{equation}\label{eq:CCBSgauge}
    ds^2=-2e^\beta dr\tau_\mu dx^\mu+g_{\mu\nu}dx^\mu dx^\nu\,,
\end{equation}
where 
\begin{equation}
    g_{\mu\nu}=-Se^{2\beta}\tau_\mu\tau_\nu+\Pi_{\mu\nu}\,,
\end{equation}
with 
\begin{equation}
    \frac{1}{2}\Pi^{\mu\nu}\partial_r\Pi_{\mu\nu}=2r^{-1}\,.
\end{equation}
The inverse metric is given by
\begin{equation}
    g^{rr}=S\,,\qquad g^{\mu r}=U^\mu\,,\qquad g^{\mu\nu}=\Pi^{\mu\nu}\,,
\end{equation}
where $\tau_\mu \Pi^{\mu\nu}=0$.

Just as in section \ref{subsec:standardBScoord} we will split the boundary coordinates as $x^\mu=(u,x^i)$. For simplicity we will require that $\tau_i=0$ but other than that the unbarred metric is allowed to be any Carroll-covariant BS gauge form of the metric. In our gauge (with $\tau_i=0$ which can always be achieved by fixing Carroll boosts) we have
\begin{equation}\label{eq:gaugechoicetaui=0}
    g_{rr}=0\,,\qquad g_{ri}=0\,,\qquad g^{ij}\partial_r g_{ij}=4r^{-1}\,.
\end{equation}
The inverse metric obeys the gauge conditions
\begin{equation}
    g^{uu}=0\,,\qquad g^{ui}=0\,.
\end{equation}

The coordinate transformation is a map $(\bar r,\bar u,\bar x^i)\mapsto (r,u,x^i)$ where 
the barred coordinates are to be determined functions of the unbarred coordinates. It will however prove rather convenient to consider the transformation in two steps. First we write
\begin{eqnarray}
    \bar r & = & \bar r\,,\label{eq:coordtrafo1}\\
    \bar u & = & \upsilon(\bar r,u,x)\,,\label{eq:coordtrafo2}\\
    \bar x^i & = & g^i(\bar r,u,x)\,,\label{eq:coordtrafo3}
\end{eqnarray}
and then we work out what $\bar r=\bar r(r,u,x)$ has to be.

\subsection{Constructing the diffeomorphism}

We will start by imposing the gauge conditions $g_{rr}=0=g_{ri}$ on the coordinate transformation. The condition $g_{rr}=0$ leads to
\begin{eqnarray}
    g_{rr} & = & -2\frac{\partial\bar r}{\partial r}\frac{\partial \bar u}{\partial r}+\bar r^2\frac{\partial\bar x^i}{\partial r}\frac{\partial\bar x^i}{\partial r}\nonumber\\
    & = & -2\left(\frac{\partial\bar r}{\partial r}\right)^2\left(\frac{\partial \upsilon}{\partial\bar r}-\frac{1}{2}\bar r^2\frac{\partial g^i}{\partial\bar r}\frac{\partial g^i}{\partial\bar r}\right)\,.
\end{eqnarray}
The Jacobian of the coordinate transformation must be non-vanishing which implies that $\frac{\partial\bar r}{\partial r}\neq 0$ (for all $r$). Since in both cases the boundary is approached for large values of $\bar r$ and $r$ we can take $\frac{\partial\bar r}{\partial r}> 0$. Using this we conclude that 
\begin{equation}\label{eq:grr=0}
    \frac{\partial \upsilon}{\partial\bar r}=\frac{1}{2}\bar r^2\frac{\partial g^i}{\partial\bar r}\frac{\partial g^i}{\partial\bar r}\,.
\end{equation}
Doing the same with $g_{ri}=0$ and using \eqref{eq:grr=0} it follows that 
\begin{equation}\label{eq:gri=0}
    \frac{\partial \upsilon}{\partial x^i}=\bar r^2\frac{\partial g^j}{\partial\bar r}\frac{\partial g^j}{\partial x^i}\,.
\end{equation}
Using both \eqref{eq:grr=0} and \eqref{eq:gri=0} we find
\begin{equation}\label{eq:gij}
    g_{ij}=\bar r^2\frac{\partial g^k}{\partial x^i}\frac{\partial g^k}{\partial x^j}\,.
\end{equation}

Equations \eqref{eq:grr=0} and \eqref{eq:gri=0} lead to a compatibility condition which follows from considering 
\begin{equation}\label{eq:compatibility}
    \frac{\partial^2 \upsilon}{\partial\bar r\partial x^i}=\frac{\partial^2 \upsilon}{\partial x^i\partial\bar r}\,.
\end{equation}
Equation \eqref{eq:gij} tells us that $\frac{\partial g^k}{\partial x^i}$ is invertible which together with equation \eqref{eq:compatibility} leads to
\begin{equation}
    \frac{\partial g^i}{\partial\bar r}+\frac{1}{2}\bar r\frac{\partial^2 g^i}{\partial{\bar r}^2}=0\,,
\end{equation}
which is solved by
\begin{equation}\label{eq:gi}
    g^i=\os{0}{g}{}^i(u,x)+\bar r^{-1}\os{1}{g}{}^i(u,x)\,.
\end{equation}
Going back to \eqref{eq:grr=0} and \eqref{eq:gri=0} we obtain
\begin{equation}\label{eq:f}
    \upsilon=\os{0}{\upsilon}-\frac{1}{2}\bar r^{-1}\os{1}{g}^i \os{1}{g}^i\,,
\end{equation}
as well as 
\begin{equation}\label{eq:g1i}
    \partial_i \os{0}{\upsilon}=-\os{1}{g}^j\frac{\partial \os{0}{g}^j}{\partial x^i}\,.
\end{equation}
We will see shortly that $\partial_i \os{0}{g}^j$ has to be invertible so that this equation can be solved for $\os{1}{g}^i$.

The gauge condition $g^{ij}\partial_r g_{ij}=4r^{-1}$ can be phrased as stating that $r^{-4}\text{det}\,g_{ij}$ is independent of $r$. From the boundary condition that $r^2g_{ij}$ is a positive definite metric we can furthermore conclude that $r^{-4}\text{det}\,g_{ij}>0$. Using \eqref{eq:gij} it thus follows that there must exist a nonzero function $Q(u,x)$ such that
\begin{equation}\label{eq:detcon}
    \frac{\bar r^2}{r^2}\text{det}\,\frac{\partial g^k}{\partial x^i}=Q\,.
\end{equation}
By taking the large $r$ limit (and thus also the large $\bar r$ limit) of this condition we conclude that 
\begin{equation}
    \text{det}\,\frac{\partial\os{0}{g}^k}{\partial x^i}\neq 0\,.
\end{equation}
For notational ease we will define the matrices 
\begin{equation}
\os{0}{\mathcal{G}}{}^k{}_i=\frac{\partial \os{0}{g}^k}{\partial x^i}\,,\qquad \os{1}{\mathcal{G}}{}^k{}_i=\frac{\partial \os{1}{g}^k}{\partial x^i}\,.  
\end{equation}
Using that for any $2\times 2$ matrix we have
\begin{equation}
    \text{det}\,(1+X)=1+\text{Tr}\,X+\text{det}\,X\,,
\end{equation}
we rewrite \eqref{eq:detcon} as
\begin{equation}
    \frac{\bar r^2}{r^2}\text{det}\,\os{0}{\mathcal{G}}\left(1+\bar r^{-1}\text{Tr}\,\os{0}{\mathcal{G}}^{-1}\os{1}{\mathcal{G}}+\bar r^{-2}\text{det}\,\left(\os{0}{\mathcal{G}}^{-1}\os{1}{\mathcal{G}}\right)\right)=Q\,.
\end{equation}
The sign of $Q$ is the sign of $\text{det}\,\frac{\partial g^k}{\partial x^i}$ which in turn is the sign of $\text{det}\,\os{0}{\mathcal{G}}$. Therefore, we can write 
\begin{equation}\label{eq:bar}
   \bar r^2-2C\bar r+D=e^{2\Phi}r^2\,,
\end{equation}
where $\Phi=\Phi(u,x)$ and where we defined
\begin{eqnarray}
    C & = & -\frac{1}{2}\text{Tr}\,\os{0}{\mathcal{G}}^{-1}\os{1}{\mathcal{G}}\,,\label{eq:C}\\
    D & = & \text{det}\,\left(\os{0}{\mathcal{G}}^{-1}\os{1}{\mathcal{G}}\right)\,.\label{eq:D}
\end{eqnarray}
Solving \eqref{eq:bar} for $\bar r$ gives
\begin{equation}\label{eq:barr2}
    \bar r=C+\sqrt{C^2-D+e^{2\Phi}r^2}=re^{\Phi}+C+\mathcal{O}(r^{-1})\,.
\end{equation}

\subsection{The metric in the new coordinates}

Next we compute the remaining components of the metric $g_{ur}, g_{ui}, g_{uu}$. Repeatedly using \eqref{eq:grr=0} and \eqref{eq:gri=0} gives us 
\begin{eqnarray}
    g_{ur} & = & -\frac{\partial\bar r}{\partial r}\left[\frac{\partial \upsilon}{\partial u}-\bar r^2\frac{\partial g^i}{\partial\bar r}\frac{\partial g^i}{\partial u}\right]\,,\label{eq:gur}\\
    g_{ui} & = & -\frac{\partial\bar r}{\partial x^i}\left[\frac{\partial \upsilon}{\partial u}-\bar r^2\frac{\partial g^k}{\partial\bar r}\frac{\partial g^k}{\partial u}\right]+\bar r^2\frac{\partial g^k}{\partial u}\frac{\partial g^k}{\partial x^i}\,,\label{eq:gui}\\
    g_{uu} & = & -2\frac{\partial\bar r}{\partial u}\left[\frac{\partial \upsilon}{\partial u}-\bar r^2\frac{\partial g^k}{\partial\bar r}\frac{\partial g^k}{\partial u}\right]+\bar r^2\frac{\partial g^k}{\partial u}\frac{\partial g^k}{\partial u}\,.\label{eq:guu}
\end{eqnarray}
Equation \eqref{eq:gur} is order $r^0$ and $g_{ui}$ and $g_{ui}$ are both order $r^2$.

It will be useful to also look at the transformation of the inverse metric. In the unbarred system of coordinates we have the following inverse metric 
\begin{eqnarray}
    g^{uu} & = & 0\,,\\
    g^{iu} & = & 0\,,\\
    g^{ur} & = & -e^{-\beta}\,,\\
    g^{ir} & = & e^{-\beta}g^{ik}g_{ku}\,,\\
    g^{rr} & = & -e^{-2\beta}g_{uu}+e^{-2\beta}g^{ij}g_{iu}g_{ju}\,,
\end{eqnarray}
where $g^{ik}$ is the inverse of $g_{kj}$, so
\begin{equation}
    g^{ik}g_{kj}=\delta^i_j\,.
\end{equation}
The coordinate transformation \eqref{eq:coordtrafo1}--\eqref{eq:coordtrafo3} maps the inverse metric in the unbarred coordinate system to the one of the barred coordinate system. We first compute $\bar g^{\bar r\bar r}$ and $\bar g^{\bar r i}$ in terms of the components of the unbarred metric and using these results we can show that 
\begin{equation}
    \bar g^{ij}=-\bar g^{\bar r\bar r}\frac{\partial g^i}{\partial\bar r}\frac{\partial g^j}{\partial\bar r}+\frac{\partial g^i}{\partial\bar r}\bar g^{\bar r j}+\frac{\partial g^j}{\partial\bar r}\bar g^{\bar r i}+\frac{\partial g^i}{\partial x^k}\frac{\partial g^j}{\partial x^l}g^{kl}\,.
\end{equation}
Now using that $\bar g^{\bar r\bar r}=0=\bar g^{\bar r i}$ and that $\bar g^{ij}=\bar r^{-2}\delta^{ij}$ we learn that
\begin{equation}\label{eq:inversesigma}
    \frac{\partial g^i}{\partial x^k}\frac{\partial g^j}{\partial x^l}g^{kl}=\bar r^{-2}\delta^{ij}\,.
\end{equation}
This relation will prove useful in what follows.

Apart from a few details that we will discuss shortly we have now constructed the diffeomorphism that takes us from 
\eqref{eq:Minkflatbdry} to \eqref{eq:CCBSgauge}, obeying the gauge choices \eqref{eq:gaugechoicetaui=0}. This transformation is given by
\eqref{eq:gi}, \eqref{eq:f}, \eqref{eq:g1i} and \eqref{eq:barr2}. We will now study what the boundary geometry is in the unbarred coordinate system. Once we have established this we will compute the shear in the unbarred coordinate system.

We will compare our results for $g_{ur}$ and $g_{\mu\nu}$ and compare them with \eqref{eq:CCBSgauge} where the fields fall off as in \eqref{eq:betaexp}--\eqref{eq:gexp}. Starting with $g_{ru}$ at order $r^0$ we read off
\begin{equation}\label{eq:conf0}
  e^{-\Phi}\tau_u=\frac{\partial \os{0}{\upsilon}}{\partial u}+\os{1}{g}^i\frac{\partial \os{0}{g}^i}{\partial u}\,.
\end{equation}
Since we set $\tau_i=0$ we now have an expression for $\tau_\mu$.
We can also check that $g_{ru}$ vanishes at order $r^{-1}$ in agreement with the fact that $\beta=\mathcal{O}(r^{-2})$. From $g_{uu}, g_{ui}, g_{ij}$ at order $r^2$ we read off that
\begin{equation}\label{eq:bdrymetricafterdiffeo}
    h_{\mu\nu}dx^\mu dx^\nu=e^{2\Phi}d\os{0}{g}^k d\os{0}{g}^k\,.
\end{equation}

We can use these results for $\tau_\mu$ and $h_{\mu\nu}$ to construct $v^\mu$ by solving $v^\mu\tau_\mu=-1$ and $v^\mu h_{\mu\nu}=0$ which results in
\begin{equation}\label{eq:v}
    v^u=-\frac{1}{\tau_u}\,,\qquad v^i=\frac{1}{\tau_u}\sigma^{ij}h_{uj}=\frac{1}{\tau_u}\left(\os{0}{\mathcal{G}}^{-1}\right){}^i{}_k\frac{\partial \os{0}{g}^k}{\partial u}\,.
\end{equation}
Using \eqref{eq:g1i} and \eqref{eq:v} we can now express \eqref{eq:conf0} as 
\begin{equation}\label{eq:Lievupsilon0}
    v^\mu\partial_\mu \os{0}{\upsilon}=-e^{-\Phi}\,.
\end{equation}

We can write the boundary metric \eqref{eq:bdrymetricafterdiffeo} as
\begin{equation}\label{eq:ZbarZbdry}
    h_{\mu\nu}dx^\mu dx^\nu=e^{2\varphi} dX^a dX^a=2e^{2\varphi}dZd\bar Z\,,
\end{equation}
where $Z=\frac{1}{\sqrt{2}}\left(X^1+iX^2\right)$ which is the most general allowable form of the boundary metric \eqref{eq:metrich}.
This implies that
\begin{eqnarray}
    dZ & = & \frac{1}{\sqrt{2}}e^h d\left(\os{0}{g}^1+i\os{0}{g}^2\right)\,,\label{eq:dZdg0}\\
    \varphi & = & \Phi-\frac{1}{2}h-\frac{1}{2}\bar h\,,\label{eq:varphi'tovarphi}
\end{eqnarray}
where $h(Z)=-\log f'(Z)$ with $f$ any holomorphic function of $Z$. We used here the freedom expressed in  \eqref{eq:varphiprime}. We thus conclude that 
\begin{equation}
    \Phi=\varphi+\zeta\,,
\end{equation}
where 
\begin{equation}
    \zeta=-\frac{1}{2}\log f'(Z)+\text{c.c.}\,,
\end{equation}
which agrees with \eqref{eq:zetatof(Z)} and \eqref{eq:DefPhi}.

The boundary Carroll geometry can be viewed as a line bundle fibred over a 2-dimensional Riemannian base manifold. The latter has a metric 
\begin{equation}
    \sigma_{ij}=h_{ij}=e^{2\Phi}\os{0}{\mathcal{G}}{}^k{}_i\os{0}{\mathcal{G}}{}^k{}_j\,.
\end{equation}
Equation \eqref{eq:inversesigma} at order $r^{-2}$ tells us that
\begin{equation}\label{eq:sigmatodelta}
    \os{0}{\mathcal{G}}{}^i{}_k\os{0}{\mathcal{G}}{}^j{}_l\sigma^{kl}=\delta^{ij}\,.
\end{equation}
We conclude from this that the inverse of $\os{0}{\mathcal{G}}$ is given by
\begin{equation}\label{eq:calG0inv}
    \left(\os{0}{\mathcal{G}}^{-1}\right){}^i{}_j=e^{2\Phi}\sigma^{il}\delta_{jk}\os{0}{\mathcal{G}}{}^k{}_l\,.
\end{equation}
The Levi-Civita connection associated with $\sigma_{ij}$ is given by
\begin{equation}\label{eq:2DLCcon}
    \Gamma^k_{ij}=\delta^k_{i}\partial_j\Phi+\delta^k_{j}\partial_i\Phi-\sigma_{ij}\sigma^{kl}\partial_l\Phi+\left(\os{0}{\mathcal{G}}^{-1}\right){}^k{}_l\partial_i \os{0}{\mathcal{G}}{}^l{}_j\,.
\end{equation}
We will denote the covariant derivative by $D_i$. Using equation \eqref{eq:g1i} and the above notion of covariant derivative we can derive the following useful identity
\begin{equation}\label{eq:calG0invcalG1}
    \left(\os{0}{\mathcal{G}}^{-1}\right){}^i{}_k\os{1}{\mathcal{G}}{}^k{}_j=-e^{2\Phi}\sigma^{ik}\left(D_j\partial_k \os{0}{\upsilon}+2\partial_{\langle j}\Phi\partial_{k\rangle}\os{0}{\upsilon}\right)\,,
\end{equation}
where $\partial_{\langle j}\Phi\partial_{k\rangle}\os{0}{\upsilon}$ is STF with respect to $\sigma^{jk}$. This can be used to express $C$ in the expression for $\bar r$ (see equation \eqref{eq:C}) as
\begin{equation}\label{eq:Csol}
    C=\frac{1}{2}e^{2\Phi}\sigma^{ij}D_i\partial_j \os{0}{\upsilon}=\frac{1}{2}e^{2\Phi}D^2 \os{0}{\upsilon}\,.
\end{equation}
When $\tau_i=0$ the shear is given by $g_{ij}$ at order $r$. Using \eqref{eq:gij} we can write
\begin{equation}\label{eq:gij2}
    g_{ij}=\bar r^2\os{0}{\mathcal{G}}{}^k{}_i\os{0}{\mathcal{G}}{}^k{}_j+\bar r\left(\os{0}{\mathcal{G}}{}^k{}_i\os{1}{\mathcal{G}}{}^k{}_j+\os{0}{\mathcal{G}}{}^k{}_j\os{1}{\mathcal{G}}{}^k{}_i\right)+\os{1}{\mathcal{G}}{}^k{}_i\os{1}{\mathcal{G}}{}^k{}_j\,.
\end{equation}
Expanding this to order $r$ and using 
\eqref{eq:calG0inv}, \eqref{eq:calG0invcalG1}, \eqref{eq:Csol}
we find the following expression for the shear
\begin{equation}\label{eq:vacshearapp}
    C_{ij}=-2e^{\Phi}\left(D_{\langle i}\partial_{j\rangle}\os{0}{\upsilon}+2\partial_{\langle i}\Phi\partial_{j\rangle}\os{0}{\upsilon}\right)\,.
\end{equation}
In terms of $C_{ij}$ we can reexpress \eqref{eq:calG0invcalG1} as
\begin{equation}
    \left(\os{0}{\mathcal{G}}^{-1}\right){}^i{}_k\os{1}{\mathcal{G}}{}^k{}_j=-C\delta^i_j+\frac{1}{2}e^{\Phi}C^{i}{}_j\,,
\end{equation}
where we raised an index of $C_{kj}$ with $\sigma^{ik}$. With the help of this result we can show that 
\begin{equation}
    C^2-D=\frac{1}{8}e^{2\Phi}C_{ij}C^{ij}\,,
\end{equation}
where $D$ is defined in \eqref{eq:D}.

We conclude that \eqref{eq:barr2} can be written as
\begin{equation}\label{eq:barrtor}
    \bar r=\frac{1}{2}e^{2\Phi}D^2 \os{0}{\upsilon}+e^{\Phi}\sqrt{r^2+\frac{1}{8}C_{ij}C^{ij}}\,.
\end{equation}
The rest of the coordinate transformation can now be written as
\begin{eqnarray}
    \upsilon & = & \os{0}{\upsilon}-\frac{1}{2}\bar r^{-1}\sigma^{ij}\partial_i \os{0}{\upsilon}\partial_j \os{0}{\upsilon}\,,\\
    g & = & \os{0}{g}^i-\bar r^{-1}e^{2\Phi}\os{0}{\mathcal{G}}{}^i{}_j\sigma^{jk}\partial_k \os{0}{\upsilon}\,,\label{eq:g}
\end{eqnarray}
where we used \eqref{eq:g1i}, \eqref{eq:calG0inv} and \eqref{eq:sigmatodelta}. We thus conclude that the coordinate transformation depends on four unspecified functions $\Phi, \os{0}{\upsilon}, \os{0}{g}^i$.

If we define $\os{0}{\upsilon}=e^{-\Phi}F$ so that \eqref{eq:Lievupsilon0} implies $\left(\mathcal{L}_v +\frac{1}{2}K\right)F=-1$, then the shear \eqref{eq:vacshearapp} becomes
\begin{equation}
    C_{ij}=2\left(-D_{\langle i}\partial_{j\rangle}F+F\left(D_{\langle i}\partial_{j\rangle}\Phi+\partial_{\langle i}\Phi\partial_{j\rangle}\Phi\right)\right)\,.
\end{equation}
This agrees with \eqref{eq:vacshearfinal} if we use $\tau_i=0$.

\section{Rewriting of the Bondi loss equations}\label{app;rewritingBondiloss}

This appendix provides calculation details for results discussed in section \ref{subsec:newBondiform}.

\subsection{Rewriting the mass loss equation}

We want to show that equation \eqref{eq:diffeoWtimeproj} which is repeated here for convenience
\begin{equation}\label{eq:oldmassloss}
     -\left(\mathcal{L}_v-\frac{3}{2}K\right)\left(\tau_\mu T^\mu\right)-\frac{1}{4}N^{\rho\sigma}N_{\rho\sigma}+\left(\mathcal{D}_\mu+a_\mu\right)\left(T^\rho h^\mu_\rho\right)=0\,,
\end{equation}
can alternatively be written as \eqref{eq:BLeq1} which reads
\begin{equation}
     -\left(\mathcal{L}_v-\frac{3}{2}K\right)\tau\cdot T'+\frac{1}{2}C^{\rho\sigma}v^\mu R(K)_{\mu\rho}{}^a e^a_\sigma+\left(\mathcal{D}_\mu+a_\mu\right)\mathcal{A}^\mu_B=0\,.
\end{equation}

To show this we start with \eqref{eq:oldmassloss} and use that the energy flux takes the form \eqref{eq:boostanom}.
We will use the following two identities\footnote{To derive the first of these identities it is useful to first show that 
\begin{equation}
    h^{\rho\sigma}\mathcal{D}_\rho N_{\sigma\nu} = -h_{\sigma\nu}\left(\mathcal{L}_v-\frac{3}{2}K\right)\left(\mathcal{D}_\rho-2a_\rho\right)C^{\rho\sigma}+C^\rho{}_\nu G_\rho\,.
\end{equation}}
\begin{eqnarray}
    \left(\mathcal{D}_\mu+a_\mu\right)\left(h^{\mu\alpha} h^{\rho\sigma}\mathcal{D}_\rho N_{\sigma\alpha}\right) & = & \left(\mathcal{D}_\mu+a_\mu\right)\left(\mathcal{D}_\rho-a_\rho\right)N^{\rho\mu}\\
    & = & -\left(\mathcal{L}_v-\frac{3}{2}K\right)\mathcal{D}_\mu\left(\mathcal{D}_\rho-2a_\rho\right)C^{\rho\mu}+C^{\mu\rho}\left(\mathcal{D}_\mu+3a_\mu\right)G_\rho\,,\nonumber\\
    N^2 & = & -\left(\mathcal{L}_v-\frac{3}{2}K\right)\left(C\cdot N\right)+C^{\rho\sigma}\mathcal{L}_v N_{\rho\sigma}\,.
\end{eqnarray}
These all follow from commuting covariant derivatives with other covariant derivatives and with Lie derivatives where we used the results from section \ref{subsec:bdrygeom}. Note that all terms have a definite Weyl weight. With the help of these two identities we can verify the rewriting with $\tau\cdot T'$ given by the third equation in \eqref{eq:redefenergyden}.

\subsection{Rewriting the angular momentum loss equation}

The rewriting of the angular momentum loss equation is considerably more work. The starting point is \eqref{eq:diffeoWIspatialproj} which is repeated here
\begin{eqnarray}
0 & = & -\left(\mathcal{L}_v-K\right) P_\kappa+ h_{\kappa\sigma}\mathcal{D}_\mu\tilde {T}^{\mu\sigma}+\frac{1}{2}h^\mu_\kappa\left(\partial_\mu+3a_\mu\right)\left(T^{\rho\sigma}h_{\rho\sigma}+\frac{1}{2}N^{\rho\sigma}C_{\rho\sigma}\right) \nonumber\\
    &&+\frac{1}{4}h_{\kappa\sigma}\mathcal{D}_\mu\left(N^{\mu\lambda}C_\lambda{}^\sigma-N^{\sigma\lambda}C_\lambda{}^\mu\right)-\frac{1}{4}N^{\mu\sigma}h^\nu_\kappa\left(\mathcal{D}_\nu+a_\nu\right) C_{\mu\sigma}+T^\sigma h^\mu_\sigma F_{\mu\kappa}\,.\nonumber\\
    &&\label{eq:diffeoWIspatialprojapp}
\end{eqnarray}
The main idea behind the rewriting process is to rewrite the terms on the second line either as $K$-curvatures or as terms that we can absorb into the terms on the first line.

To this end consider the first two terms on the second line of \eqref{eq:diffeoWIspatialproj}, i.e. 
\begin{equation}
    \frac{1}{4}h_{\kappa\sigma}\mathcal{D}_\mu\left(N^{\mu\lambda}C_\lambda{}^\sigma-N^{\sigma\lambda}C_\lambda{}^\mu\right)-\frac{1}{4}N^{\mu\sigma}h^\nu_\kappa\left(\mathcal{D}_\nu+a_\nu\right) C_{\mu\sigma}\,.
\end{equation}
This is not written in a manifestly Weyl covariant manner\footnote{The term $N^{\mu\lambda}C_\lambda{}^\sigma-N^{\sigma\lambda}C_\lambda{}^\mu$ has Weyl weight $-5$ and so Weyl covariant would be $\left(\mathcal{D}_\mu+2a_\mu\right)\left(N^{\mu\lambda}C_\lambda{}^\sigma-N^{\sigma\lambda}C_\lambda{}^\mu\right)$. We refer to sections 6.7 and B.6 of \cite{Hartong:2026rbr} for more details on Weyl covariant derivatives.}. To remedy this we write 
\begin{equation}
    N^{\mu\sigma}h^\nu_\kappa\mathcal{D}_\nu C_{\mu\sigma}=    N^{\mu\sigma}h^\nu_\kappa\left(\mathcal{D}_\nu C_{\mu\sigma}-\mathcal{D}_\mu C_{\nu\sigma}\right)+N^{\mu\sigma}h^\nu_\kappa\mathcal{D}_\mu C_{\nu\sigma}\,.
\end{equation}
We then use \eqref{eq:DSTFid} to rewrite the first term on the RHS of this equation leading to 
\begin{equation}
    N^{\mu\sigma}h^\nu_\kappa\mathcal{D}_\nu C_{\mu\sigma}= N^\rho{}_\kappa h^{\alpha\beta}\mathcal{D}_\alpha C_{\beta\rho}+N^{\mu\sigma}h^\nu_\kappa\mathcal{D}_\mu C_{\nu\sigma}\,.
\end{equation}
Using the product rule in reverse on the second term and writing 
\begin{equation}
    N^{\mu\sigma}C_{\sigma}{}^\rho=\frac{1}{2}\left(N^{\mu\sigma}C_{\sigma}{}^\rho-N^{\rho\sigma}C_{\sigma}{}^\mu\right)+\frac{1}{2}h^{\mu\rho}N\cdot C\,,
\end{equation}
where we used \eqref{eq:STFSTF} we obtain
\begin{align}
    N^{\mu\sigma}h^\nu_\kappa\mathcal{D}_\nu C_{\mu\sigma}= &\frac{1}{2}h_{\kappa\alpha}\mathcal{D}_\mu\left(N^{\mu\sigma}C_\sigma{}^\alpha-N^{\alpha\sigma}C_\sigma{}^\mu\right)+\frac{1}{2}h_{\kappa\alpha}\mathcal{D}_\mu\left(h^{\mu\alpha} N\cdot C\right)\nonumber\\
  & -C_{\sigma\kappa}\mathcal{D}_\mu N^{\mu\sigma}+N^\rho{}_\kappa h^{\alpha\beta}\mathcal{D}_\alpha C_{\beta\rho}\,.
\end{align}
With the help of this result we can write for the first two terms on the second line of \eqref{eq:diffeoWIspatialprojapp} the following
\begin{align}
    &\frac{1}{4}h_{\kappa\sigma}\mathcal{D}_\mu\left(N^{\mu\lambda}C_\lambda{}^\sigma-N^{\sigma\lambda}C_\lambda{}^\mu\right)-\frac{1}{4}N^{\mu\sigma}h^\nu_\kappa\left(\mathcal{D}_\nu+a_\nu\right) C_{\mu\sigma}=\nonumber\\
    &\frac{1}{8}h_{\kappa\sigma}\left(\mathcal{D}_\mu+2a_\mu\right)\left(N^{\mu\lambda}C_\lambda{}^\sigma-N^{\sigma\lambda}C_\lambda{}^\mu\right)-\frac{1}{8}h^\mu_\kappa\left(\partial_\mu+3a_\mu\right)N\cdot C\nonumber\\
    &+\frac{1}{4}C^\rho{}_\kappa h^{\alpha\beta}\mathcal{D}_\alpha N_{\beta\rho}-\frac{1}{4}N^\rho{}_\kappa h^{\alpha\beta}\left(\mathcal{D}_\alpha-a_\alpha\right)C_{\beta\rho}\,.\label{eq:rewritinglast2terms}
\end{align}
Every term on the right-hand side is now Weyl covariant.

The last term can be written as
\begin{align}
    &-\frac{1}{4}N^\rho{}_\kappa h^{\alpha\beta}\left(\mathcal{D}_\alpha-a_\alpha\right)C_{\beta\rho}\nonumber\\
    =&\left(\mathcal{L}_v-K\right)\left(\frac{1}{4}C^\rho{}_\kappa h^{\alpha\beta}\left(\mathcal{D}_\alpha-a_\alpha\right)C_{\beta\rho}\right)-\frac{1}{4}C^\rho{}_\kappa\left(\mathcal{L}_v-\frac{1}{2}K\right)\left(h^{\alpha\beta}\left(\mathcal{D}_\alpha-a_\alpha\right)C_{\beta\rho}\right)\nonumber\\
    =&\left(\mathcal{L}_v-K\right)\left(\frac{1}{4}C^\rho{}_\kappa h^{\alpha\beta}\left(\mathcal{D}_\alpha-a_\alpha\right)C_{\beta\rho}\right)+\frac{1}{2}C^\rho{}_\kappa\left(\mathcal{L}_v-\frac{1}{2}K\right)\left(\overset{(0)}{P}_\rho-\frac{1}{2}\mathcal{D}_\alpha F^\alpha{}_\rho\right)\nonumber\\
    =&\left(\mathcal{L}_v-K\right)\left(\frac{1}{4}C^\rho{}_\kappa h^{\alpha\beta}\left(\mathcal{D}_\alpha-a_\alpha\right)C_{\beta\rho}\right)+\frac{1}{2}C^\rho{}_\kappa\left(-\frac{1}{2}C^\mu{}_\rho G_\mu+\frac{1}{2}h^{\alpha\beta}\mathcal{D}_\alpha N_{\beta\rho}\right)\,.
\end{align}
This allows us to write \eqref{eq:rewritinglast2terms} as
\begin{align}
    &\frac{1}{4}h_{\kappa\sigma}\mathcal{D}_\mu\left(N^{\mu\lambda}C_\lambda{}^\sigma-N^{\sigma\lambda}C_\lambda{}^\mu\right)-\frac{1}{4}N^{\mu\sigma}h^\nu_\kappa\left(\mathcal{D}_\nu+a_\nu\right) C_{\mu\sigma}\nonumber\\
    =&\frac{1}{8}h_{\kappa\sigma}\left(\mathcal{D}_\mu+2a_\mu\right)\left(N^{\mu\lambda}C_\lambda{}^\sigma-N^{\sigma\lambda}C_\lambda{}^\mu\right)-\frac{1}{8}h^\mu_\kappa\left(\partial_\mu+3a_\mu\right)N\cdot C\nonumber\\
    &+\frac{1}{2}C^\rho{}_\kappa h^{\alpha\beta}\mathcal{D}_\alpha N_{\beta\rho}-\frac{1}{8}C^2 G_\kappa+\left(\mathcal{L}_v-K\right)\left(\frac{1}{4}C^\rho{}_\kappa h^{\alpha\beta}\left(\mathcal{D}_\alpha-a_\alpha\right)C_{\beta\rho}\right)\,,
\end{align}
where we used $C^\rho{}_\kappa C^\mu{}_\rho=\frac{1}{2}C^2 h^\mu_\kappa$.

The first term on the right-hand side can be rewritten using \eqref{eq:eeRK} which leads to
\begin{align}
    &\frac{1}{8}h_{\kappa\sigma}\left(\mathcal{D}_\mu+2a_\mu\right)\left(N^{\mu\lambda}C_\lambda{}^\sigma-N^{\sigma\lambda}C_\lambda{}^\mu\right)\nonumber\\
    =&\frac{1}{2}h_{\kappa\nu}\left(\mathcal{D}_\mu+2a_\mu\right) R(K)^{\mu\nu}-\frac{1}{2}F^{\mu}{}_\kappa\left(\partial_\mu+2a_\mu\right)\left(\overset{(0)}{S}-a^2\right)-\frac{1}{2}\left(\overset{(0)}{S}-a^2\right)\mathcal{D}_\mu F^\mu{}_\kappa\nonumber\\
    &-\frac{1}{2}h_{\kappa\nu}\left(\mathcal{D}_\mu+2a_\mu\right)\left[h^{\mu\alpha}h^{\nu\beta}\left(\left(\mathcal{D}_\alpha+a_\alpha\right)\overset{(0)}{P}_\beta-\left(\alpha\leftrightarrow\beta\right)\right)\right]\,,
\end{align}
where we defined $R(K)^{\mu\nu}=e^\mu_a e^\nu_b R(K)_{ab}$. We then use the identity
\begin{align}
    &-\frac{1}{2}h_{\kappa\nu}\left(\mathcal{D}_\mu+2a_\mu\right)\left[h^{\mu\alpha}h^{\nu\beta}\left(\left(\mathcal{D}_\alpha+a_\alpha\right)\overset{(0)}{P}_\beta-\left(\alpha\leftrightarrow\beta\right)\right)\right]\nonumber\\
    =& 2\overset{(0)}{P}{}^\rho\left(\mathcal{L}_v+\frac{1}{2}K\right)F_{\rho\kappa}+F^\rho{}_\kappa\left(\mathcal{L}_v-\frac{1}{2}K\right)\overset{(0)}{P}_\rho+\left(\overset{(0)}{S}-a^2\right)\overset{(0)}{P}_\kappa\nonumber\\
    &-h_{\kappa\nu}\mathcal{D}_\mu\left(h^{\alpha\langle\mu}h^{\nu\rangle\beta}\left(\mathcal{D}_\alpha+3a_\alpha\right)\overset{(0)}{P}_\beta\right)+\frac{1}{2}h^\rho_\kappa\left(\partial_\rho+3a_\rho\right)\left(\mathcal{D}_\sigma\overset{(0)}{P}{}^\sigma\right)\,.
\end{align}
This identity follows essentially from using 
\begin{equation}
    \mathcal{D}_\mu\left(\mathcal{D}_\alpha\overset{(0)}{P}_\beta-\mathcal{D}_\beta\overset{(0)}{P}_\alpha\right)=\mathcal{D}_\mu\left(\mathcal{D}_\alpha\overset{(0)}{P}_\beta+\mathcal{D}_\beta\overset{(0)}{P}_\alpha\right)-2\left[\mathcal{D}_\mu\,,\mathcal{D}_\beta\right]\overset{(0)}{P}_\alpha-2\mathcal{D}_\beta \mathcal{D}_\mu\overset{(0)}{P}_\alpha\,,
\end{equation}
and uses what we have learnt about the Riemann tensor in appendix \ref{app:bdryRiem}.

Next we consider all the terms on the second row of \eqref{eq:diffeoWIspatialprojapp}. Using what have just derived as well  
\begin{equation}
    T^\mu F_{\mu\kappa}=F^\rho{}_\kappa\mathcal{A}_\rho^B+\frac{1}{2}F^\rho{}_\kappa h^{\alpha\beta}\mathcal{D}_\alpha N_{\beta\rho}\,,
\end{equation}
we find
\begin{align}
    &\frac{1}{4}h_{\kappa\sigma}\mathcal{D}_\mu\left(N^{\mu\lambda}C_\lambda{}^\sigma-N^{\sigma\lambda}C_\lambda{}^\mu\right)-\frac{1}{4}N^{\mu\sigma}h^\nu_\kappa\left(\mathcal{D}_\nu+a_\nu\right) C_{\mu\sigma}+T^\mu F_{\mu\kappa}\nonumber\\
    =&\frac{1}{2}h^\rho_\kappa\left(\partial_\rho+3a_\rho\right)\left(\mathcal{D}_\sigma\overset{(0)}{P}{}^\sigma-\frac{1}{4}N\cdot C\right)-h_{\kappa\nu}\mathcal{D}_\mu\left(h^{\alpha\langle\mu}h^{\nu\rangle\beta}\left(\mathcal{D}_\alpha+3a_\alpha\right)\overset{(0)}{P}_\beta\right)\nonumber\\
    &+\left(\mathcal{L}_v-K\right)\left(\frac{1}{4}C^\rho{}_\kappa h^{\alpha\beta}\left(\mathcal{D}_\alpha-a_\alpha\right)C_{\beta\rho}+2\overset{(0)}{P}{}^\rho F_{\rho\kappa}\right)+\frac{1}{2}h_{\kappa\nu}\left(\mathcal{D}_\mu+2a_\mu\right) R(K)^{\mu\nu}\nonumber\\
    &+\frac{1}{2}C^\rho{}_\kappa h^{\alpha\beta}\mathcal{D}_\alpha N_{\beta\rho}-\frac{1}{8}C^2 G_\kappa+\frac{1}{2}F^\rho{}_\kappa h^{\alpha\beta}\mathcal{D}_\alpha N_{\beta\rho}+\left(\overset{(0)}{S}-a^2\right)\left[\overset{(0)}{P}_\kappa-\frac{1}{2}\mathcal{D}_\mu F^\mu{}_\kappa\right]\nonumber\\
    &+F^\rho{}_\kappa\left[\mathcal{A}_\rho^B-\left(\mathcal{L}_v-\frac{1}{2}K\right)\overset{(0)}{P}_\rho-\frac{1}{2}\left(\partial_\rho+2a_\rho\right)\left(\overset{(0)}{S}-a^2\right)\right]\,,\label{eq:intermedreqritingdiffWI}
\end{align}
where we used
\begin{equation}
    2\overset{(0)}{P}{}^\rho\left(\mathcal{L}_v+\frac{1}{2}K\right)F_{\rho\kappa}=\left(\mathcal{L}_v-K\right)\left(2\overset{(0)}{P}{}^\rho F_{\rho\kappa}\right)-2F^\rho{}_\kappa\left(\mathcal{L}_v-\frac{1}{2}K\right)\overset{(0)}{P}_\rho\,.
\end{equation}
The first two lines on the right-hand side of \eqref{eq:intermedreqritingdiffWI} are of the required form, i.e. either $K$-curvatures or terms that we can absorb into the first line of  \eqref{eq:diffeoWIspatialprojapp}. The terms on the last two lines of \eqref{eq:intermedreqritingdiffWI} still need further work.

There are two expressions for the boost anomaly which we repeat here for convenience
\begin{eqnarray}
    \mathcal{A}^B_\rho & = & \frac{1}{2}h^{\alpha\beta}\mathcal{D}_\alpha N_{\beta\rho}+\frac{1}{2}h^\mu_\rho\left(\partial_\mu+2a_\mu\right)\left(\overset{(0)}{S}-a^2\right)-\frac{1}{2}F^\mu{}_\rho G_\mu\nonumber\\
    &&+\frac{1}{2}\left(\mathcal{L}_v-\frac{1}{2}K\right)\mathcal{D}_\kappa F^\kappa{}_\rho\\
    &= & \left(\mathcal{L}_v-\frac{1}{2}K\right)\overset{(0)}{P}_\rho+\frac{1}{2}h^\mu_\rho\left(\partial_\mu+2a_\mu\right)\left(\overset{(0)}{S}-a^2\right)-\frac{1}{2}F^\mu{}_\rho G_\mu+\frac{1}{2}C^\mu{}_\rho G_\mu\,.\nonumber
\end{eqnarray}
The two are related via the identity
\begin{equation}
    \left(\mathcal{L}_v-\frac{1}{2}K\right)\left(\overset{(0)}{P}_\rho-\frac{1}{2}\mathcal{D}_\kappa F^\kappa{}_\rho\right)=-\frac{1}{2}C^\mu{}_\rho G_\mu+\frac{1}{2}h^{\alpha\beta}\mathcal{D}_\alpha N_{\beta\rho}\,,
\end{equation}
which is just a result of commuting a covariant derivative with a Lie derivative along $v^\mu$. We now use these two forms of the anomaly to rewrite the last two lines of \eqref{eq:intermedreqritingdiffWI} as
\begin{align}
    &+\frac{1}{2}C^\rho{}_\kappa h^{\alpha\beta}\mathcal{D}_\alpha N_{\beta\rho}-\frac{1}{8}C^2 G_\kappa+\frac{1}{2}F^\rho{}_\kappa h^{\alpha\beta}\mathcal{D}_\alpha N_{\beta\rho}+\left(\overset{(0)}{S}-a^2\right)\left[\overset{(0)}{P}_\kappa-\frac{1}{2}\mathcal{D}_\mu F^\mu{}_\kappa\right]\nonumber\\
    &+F^\rho{}_\kappa\left[\mathcal{A}_\rho^B-\left(\mathcal{L}_v-\frac{1}{2}K\right)\overset{(0)}{P}_\rho-\frac{1}{2}\left(\partial_\rho+2a_\rho\right)\left(\overset{(0)}{S}-a^2\right)\right]\nonumber\\
    =&C^\rho{}_\kappa\mathcal{A}^B_\rho -\frac{1}{2}h_{\kappa\beta}\mathcal{D}_\alpha\left[\left(\overset{(0)}{S}-a^2\right)C^{\alpha\beta}\right]+\left(\mathcal{L}_v-K\right)\left(-\frac{1}{2}C^\rho{}_\kappa\mathcal{D}_\sigma F^\sigma{}_\rho\right)\nonumber\\
    &+\frac{1}{4}F^2 G_\kappa-\frac{1}{8}C^2 G_\kappa-\frac{1}{2}N^\rho{}_\kappa\mathcal{D}_\sigma F^\sigma{}_\rho+\frac{1}{2}F^\rho{}_\kappa h^{\alpha\beta}\mathcal{D}_\alpha N_{\beta\rho}\,,
\end{align}
where we used
\begin{equation}
    F^\rho{}_\kappa C^\mu{}_\rho G_\mu+C^\rho{}_\kappa F^\mu{}_\rho G_\mu=0\,,
\end{equation}
as well as
\begin{equation}
    -\frac{1}{2}C^\rho{}_\kappa\left(\mathcal{L}_v-\frac{1}{2}K\right)\mathcal{D}_\sigma F^\sigma{}_\rho=\left(\mathcal{L}_v-K\right)\left(-\frac{1}{2}C^\rho{}_\kappa\mathcal{D}_\sigma F^\sigma{}_\rho\right)-\frac{1}{2}N^\rho{}_\kappa\mathcal{D}_\sigma F^\sigma{}_\rho\,.
\end{equation}
This allows us to write \eqref{eq:intermedreqritingdiffWI} as
\begin{align}
    &\frac{1}{4}h_{\kappa\sigma}\mathcal{D}_\mu\left(N^{\mu\lambda}C_\lambda{}^\sigma-N^{\sigma\lambda}C_\lambda{}^\mu\right)-\frac{1}{4}N^{\mu\sigma}h^\nu_\kappa\left(\mathcal{D}_\nu+a_\nu\right) C_{\mu\sigma}+T^\mu F_{\mu\kappa}\nonumber\\
    =&\frac{1}{2}h^\rho_\kappa\left(\partial_\rho+3a_\rho\right)\left(\mathcal{D}_\sigma\overset{(0)}{P}{}^\sigma-\frac{1}{4}N\cdot C\right)\nonumber\\
    &-h_{\kappa\nu}\mathcal{D}_\mu\left(h^{\alpha\langle\mu}h^{\nu\rangle\beta}\left(\mathcal{D}_\alpha+3a_\alpha\right)\overset{(0)}{P}_\beta+\frac{1}{2}\left(\overset{(0)}{S}-a^2\right)C^{\mu\nu}\right)\nonumber\\
    &+\left(\mathcal{L}_v-K\right)\left(\frac{1}{4}C^\rho{}_\kappa h^{\alpha\beta}\left(\mathcal{D}_\alpha-a_\alpha\right)C_{\beta\rho}+2\overset{(0)}{P}{}^\rho F_{\rho\kappa}-\frac{1}{2}C^\rho{}_\kappa\mathcal{D}_\sigma F^\sigma{}_\rho\right)\nonumber\\
    &+\frac{1}{2}h_{\kappa\nu}\left(\mathcal{D}_\mu+2a_\mu\right) R(K)^{\mu\nu}+C^\rho{}_\kappa\mathcal{A}^B_\rho \nonumber\\
    &+\frac{1}{4}F^2 G_\kappa-\frac{1}{8}C^2 G_\kappa-\frac{1}{2}N^\rho{}_\kappa\mathcal{D}_\sigma F^\sigma{}_\rho+\frac{1}{2}F^\rho{}_\kappa h^{\alpha\beta}\mathcal{D}_\alpha N_{\beta\rho}\,.\label{eq:intermedreqritingdiffWI2}
\end{align}
Note that the three lines on the right-hand side are all of the form of terms we encounter on the first line of \eqref{eq:diffeoWIspatialprojapp} whereas as the fourth line consists of $K$-curvatures. 
It is left to rewrite the last line of this expression such that it can be included in these two types of terms. 

Using \eqref{eq:DSTFid} we can show that
\begin{equation}
    -\frac{1}{2}h_{\kappa\nu}\mathcal{D}_\mu\left(F^{\mu\rho}N_\rho{}^\nu\right)=-\frac{1}{2}N^\rho{}_\kappa\mathcal{D}_\sigma F^\sigma{}_\rho+\frac{1}{2}F^\rho{}_\kappa h^{\alpha\beta}\mathcal{D}_\alpha N_{\beta\rho}\,.
\end{equation}
Finally, the $F^2 G_\kappa$ and $C^2 G_\kappa$ terms can be dealt with using the following identity
\begin{equation}
    h^\rho_\kappa\left(\partial_\rho+3a_\rho\right)\left(\mathcal{L}_v-K\right)X-\left(\mathcal{L}_v-K\right)\left(h^\rho_\kappa\left(\partial_\rho+2a_\rho\right)X\right)=2XG_\kappa\,,
\end{equation}
where $X$ is any function (with Weyl weight $-2$). We then apply this to $X=F^2$ and $C^2$. Note that $\left(\mathcal{L}_v-K\right)C^2=-2N\cdot C$. In the end we obtain for \eqref{eq:intermedreqritingdiffWI2} the following
\begin{align}
    &\frac{1}{4}h_{\kappa\sigma}\mathcal{D}_\mu\left(N^{\mu\lambda}C_\lambda{}^\sigma-N^{\sigma\lambda}C_\lambda{}^\mu\right)-\frac{1}{4}N^{\mu\sigma}h^\nu_\kappa\left(\mathcal{D}_\nu+a_\nu\right) C_{\mu\sigma}+T^\mu F_{\mu\kappa}\nonumber\\
    =&\frac{1}{2}h^\rho_\kappa\left(\partial_\rho+3a_\rho\right)\left(\mathcal{D}_\sigma\overset{(0)}{P}{}^\sigma+\frac{1}{4}\left(\mathcal{L}_v-K\right)F^2\right)\nonumber\\
    &-h_{\kappa\nu}\mathcal{D}_\mu\left(h^{\alpha\langle\mu}h^{\nu\rangle\beta}\left(\mathcal{D}_\alpha+3a_\alpha\right)\overset{(0)}{P}_\beta+\frac{1}{2}\left(\overset{(0)}{S}-a^2\right)C^{\mu\nu}+\frac{1}{2}F^{\mu\rho}N_\rho{}^\nu\right)\nonumber\\
    &+\left(\mathcal{L}_v-K\right)\left(\frac{1}{4}C^\rho{}_\kappa h^{\alpha\beta}\left(\mathcal{D}_\alpha-a_\alpha\right)C_{\beta\rho}+2\overset{(0)}{P}{}^\rho F_{\rho\kappa}-\frac{1}{2}C^\rho{}_\kappa\mathcal{D}_\sigma F^\sigma{}_\rho\right.\nonumber\\
    &\left.-\frac{1}{8}h^\rho_\kappa\left(\partial_\rho+2a_\rho\right)F^2+\frac{1}{16}h^\rho_\kappa\left(\partial_\rho+2a_\rho\right)C^2\right)\nonumber\\
    &+\frac{1}{2}h_{\kappa\nu}\left(\mathcal{D}_\mu+2a_\mu\right) R(K)^{\mu\nu}+C^\rho{}_\kappa\mathcal{A}^B_\rho\,.\label{eq:intermedreqritingdiffWI3}
\end{align}

Next consider the 3rd and 4th row, in particular the term in parenthesis that is acted on by $\mathcal{L}_v-K$. We will rewrite this as follows. First using the definition of $\overset{(0)}{P}_\mu$ we can write
\begin{equation}
    \frac{1}{4}C^\rho{}_\kappa h^{\alpha\beta}\left(\mathcal{D}_\alpha-a_\alpha\right)C_{\beta\rho}=-\frac{1}{2}C^\rho{}_\kappa\overset{(0)}{P}_\rho+\frac{1}{4}C^\rho{}_\kappa\mathcal{D}_\sigma F^\sigma{}_\rho\,.
\end{equation}
Next, using \eqref{eq:ASSTF} and \eqref{eq:DSTFid} we can show that 
\begin{equation}
    C^\rho{}_\kappa\mathcal{D}_\sigma F^\sigma{}_\rho=-2F^\rho{}_\kappa\overset{(0)}{P}_\rho+F^\rho{}_\kappa\mathcal{D}_\sigma F^\sigma{}_\rho-h^{\sigma\nu}\mathcal{D}_\sigma\left(C_\kappa{}^\rho F_{\rho\nu}\right)
\end{equation}
Finally,using \eqref{eq:DASA} and \eqref{eq:AtildeA} we can show that
\begin{equation}
    F^\rho{}_\kappa\mathcal{D}_\sigma F^\sigma{}_\rho=-\frac{1}{4}h^\rho_\kappa\left(\partial_\rho+2a_\rho\right)F^2\,.
\end{equation}
These three results allow us to write the terms in parenthesis that are acted upon by $\mathcal{L}_v-K$ as
\begin{align}
    &\frac{1}{4}C^\rho{}_\kappa h^{\alpha\beta}\left(\mathcal{D}_\alpha-a_\alpha\right)C_{\beta\rho}+2\overset{(0)}{P}{}^\rho F_{\rho\kappa}-\frac{1}{2}C^\rho{}_\kappa\mathcal{D}_\sigma F^\sigma{}_\rho+\frac{1}{16}h^\rho_\kappa\left(\partial_\rho+2a_\rho\right)\left(C^2-2F^2\right)\nonumber\\
    =&-\frac{1}{2}C^\rho{}_\kappa\overset{(0)}{P}_\rho+\frac{5}{2}F^\rho{}_\kappa\overset{(0)}{P}_\rho-\frac{1}{16}h^\rho_\kappa\left(\partial_\rho+2a_\rho\right)\left(F^2-C^2\right)+\frac{1}{4}h^{\sigma\nu}\mathcal{D}_\sigma\left(C_\kappa{}^\rho F_{\rho\nu}\right)\,.\label{eq:Lv-Kterms}
\end{align}

Using \eqref{eq:intermedreqritingdiffWI3} and \eqref{eq:Lv-Kterms} we can write \eqref{eq:diffeoWIspatialprojapp} as
\begin{eqnarray}
    0 & = &-\left(\mathcal{L}_v-K\right)\left(P_\kappa+\frac{1}{2}C^\rho{}_\kappa\overset{(0)}{P}_\rho-\frac{5}{2}F^\rho{}_\kappa\overset{(0)}{P}_\rho+\frac{1}{16}h^\rho_\kappa\left(\partial_\rho+2a_\rho\right)\left(F^2-C^2\right)-\frac{1}{4}h^{\sigma\nu}\mathcal{D}_\sigma\left(C_\kappa{}^\rho F_{\rho\nu}\right)\right)\nonumber\\
    &&+ h_{\kappa\nu}\mathcal{D}_\mu\left(\tilde {T}^{\mu\nu}-h^{\alpha\langle\mu}h^{\nu\rangle\beta}\left(\mathcal{D}_\alpha+3a_\alpha\right)\overset{(0)}{P}_\beta-\frac{1}{2}\left(\overset{(0)}{S}-a^2\right)C^{\mu\nu}-\frac{1}{2}F^{\mu\rho}N_\rho{}^\nu\right)\nonumber\\
    &&-\frac{1}{2}h^\mu_\kappa\left(\partial_\mu+3a_\mu\right)\left(\tau_\rho T^\rho-\frac{1}{4}N^{\rho\sigma}C_{\rho\sigma}-\mathcal{D}_\sigma\overset{(0)}{P}{}^\sigma-\frac{1}{4}\left(\mathcal{L}_v-K\right)F^2\right)\nonumber\\
    &&+\frac{1}{2}h_{\kappa\nu}\left(\mathcal{D}_\mu+2a_\mu\right) R(K)^{\mu\nu}+C^\rho{}_\kappa\mathcal{A}^B_\rho\,,\label{eq:spatialDiffWIintermed}
\end{eqnarray}
where we used the Weyl Ward identity \eqref{eq:WeylWI} to write
\begin{align}
    &T^{\rho\sigma}h_{\rho\sigma}+\frac{1}{2}N^{\rho\sigma}C_{\rho\sigma}+\mathcal{D}_\sigma\overset{(0)}{P}{}^\sigma+\frac{1}{4}\left(\mathcal{L}_v-K\right)F^2\nonumber\\
    =&-\tau_\rho T^\rho+\frac{1}{4}N^{\rho\sigma}C_{\rho\sigma}+\mathcal{D}_\sigma\overset{(0)}{P}{}^\sigma+\frac{1}{4}\left(\mathcal{L}_v-K\right)F^2\,,
\end{align}
which is the combination of terms that appears on the 3rd line of \eqref{eq:spatialDiffWIintermed}.

Next we define
\begin{eqnarray}
    P'_\kappa & = & P_\kappa+\frac{1}{2}C^\rho{}_\kappa\overset{(0)}{P}_\rho-\frac{5}{2}F^\rho{}_\kappa\overset{(0)}{P}_\rho+\frac{1}{16}h^\rho_\kappa\left(\partial_\rho+2a_\rho\right)\left(F^2-C^2\right)-\frac{1}{4}h^{\sigma\nu}\mathcal{D}_\sigma\left(C_\kappa{}^\rho F_{\rho\nu}\right)\,,\nonumber\\
    &&\\
    \tilde {T}'^{\mu\nu} & = & \tilde {T}^{\mu\nu}-h^{\alpha\langle\mu}h^{\nu\rangle\beta}\left(\mathcal{D}_\alpha+3a_\alpha\right)\overset{(0)}{P}_\beta-\frac{1}{2}\left(\overset{(0)}{S}-a^2\right)C^{\mu\nu}-\frac{1}{2}F^{\mu\rho}N_\rho{}^\nu\,,\\
    \tau_\rho T'^\rho & = & \tau_\rho T^\rho-\frac{1}{4}N^{\rho\sigma}C_{\rho\sigma}-\mathcal{D}_\sigma\overset{(0)}{P}{}^\sigma-\frac{1}{4}\left(\mathcal{L}_v-K\right)F^2\,,
\end{eqnarray}
so that the angular momentum Ward Identity now reads
\begin{eqnarray}
    0 & = &-\left(\mathcal{L}_v-K\right)P'_\kappa+ h_{\kappa\sigma}\mathcal{D}_\mu\tilde {T}'^{\mu\sigma}-\frac{1}{2}h^\mu_\kappa\left(\partial_\mu+3a_\mu\right)\left(\tau_\rho T'^\rho\right)\nonumber\\
    &&+\frac{1}{2}h_{\kappa\nu}\left(\mathcal{D}_\mu+2a_\mu\right) R(K)^{\mu\nu}+C^\rho{}_\kappa\mathcal{A}^B_\rho\,.\label{eq:spatialDiffWInewformapp}
\end{eqnarray}
In here we have
\begin{eqnarray}
    P'_\kappa & = & \frac{3}{2}\overset{(1)}{P}_\kappa-\frac{3}{32}a_\kappa\left(F^2-C^2\right)+\frac{3}{32}h^\sigma_\kappa\left(\partial_\sigma+2a_\sigma\right)\left(F^2-C^2\right)\nonumber\\
    &&-a_\sigma D^\sigma{}_{\kappa}+\frac{1}{2}h^{\sigma\nu}\mathcal{D}_\sigma\tilde\chi_{\nu\kappa}\,,\\
    \tilde {T}'^{\mu\nu} & = & \frac{1}{2}K D^{\mu\nu}+\frac{1}{2}h^{\mu\rho}h^{\nu\sigma}\mathcal{L}_v\tilde\chi_{\rho\sigma}\,,\\
    \tau_\rho T'^\rho & = & -v^\rho v^\sigma g^{(1)}_{\rho\sigma}-\frac{1}{16}K\left(F^2-C^2\right)-\frac{1}{8}\left(\mathcal{L}_v-K\right)F^2-\frac{1}{4}C\cdot N\,,
\end{eqnarray}
where we defined
\begin{equation}
    \tilde\chi_{\rho\sigma}=\chi_{\rho\sigma}-\frac{1}{2}C_\rho{}^\alpha F_{\alpha\sigma}\,.
\end{equation}
The $\tilde\chi_{\mu\nu}$ terms drop out of the diffeomorphism Ward Identity \eqref{eq:spatialDiffWInewformapp}. This is the final result.

\newpage
\bibliographystyle{JHEP}
\bibliography{Masterbibliography}

\end{document}